\documentclass[%
 reprint,
 amsmath,amssymb,
 aps,
 prb,
floatfix,
]{revtex4-2}

\usepackage{graphicx}
\usepackage{dcolumn}
\usepackage{bm}
\usepackage{siunitx}
\usepackage{amsmath}
\usepackage{multirow}
\newcommand*{\rom}[1]{\expandafter\@slowromancap\romannumeral #1@}

\begin{document}

\newcommand{\jwn}[1]{\color{black}#1 \color{black}}
\newcommand{\jwnresolved}[1]{\color{black}#1 \color{black}}
\newcommand{\jwncom}[1]{\color{black}[#1]\color{black}}
\newcommand{\remove}[1]{}
\newcommand{\ts}{\textsuperscript}
\newcommand{\dac}[1]{\color{black}#1 \color{black}}
\preprint{APS/123-QED}

\title{Three-dimensional receptivity of hypersonic sharp and blunt cones to free-stream planar waves using hierarchical input-output analysis}

\author{David A. Cook}
 \email{Corresponding author: cookx894@umn.edu}
\author{Joseph W. Nichols}%

\affiliation{%
 Department of Aerospace Engineering and Mechanics, University of Minnesota, Minneapolis, MN 55414, USA
}%


\date{\today}

\begin{abstract}
Understanding the receptivity of hypersonic flows to free-stream disturbances is crucial for predicting laminar to turbulent boundary layer transition. Input-output analysis as a receptivity tool considers which free-stream disturbances lead to the largest response from the boundary layer using the global linear dynamics. Two technical challenges are addressed. First, we extend recent work by Kamal et al.~\cite{Kamal2023} and restrict the allowable forcing to physically realizable inputs via a free-stream boundary modification to the classic input-output formulation. Second, we develop a hierarchical input-output (H-IO) analysis which allows us to solve the three-dimensional problem at a fraction of the computational cost otherwise associated with directly inverting the fully three-dimensional resolvent operator. Next, we consider Mach 5.8 flows over a sharp cone and \dac{two blunt cones with 3.6 \unit{mm} and 7.2 \unit{mm} spherically blunt tips.} H-IO correctly predicts that the sharp cone boundary layer is most receptive to slow acoustic waves at an optimal incidence angle of \ang{10}, validating the method. We then investigate the effect of free-stream disturbances on the blunt cone boundary layer, and identify two distinct vorticity-dominated receptivity mechanisms for the oblique first mode instability at 10 \unit{\kilo\hertz} and an entropy layer instability at \dac{40 and }70 \unit{\kilo\hertz}.  Our results reveal these receptivity processes to be highly three-dimensional in nature, involving both the nose-tip and excitation along narrow bands at certain azimuthal angles along the oblique shock downstream. We interpret these processes in terms of critical angles from linear shock/perturbation interaction theory. \dac{Finally, we show how these novel receptivity processes vary with frequency and nose tip bluntness, and demonstrate how this methodology might be applied to transition prediction from first principles.} 
\end{abstract}

\maketitle

\section{\label{sec:introduction}Introduction}
Laminar to turbulent boundary transition continues to be a critical area of research for \jwnresolved{accurate prediction of} aerodynamic performance during high speed flight. The increased heating and skin friction from turbulence in the boundary layer makes delaying transition, or at least understanding where the transition front will most likely occur, a priority. 

\jwnresolved{It is well known that boundary layer instabilities have significant influence on transition to turbulence.}  Linear stability analysis \cite{Mack1984} decomposes the \jwnresolved{dynamics of small fluctuations about a } locally parallel flow into wall-normal eigenfunctions, \jwnresolved{or modes,} each of which may grow or decay exponentially downstream according to their eigenvalues. The development of the parabolized equations \cite{Herbert1997} relaxed the parallel assumption such that the mean boundary layer can be treated as slowly growing in the streamwise direction, a good assumption for high speed boundary layers well away from complex geometry and shock waves. \dac{Global methods depart from the local framework and adopt a BiGlobal or TriGrobal framework \cite{Theofilis2011a,Theofilis2011b}, depending on whether the eigenvalue problem or initial value problem is posed in two or three spatial dimensions. Resolvent analysis \cite{Jovanovic2005,Bagheri2009,McKeon2010,Sipp2010} is a type of global analysis which decomposes the global linear dynamics into an orthogonal set of modes (also termed directions) which optimally describe the linear growth of the harmonic linearized equations. A special case of resolvent analysis---input-output analysis---specifies input and output maps to restrict the allowable types of forcing and responses. This is most often done by restriction to subspaces of the state space \cite{Jeun2016,Cook2018,Cook2019,Cook2022}. Extension of input-output analysis to compressible flows and hypersonic flows has also shed valuable insight on the worst case linear instabilities which may exist in the flow \cite{Barret2018,Hildebrand2019,Jeun2018,Dwivedi2020b}. Recent work by Kamal et al. restricted the input to physically realizable forcing types, for example, free-stream planar waves \cite{Kamal2023}, has made made input-output analysis a more useful tool to study the receptivity of the boundary layer to free-stream disturbances that could occur naturally in a wind tunnel or atmospheric environment. Incorporation of the receptivity process into stability analysis is a critical step to move away from empiricism and toward a first-principles approach to transition prediction. }

Since Stetson's seminal experiments \cite{Stetson1983}, much attention has been focused on the effect of nose bluntness on transition over conical test articles. Small nose-tip bluntness delays transition when compared to sharp articles, and in this regime, increasing bluntness leads to increasing transition delay.  At some point, \jwnresolved{however,} the trend reverses and an increase in nose-tip bluntness causes the transition front to revert upstream. Analysis of the Stetson boundary layers via modal methods successfully predicted transition delay, but not transition reversal \cite{Jewell2017,Malik1990,Rufer2005,Robarge2005,Li2008,Lei2012,Zhong2012}. \dac{The delay of transition from nose-tip bluntness arises from the generation of an entropy layer by the blunt tip. Tip bluntness creates a strong curved bow shock and a region of rotational, high entropy fluid in the inviscid flow region above the leading edge boundary layer. This entropy layer persists for a streamwise extent before it is swallowed by the slowly growing boundary layer. The entropy layer swallowing length \cite{Rotta1966,Han2022} is a key feature for assessing the stability of boundary layer with blunt tips or leading edges. Early analysis of entropy layer instabilities \cite{Dietz1999,Fedorov1990} showed that entropy layer instabilities in compressible flows over blunted plates were dominated by temperature and density fluctuations. These fluctuations amplify in the inviscid region, outside the boundary layer, along the generalized inflection point. Further analysis showed that while linear stability theory could predict an entropy layer instability, its amplification rate was very small \cite{Fedorov2004}. The identification of non-modal growth as a possible mechanism has been investigated \cite{Paredes2019}, and has demonstrated---via optimal growth methods---that the entropy layer can, under optimal forcing, amplify travelling waves far more than modal analysis predicted. The underlying mechanism was found to strongly resemble the Orr mechanism \cite{Orr1907a,Orr1907b}, a well-known non-modal growth mechanism in low-speed shear flows \cite{Butler1992,Schmid2007}. The receptivity of these travelling structures to free-stream disturbances, however, was not addressed. The same mechanisms were also identified and connected to the receptivity process using the input-output framework \cite{Cook2022}. The worst-case free-stream disturbances were found to impinge on the shock in a compact region above the entropy layer, generating entropy and vorticity waves post-shock. Furthermore, it was found that the entropy layer could amplify these axisymmetric disturbances by an order of magnitude via a rotation and deceleration mechanism as the disturbances convected downstream.}

In order for transition models to be predictive, \jwnresolved{it is imperative to understand} the receptivity of high-speed boundary layers. In other words, \jwnresolved{a predictive model must} meaningfully connect realistic disturbance environments (e.g., wind tunnels, atmospheric turbulence) to the initiation of modal mechanisms (if they exist), as well as other non-modal growth mechanisms in the boundary layer. Early receptivity studies of the compressible boundary layer showed a compelling connection between slow acoustic waves in the boundary layer edge vicinity and the efficient destabilization of boundary layer modes~\cite{Ma2003a,Ma2003b,Fedorov2001}. Synchronization of the fast and slow acoustic boundary layer modes destabilizes the slow acoustic mode at the upstream neutral point, initiating exponential growth as the mode resonates between critical layers in the boundary layer \jwnresolved{(this is the well-known Mack second-mode instability~\cite{Mack1963,Mack1984,Malik1990,Malik1991}).} For flows over flat plates and sharp cones, this work has been extended to predict that slow acoustic waves in the free-stream, outside of the shock, are the most important waves for activation of the Mack mode. However, free-stream vortical waves also play a role, and can also activate modal boundary layer growth~\cite{Ma2005,Balakumar2011,Balakumar2015}. \jwnresolved{Recent experiments involving Schlieren imaging of Mach 6 flow over ogive-cylinder geometries revealed the presence of low-frequency instabilities in addition to the high-frequency Mack 2\ts{nd} mode~\cite{Hill2021}.  PSE calculations suggested these observations could be explained by oblique first-mode instability~\cite{Scholten2022}, although this is difficult to confirm using Schlieren imaging alone. The PSE calculations, however, do not take into account the presence of the shock nor the receptivity to the free-stream, and consider growth only on the cylinder portion of the geometry, downstream of the ogive and nose-tip.  To include the effect of the shock wave and its receptivity, axisymmetric I/O analysis of sharp and blunt cones at Mach 6 revealed the presence of a new type of low-frequency instability that depended on acoustic reflection between the boundary layer at the surface of the cone and the underside of the shock~\cite{Cook2022}, providing an alternate explanation of the experimental observations.} \dac{While much of the previous work has been focused on axisymmetric disturbances, recent work by Buchta and Zaki showed that two-dimensional waves are not sufficient to interpret experimental measurements, and that three-dimensional waves must be included \cite{Buchta2022}. Therefore, a predictive model must also include the effects of three-dimensionality in the disturbance field.}

\dac{Inclusion of three-dimensionality dramatically increases the computational cost associated with global methods, including input-output analysis, especially in terms of overhead memory. One approach is to use iterative methods, which significantly relax the large overhead memory requirements at the cost of several iterative steps. One class of matrix-free iterative methods, time-stepping methods \cite{Monokrousos2010,Martini2020}, has been successfully applied within the resolvent analysis framework. These methods do not require explicit formulation of the discrete matrix problem, and use far less memory and computational time for test problems and low-speed flow applications \cite{Martini2021,Farghadan2023}, especially for low-frequency dynamics of systems for which there is a large separation in gain between the leading and sub-optimal resolvent modes. For problems without these features, time-stepping methods can be expensive due to the CFL condition and multiple power iterations required to resolve gains without a large separation. 

Another approach is to use direct methods are built on the explicit formation of the discrete operator, which is then directly factored, usually by $LU$ decomposition. Once these factors are computed, the resolvent action can be efficiently applied to a vector by two back-solve operations with the lower and upper factors. Direct methods are very efficient for one-dimensional and small two-dimensional problems, but scale poorly with respect to computational time and overhead memory requirements as the number of discrete degrees of freedom increase. One of the more commonly used direct solvers, the MUMPS package \cite{Amestoy2001,Amestoy2019} has been developed in order to exploit sparsity and domain decomposition in the computation of the direct solution. One included feature of this software is the Block Low-Rank (BLR), which can reduce the overhead memory footprint. Hierarchical methods built on exploiting low-rank behavior are a promising approach because they are designed to use memory efficiently to reduce computational cost, while still avoiding the full direct matrix factorization. }

In this paper, we apply I/O analysis to the interaction of fully three-dimensional free-stream disturbances to sharp and blunt cones at Mach 5.8.  To overcome the computational expense of fully 3D I/O analysis, we have developed a new ``hierarchical I/O analysis" methodology that has enabled the calculations described below. Importantly, our approach accurately captures the interaction of free-stream disturbances with the shock wave using a shock-kinematic boundary condition (SKBC)~\cite{Cook2022} so that transmission and reflection amplitudes match that of theory~\cite{McKenzie1968}.  The SKBC, which we extend in this paper to three-dimensional interactions, bears some similarity to shock-fitting methods~\cite{Rawat2010}, although it takes advantage of the separation between baseflow and perturbation quantities in keeping with theory~\cite{McKenzie1968}.  While shock-capturing methods may be efficient at computing steady baseflows, they are problematic for delicate receptivity calculations because they are known to generate spurious waves when unsteady perturbations interact with the shock~\cite{Melander2021,Melander2022,Johnston2022}.  Lastly, in order to study sensitivity to realizable free-stream disturbances (which satisfy the free-stream dispersion relation), we follow the approach of Kamal et al.~\cite{Kamal2023} and restrict the allowable inputs to our calculations to correspond to a decomposition of acoustic, vortical and entropic plane-waves at various angles of incidence.  Besides providing a realizable basis onto which realistic free-stream disturbance fields can be projected, I/O analysis starting from the plane-wave decomposition allows us to identify to which types of waves the cone is most receptive and at which angles.  The interaction of oblique plane waves with a round cone is a fully three-dimensional problem, and as we will see, reveals the importance of highly three-dimensional receptivity processes.

The remainder of this paper is organized as follows. In \S \ref{sec:METHODOLOGY}, we outline the governing equations, adapt the realizable input/output analysis framework \cite{Kamal2023} for receptivity to realizable disturbances in the free-stream, describe a \jwnresolved{hierarchical} approach to efficiently computing 3D I/O analysis, and comment on the numerical methods used. In \S \ref{sec:RESULTS}, we present a verification case using a sharp cone boundary layer and \dac{extend the method to blunt-tipped cones over a range of frequencies, highlighting novel insights into the receptivity process.} \dac{Finally, in \S \ref{sec:CONCLUSION}, we summarize our findings and outline a path for future work. }
\section{\label{sec:METHODOLOGY} Methodology}
\subsection{\label{sec:compressible_ns}Compressible Navier--Stokes equations with spanwise/azimuthal parameterization}
We begin by solving the axisymmetric compressible Navier--Stokes equations to obtain steady baseflows.  The general form is 
\begin{equation}
   \label{eq: finite_volume_form}
    \frac{\partial U}{\partial t} + \nabla \cdot  (\vec{F_I} + \vec{F_V}) = 0, 
\end{equation}
where $U = [\rho, \rho u, \rho v, \rho w, E]^T$ is the vector of conservative state variables, $\vec{F}_I$, $\vec{F}_V$ are the inviscid and viscous flux vectors, respectively. We \jwnresolved{then} expand the problem in terms of a mean and fluctuating component \dac{in a cylindrical ($x$, $r$, $\theta$) coordinate system
\begin{equation}
   \label{eq: mean_and_fluctuating}
    U(x,r,\theta) = \bar{U}(x,r,\theta)  + \tilde{U} (x,r) e^{i m \theta}, 
\end{equation}}
where $\bar{U}$ is a \jwnresolved{steady} solution to Eq. \ref{eq: finite_volume_form}, and $\tilde{U}$ is a small amplitude perturbation to the mean flow, parameterized with an azimuthal wavenumber $m$. \jwnresolved{Owing to the axisymmetry of the baseflow, substitution of} this expression \jwnresolved{into Eq. \ref{eq: finite_volume_form} decouples azimuthal derivatives according to wavenumber $m$.  In other words, the problem is reduced to a series of two-dimensional $x\text{--}r$ planes, each having a different wavenumber.}  Once we have this parameterization, we extract the resulting global system Jacobian from the numerical solver using complex step differentiation at each discrete wavenumber.

\subsection{Shock-kinematic boundary condition}

In order to study the receptivity of hypersonic flows to free-stream disturbances, perturbations must pass through a vehicle's bow shock. While standard shock-capturing schemes model steady shock waves efficiently, they introduce errors in unsteady perturbation fields interacting with the shock~\cite{Melander2021,Johnston2022}. Owing to the hyperbolic nature of the flow, unsteady numerical errors created at the shock propagate and contaminate the downstream domain. This is especially problematic for boundary layer transition prediction, as instabilities in the downstream flow amplify small disturbances until they eventually cause transition. It is therefore important for these small disturbances to be of physical rather than numerical origin. To avoid errors introduced by standard shock-capturing schemes, the authors have previously developed a shock-kinematic boundary condition (SKBC)~\cite{Cook2022} which ensures shock/perturbation interaction in our calculations remains consistent with theory~\cite{McKenzie1968}. The SKBC is similar to shock-fitting methods~\cite{Zhong2006,Zhong2012}, but with explicit terms added to capture the interaction of the shock with small perturbations.  While our original formulation was valid for two-dimensional and axisymmetric problems, we extend it here to handle the interaction of three-dimensional disturbances with shock waves.
\begin{figure}[htb]
    \centering
    \includegraphics[width=0.45\textwidth]{ 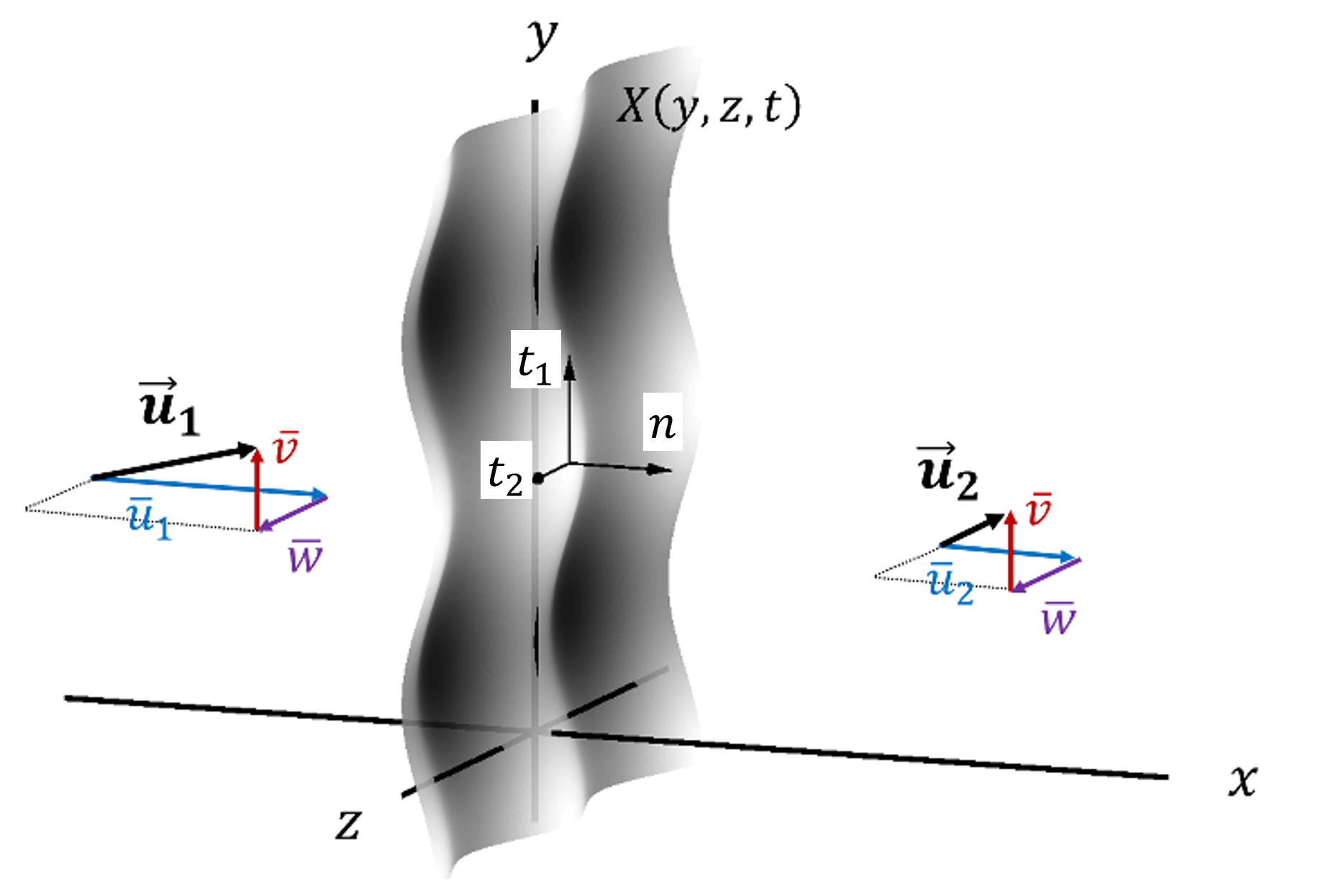}
    \caption{A schematic of the shock-kinematic boundary condition.
The function X(y,z,t) describes the displacement of the shock as
a function of space and time.
    }\label{fig:skbc_schematic}
\end{figure}
\dac{

To model shock/perturbation interaction in a Cartesian frame, consider a stationary shock aligned normal to the $x$-axis, subject to small perturbations, as depicted in Fig.~\ref{fig:skbc_schematic}. The baseflow passes through the shock from left to right, although there may be an oblique component with respect to the $y$ and $z$ directions. For the time-domain problem, the SKBC can be written compactly as
\begin{equation}
    \mathbf{A_3} \Psi = \mathbf{A_1 Q_1} \Phi^+_1 - \mathbf{A_2 Q_2} \Phi^-_2 + \zeta X_y + \beta X_z,
\end{equation}
where $\Psi$ contains the outgoing characteristics with respect to the shock, along with the instantaneous shock velocity. The vectors $\Phi^+_1$ and $\Phi^-_2$ contain those characteristics which are incident upon the shock wave, and the terms $X_y$ and $X_z$ are the local shock inclinations (partial derivatives) with respect to the $y$ and $z$ coordinate directions. The matrices $A_1$, $A_2$, and $A_3$ are built from the linearization of the Rankine--Hugoniot equations in the reference frame of three-dimensionally moving shock, and the matrices $Q_1$ and $Q_2$ perform a change of variables from primitives to characteristics. While the derivation for the two-dimensional case has been previously published by Cook \& Nichols \cite{Cook2022}, the full three-dimensional derivation is provided for completeness in Appendix \ref{AppendixA}.

Practically, disturbances in the pre-shock region near the shock are measured from the pre-shock grid cells and used to compute the shock displacement at each cell face along the shock. Similarly, the post-shock disturbances are used to measure the slow acoustic characteristic incident from the post-shock side. The shock inclinations $X_y$ and $X_z$ can also be measured at each cell along the shock. Together, the SKBC uses these measured quantities to solve for the shock velocity along with the characteristics outgoing from the shock. The shock position is then updated from the shock velocity together with a numerical time-stepping scheme. The outgoing characteristics are added to the post-shock fluxes in the post-shock cell, at which point the regular numerical method of the simulation is used to solve for the remainder of the post-shock cells. One way to understand the SKBC is as a coupled supersonic outflow boundary condition and a subsonic inflow boundary condition with some extra degrees of freedom to store and update the shock position.}

\subsection{\label{sec:inputoutput} Input-output analysis for receptivity}

\dac{
Input-output (I/O) analysis, which is based on resolvent analysis \cite{Jovanovic2005,Bagheri2009,McKeon2010,Sipp2010}, considers the linear gain between arbitrary input forcing and output response with respect to some norm \cite{Dwivedi2020,Jovanovic2021}. Unlike other stability analyses, I/O analysis does not assume anything about the spatial structure of the flow, and so can handle complex flow features in a natural way. In the present flow configuration, the total system response directly incorporates the interaction of perturbations with the bow shock, the stagnation region and strongly accelerating flow around the blunt tip, as well as the rest of the flow downstream on the main body of the cone. Assuming that small disturbances to the flow are time-harmonic, we can obtain the transfer function 
\begin{equation}
\hat{y}(x,r,\theta) = C(-i\omega I - A)^{-1} B \hat{f}= H(\omega)\hat{f}.
\label{eq:transferfunctionform}
\end{equation}
This transfer function maps input into the system via the linear mapping $B$, and measures a system output via the linear mapping $C$. With a proper construction of input/output operators, discretization, and norms, we compute a subset of the singular value decomposition of the transfer function, such that $H(\omega) = U\Sigma V^H$, where the columns of $U$ and $V$ contain unitary basis directions of the outputs and inputs, respectively. The inputs are scaled by $\Sigma$, which contain the gains in descending order. This provides a very natural way for understanding the mechanisms present in the flow as well as the types of forcing to which those mechanisms are receptive. 

In order for this analysis to be both feasible and informative, we have to carefully consider how to structure the inputs and outputs according to the particular questions we want to investigate, as well as consider efficient computational techniques. In the following subsections, we describe (1) the formulation of I/O analysis for receptivity to the physical state versus receptivity to volumetric forcing, (2) proper construction of the input and output matrices to reflect boundary layer receptivity to physically realizable free-stream disturbances, and (3) a hierarchical input-output (H-IO) method for efficiently and feasibly computing the full three-dimensional I/O analysis over axisymmetric flows. 
}

\subsubsection{\label{sec:state_forcing} Receptivity to state vs. receptivity to volumetric forcing}

When constructing I/O analysis for receptivity applications, it is important to carefully consider how to construct the input such that it corresponds to our particular research question. Eq. \ref{eq:transferfunctionform} considers the inputs to be non-linear volumetric forcing terms to the Navier-Stokes equations; however, when considering boundary layer transition applications for I/O analysis, we are less interested in the sensitivity of a flow to the non-linear terms, and more interested in how some spatial regions of the state (e.g., the state of the boundary layer) are sensitive to other regions of the state (e.g., the free-stream disturbance state). In other words, we wish to define our linear gain optimization in a framework in which the input norm has the same units as our output norm, such that the gain between the input and output directions takes on a more physically interpretative meaning. One way to accomplish this goal is to re-frame the governing equations such that the forcing terms are directly summed to the state prior to their multiplication by the Jacobian:
\begin{equation}
\dot{q}(x,r,\theta,t) = A \left(q(x,r,\theta,t) + B f \right).
\label{eq:reframed_gov_eq}
\end{equation}
Distributing the matrix multiplication yields 
\begin{equation}
\dot{q}(x,r,\theta,t) = A q(x,r,\theta,t) + A B f. 
\end{equation}
Another simplification occurs if the time-oscillating forcing terms $Bf$ satisfy the Navier-Stokes equations. After the Fourier transform in time and output measurement, we have $-i \omega B \hat{f} = A B \hat{f}$, and Eq. \ref{eq:transferfunctionform} becomes  
\begin{equation}
\hat{y}(x,r,\theta) = C(-i\omega I - A)^{-1} (-i \omega) B \hat{f} = H(\omega)\hat{f}.
\label{eq:modified_io_form}
\end{equation}

Forcing that satisfies the linearized equations does not occur automatically. Consider the case  where, after discretization, $B$ is a matrix that spatially applies the forcing in the pre-shock region only. Furthermore, let us assume that $B$ ensures that the multiplied forcing term $Bf$ is a superposition of propagating acoustic waves. If we compute the I/O analysis and examine the forcing directions, while the units of the forcing are now consistent with the output response directions and the \textit{forcing terms} are physically realizable, the \textit{free-stream state} that results from this forcing is not, in fact, a superposition of propagating acoustic waves. This happens because the pre-shock state acts as an accumulator for the forcing terms. In this case the problem we have posed is the receptivity to a free-stream acoustic source, which is not the same as receptivity to free-stream acoustic waves. The questions we want to pose is how the flow is receptive to disturbances that already exist in the state, irrespective of their source, e.g., atmospheric turbulence, wind-tunnel acoustics.

One way to accomplish this is to treat the pre-shock forcing terms as a boundary condition instead of volumetric forcing. We define a state-decoupling matrix $D$ such that we remove the rows of $A$ which couple the free-stream degrees of freedom to itself. This can be accomplished by left multiplying $A$ with 
\begin{equation}
D = I_n - B_{d} I_d B_{d}^T, 
\label{state_decoupling_matrix}
\end{equation}
where $I_n$ and $I_d$ are identity matrices of appropriate dimensions. The matrix $B_d$ is the subset of the identity matrix which maps to the free-stream degrees of freedom. With this decoupling accomplished, the final formulation of the I/O problem is 
\begin{equation}
\hat{y}(x,r,\theta) = C(-i\omega I - DA)^{-1} (-i \omega) B \hat{f} = H(\omega)\hat{f}.
\label{eq:modified_io_form_D}
\end{equation}
This applies the forcing term as a boundary condition to the governing equations, which ensures that pre-shock state contains the spatially mapped $Bf$ forcing. In other words, if we construct $B$ such that forcing is a superposition of free-stream waves, the pre-shock state is also a superposition of free-stream waves, applied as a boundary condition with respect to the post-shock flow. 

\subsubsection{\label{sec:io_matrix_construction} Input/output matrix construction}
We will now turn our attention to the construction of the $B$ and $C$ matrices. \dac{Recent work in realizable I/O analysis by Kamal et al. \cite{Kamal2023}, considered a restriction of the input-output formulation to physically realizable inputs, e.g., a superposition of two-dimensional planar waves of various types. They accomplish this by adopting a formalism which allows them to solve for the scattered solution in terms of incident waves. In this section, we adopt this approach, but for three-dimensional waves restricted to a uniform free-stream. In contrast to the scattering formalism approach, we instead consider how the input forcing might be applied as a boundary forcing term in the free-stream. Following Kamal et al. \cite{Kamal2023}, we consider the construction of the matrix $B$ such that the forcing is a superposition of planar waves which satisfy the Euler equations, but restricted to a uniform free-stream only.} We begin by defining our forcing as wave amplitudes for five types of three-dimensional free-stream waves as functions of wave angle $\psi$ with respect to the streamwise direction,
\begin{equation}
\hat{f}(\psi) = [a^-(\psi), a^s(\psi), a^{uv}(\psi), a^w(\psi), a^+(\psi)]^T ,
\label{wave_amplitudes_definition}
\end{equation}
where $a^-$ and $a^+$ are the amplitudes of the slow and fast acoustic waves, $a^s$ are the amplitudes of entropy waves, and $a^{uv}$ and $a^w$ are the vortical waves. Note that we have two relevant coordinate systems. We want to describe the receptivity of an axisymmetric flow ($x$-$r$-$\theta$ coordinates) to 3D free-stream waves ($x$-$y$-$z$ coordinates). In general, we need two wave angles to describe 3D planar waves; however, because the flows we consider are axisymmetric, all 3D wave angles can be mapped onto 2D waves in the flow coordinate system through simple rotation around the symmetry axis, and so we only consider a single free-stream wave angle. 

We can now describe the input matrix as a decomposition of linear mappings from these amplitudes into the state space
\begin{equation}
B\hat{f}(\psi) = P Q S N^{-1} \hat{f}(\psi) , 
\label{Blinmaps}
\end{equation}
where $S$ is the spatial distributor matrix, $Q$ is the wave decomposition matrix, $P$ is the rotation matrix, and $N$ is the normalization matrix. 

The spatial distributor matrix $S$ can be defined as 
\begin{equation*}
    S = \exp{
    \begin{pmatrix}
    i (\mathbf{k}_- \cdot \mathbf{x})  & 0 & 0 & 0 & 0 \\
   0 & i (\mathbf{k}_c \cdot \mathbf{x}) & 0 & 0 & 0 \\
    0 & 0 &i (\mathbf{k}_c \cdot \mathbf{x}) & 0 & 0 \\
    0 & 0 & 0 & i (\mathbf{k}_c \cdot \mathbf{x}) & 0 \\
    0 & 0 & 0 & 0 & i (\mathbf{k}_+ \cdot \mathbf{x})
    \end{pmatrix}},   
\end{equation*}
where the wavenumber vectors $k_{\pm,c} = k_{\pm,c}(\psi)$ must satisfy their corresponding dispersion relations for propagating waves in a uniform flow: $\omega = \mathbf{u} \cdot \mathbf{k}_{\pm} \pm c \left|\mathbf{k}_{\pm}\right|$  for acoustic waves and $\omega = \mathbf{u} \cdot \mathbf{k}_{c}$ for purely convected waves. The matrix $Q$ decomposes the spatially distributed waveform into the proper amplitudes of primitive variables corresponding to different wave types such that $\hat{\phi}(x,r,\theta) = QS\hat{f}(\psi)$, where $\phi = [p, u, v, w, \rho]^T$: 
\begin{equation}
    Q = 
    \begin{pmatrix}
    1 & 0 & 0 & 0 & 1 \\
    \frac{-\cos{\psi}}{\rho a} & 0 & -\sin{\psi} & 0 & \frac{\cos{\psi}}{\rho a}  \\
    \frac{-\sin{\psi}}{\rho a} & 0 & \cos{\psi} & 0 & \frac{\sin{\psi}}{\rho a} \\
    0 & 0 & 0 & 1 & 0 \\
    \frac{1}{a^2} & 1 & 0 & 0 & \frac{1}{a^2}
    \end{pmatrix}.
    \label{eq:Qmatrix}
\end{equation}

At this point, we apply the Chu norm \cite{Chu1965,George2011} as a normalization via such that a unit amplitude in $f(\psi)$ corresponds to wave with a unit energy density. The normalization is applied via $N = L_C^H L_C \delta_{ij}$, where $d_{ij}$ only retains the diagonal terms corresponding to the column-to-column inner products of $L_C$. The Cholesky factor $L_C$ can be written as 
\begin{equation}
L_C =   W_{in} Z Q S, 
\label{normalization matrix}
\end{equation}
where $Z$ is the change of variables matrix accounting for the adjustment from $\hat{\phi}$ variables to $\hat{z}$ variables suitable for the application of the Chu norm: 

\begin{equation}
Z= 
    \begin{pmatrix}
    0 & 0 & 0 & 0 & 1 \\
    0 & 1 & 0 & 0 &0  \\
    0 & 0 & 1 & 0 &0  \\
    0 & 0 & 0 & 1 & 0 \\
    \frac{1}{R_a \bar{\rho}} & 0 & 0 & 0 & \frac{-\bar{p}}{R_a\bar{\rho} ^ 2}
    \end{pmatrix}.
    \label{eq:cov matrix}
\end{equation}
Here, $R_a$ is the gas constant for air, and $\hat{z} = Z \hat{\phi} = [\rho, u, v, w, T]^H$. The matrix $W_{in}$ applies the Chu energy weighting and is given as 
\begin{equation}
W_{in} ^ 2= 
    \frac{\Delta V_{i,j,k}}{V_{in}}
    \begin{pmatrix}
    \frac{a^2}{\gamma  \bar{\rho}} & 0 & 0 & 0 & 0 \\
    0 & \bar{\rho} & 0 & 0 &0  \\
    0 & 0 & \bar{\rho} & 0 &0  \\
    0 & 0 & 0 & \bar{\rho} & 0 \\
    0 & 0 & 0 & 0 & \frac{\bar{\rho} C_v}{\bar{T}}
    \end{pmatrix},
    \label{eq:chu matrix}
\end{equation}
where $\Delta V_{i,j,k}$ is the cell volume quadrature. We also choose the normalization by the total volume of the input region, such that the norm yields the input energy density instead of the input total energy. We prefer this choice, as it gives the I/O gains a more intuitive physical meaning. 

The rotation matrix $P$ is necessary to rotate the velocity components from a Cartesian to a cylindrical coordinate system. Remember that the free-stream waves have been defined in a Cartesian frame ($x$-$y$-$z$). However, the physics and governing equations have been expressed in a cylindrical ($x$-$r$-$\theta$) frame. Therefore, we have to apply a rotation to the $y$ and $z$ components of the velocity such that they become $r$ and $\theta$ components. This is accomplished through application of the rotation matrix
\begin{equation}
P = \begin{pmatrix}
    1 & 0 & 0 & 0 & 0 \\
    0 & 1 & 0 & 0 &0  \\
    0 & 0 & \cos{\theta} & \sin{\theta} &0  \\
    0 & 0 & -\sin{\theta} & \cos{\theta} & 0 \\
    0 & 0 & 0 & 0 & 1
    \end{pmatrix}.
    \label{eq:rotation_matrix}
\end{equation}

With this final matrix definition, we now have a mapping in Eq. \ref{wave_amplitudes_definition} from input amplitudes to three-dimensional free-stream waves, each with unit energy density. The construction of the output matrix $C$ is straightforward, and we take it to be a subset of the identity matrix in a spatial region of interest. The output norm is chosen to mirror the choice of input norm, and is also normalized by the total output volume, e.g., replace $V_{in}$ with $V_{out}$ in Eq. \ref{eq:chu matrix}. The final definition of the gain more explicitly becomes 
\begin{equation}
G^2 = \frac{\langle \hat{y}, \hat{y} \rangle ^2 }{\langle \hat{f}, \hat{f} \rangle ^2} = \frac{\hat{y}^H W_{out} \hat{y}}{\hat{f}(\psi)^H\hat{f}(\psi)}, 
\label{eq:final_gain}
\end{equation}
where the normalization included in the construction of $B$ ensures that a unit two-norm of the input corresponds to a unit energy density of waves in the free-stream. This is a helpful choice of norm for transition related problems and has a clear interpretation. $G>1$ implies an increase in energy density from the input region to the output region and thus there are disturbance amplifying physics. Conversely, $G<1$ implies that the energy density decreases from input to output and the disturbances are spatially damped between input and output location. 

\subsubsection{\label{sec:3D_IO_f-rom_method} Hierarchical I/O analysis}

Theoretically, at this point in the formulation, we could proceed with the analysis via Eq. \ref{eq:modified_io_form_D}. Practically, the three-dimensional problem is computationally expensive. In particular, the inversion of the resolvent operator $R = (-i \omega I - DA)$ is costly. Even for the two-dimensional case, the memory and computational costs of directly factoring the resolvent are high. Furthermore, the resolvent is highly non-normal and poorly conditioned such that iterative methods are slow to converge, even with modern preconditioners. Not only is the 3D matrix much larger, but it is more dense, increasing the computational expense yet further. \dac{We choose to focus solution efforts on solving the linear problem with the explicitly constructed matrix via hierarchical direct methods, for three reasons. Firstly, if they can be constructed, hierarchical methods are computationally more efficient that iterative methods, especially for transition related problems in which there is not large separation in the leading resolvent gains. For these problems, several power iterations are required to converge the leading singular triplets, and so it is important to minimize the cost of applying the resolvent operator to a vector. Secondly, direct matrix methods allow much lower dissipation than time-stepping schemes, which allow higher frequency components to be resolved with less grid points per wavelength. Thirdly, direct methods are not inherently constrained by the CFL condition, which is increasingly important for high speed, high Reynolds number flows, which require small grid cells to resolve very thin boundary layers. }

In many cases, we can examine some of the three-dimensional effects if we parameterise the governing equations using a spanwise/azimuthal wavenumber such that disturbances have the form $\hat{q}(x,r,\theta) = \tilde{q}(x,r) e^{i m \theta}$. Then, if we wish, we can do a separate I/O analysis at each wavenumber to get a sense for how the receptivity is sensitive to different levels of obliquity in the flow disturbances. First, we can discretize the azimuthal coordinate $\theta$ and take the discrete Fourier transform (DFT) of the free-stream forcing with respect to $\theta$
\begin{equation}
\tilde{y}(x,r,m_k) = C(-i\omega I - D\tilde{A})^{-1} \mathcal{F}\bigl\{(-i \omega) B \hat{f} \bigl\}_k.
\label{eq:modified_io_dft}
\end{equation}
Here, the modified Jacobian $\tilde{A}$ has been parameterized by the azimuthal wavenumber, and $\mathcal{F}$ is the unitary discrete Fourier transform. If we take the variable $\zeta_n = -i \omega B_n \hat{f}$ as slices through the 3D forcing field at the discrete azimuthal angle $\theta_n$ then we can express the DFT as 
\begin{equation}
\mathcal{F}\{\cdot\}_k = \frac{1}{\sqrt{N}}\sum_{n=0}^{N-1} \zeta_n \cdot e^{-\frac{2 \pi i}{N} k n}. 
\label{dft_def}
\end{equation}
With Eq.  \ref{eq:modified_io_dft}, we have the option to perform a separate I/O analysis at each discrete wavenumber. However, one drawback of this parameterization is that the I/O analysis at each wavenumber yields different input/output bases for the free-stream and response. We have lost the azimuthal coherence of the free-stream environment in decoupling the global optimization. If we want to reconstruct azimuthally coherent I/O bases, we would need to compute all of the I/O analyses simultaneously so we could invert the Fourier transform and take the full 3D norm. While the Fourier-decoupled problem is cheaper than the full 3D problem, it still requires access to many computational resources simultaneously. 

\begin{figure}[hbt!]
\begin{tabular}{l}
    (a) Step 1: 2-D I/O analysis at each $m_k$ for $k = 1\text{ to }N$\\
   \includegraphics[trim=4 4 4 4, clip, width = 0.45\textwidth]{ 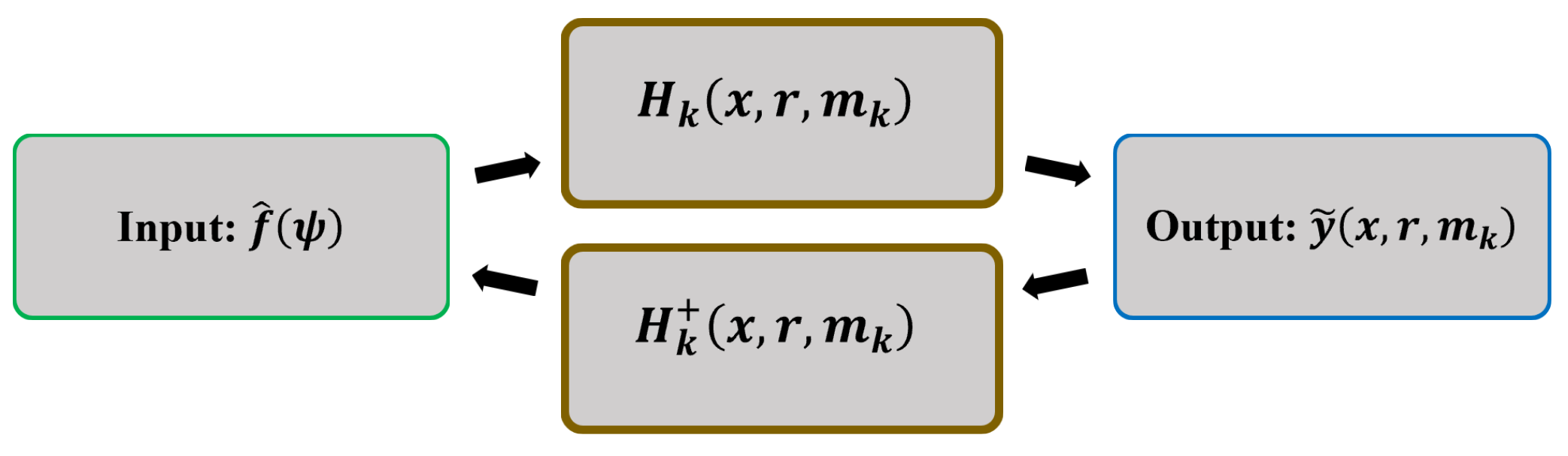} \\
   (b) Step 2: build reduced-order model for each $m_k$\\
   \includegraphics[trim=4 4 4 4, clip, width = 0.45\textwidth]{ 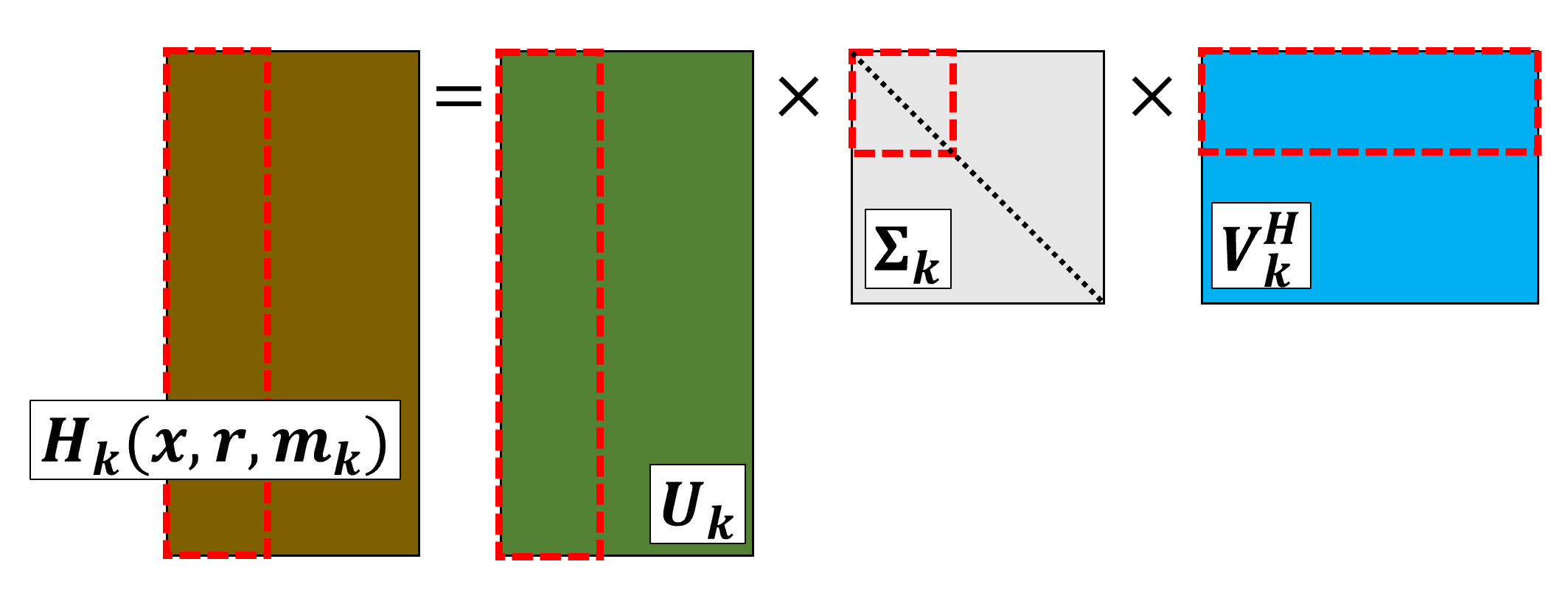} \\
   (c) Step 3: reconstruct 3D reduced-order transfer function\\
   \includegraphics[trim=4 4 4 4, clip, width = 0.45\textwidth]{ 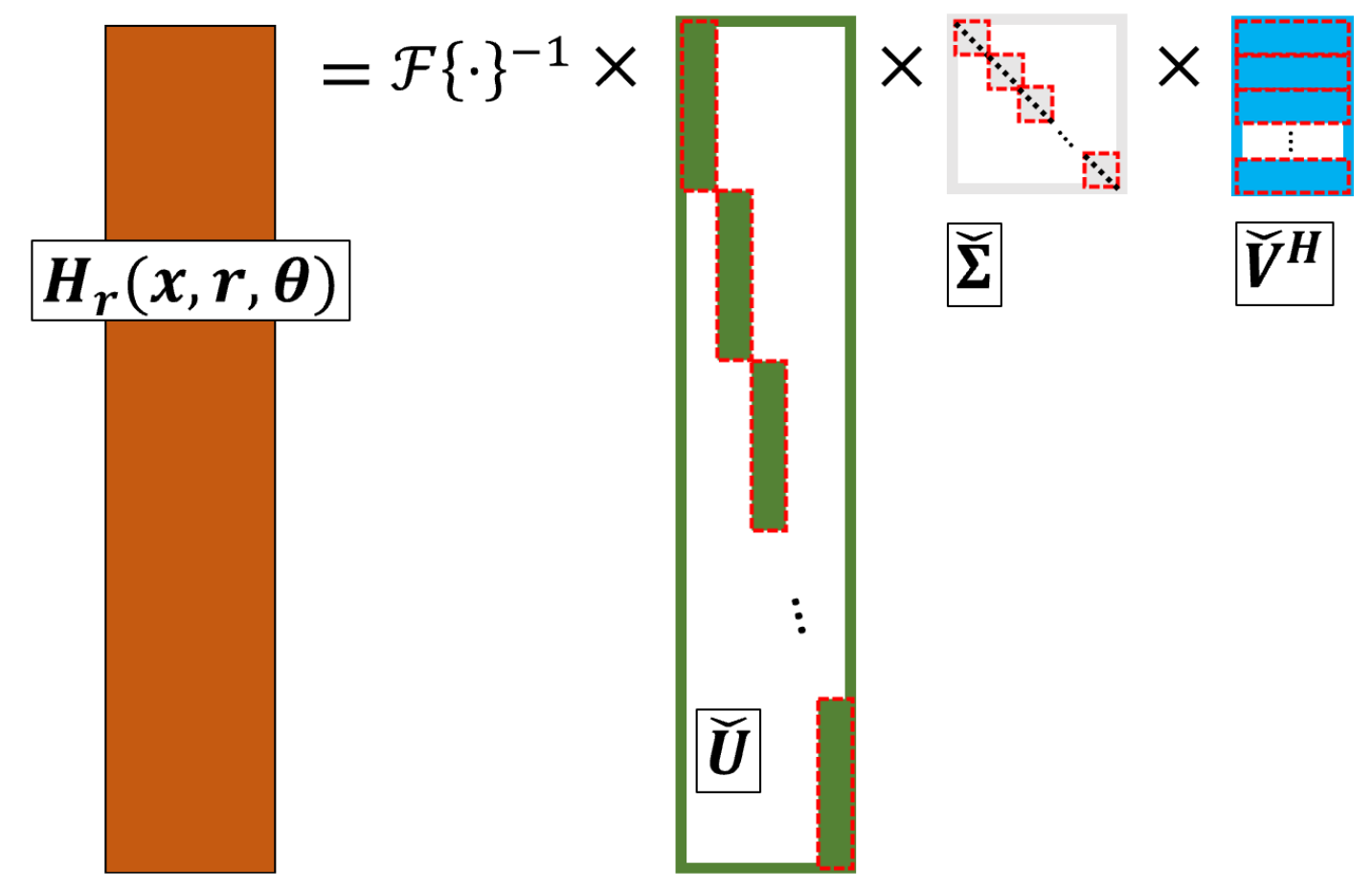} \\
   (d) Step 4: 3D I/O analysis using reconstructed model\\
   \includegraphics[trim=4 4 4 4, clip, width = 0.45\textwidth]{ 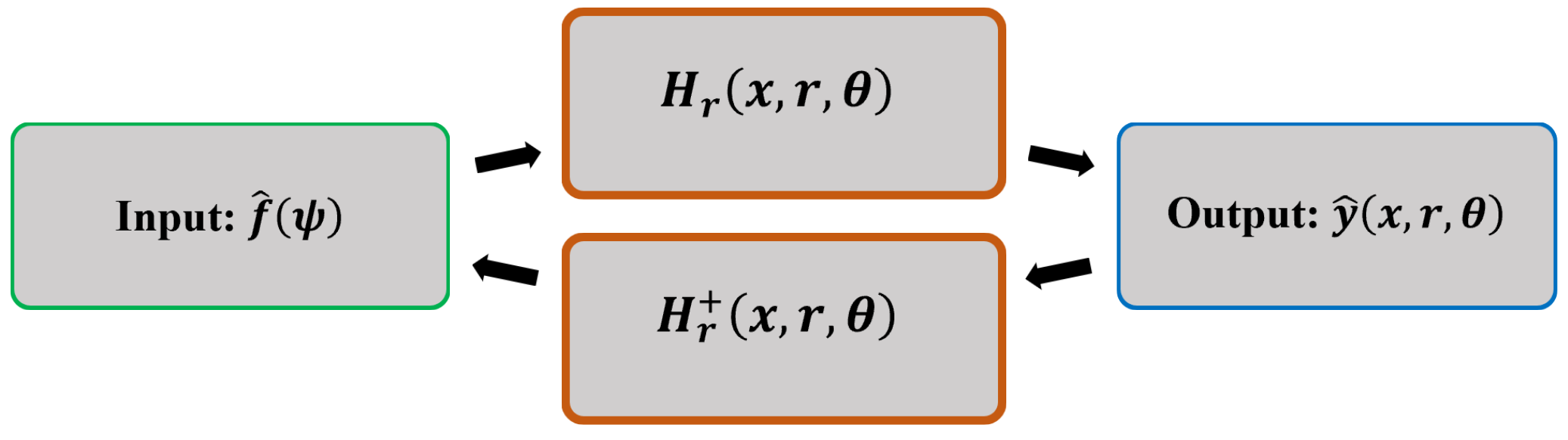} 
\end{tabular}
\caption{\label{fig:F-ROM method} Illustration of the main steps in the hierarchical approach to 3D I/O analysis.}
\end{figure}

Instead we use a reduced-order model based reconstruction to obtain azimuthally coherent I/O directions much more efficiently. The method consists of four steps, with an optional fifth step if we want to examine the full state. These steps are illustrated graphically in Fig. \ref{fig:F-ROM method}.

Step 1, depicted in Figure \ref{fig:F-ROM method}(a), is to compute the I/O analysis of Eq. \ref{eq:modified_io_dft} at each discrete wavenumber. This can be done sequentially or in parallel as computational resources are available; each of these are independent computations. The result of each of these computations is a decomposition in terms of the singular values and singular vectors such that 
\begin{equation}
\tilde{y}(x,r,m_k) = U_k \Sigma_k V_k^H \hat{f}(\psi). 
\label{svd_rom}
\end{equation}

Step 2, depicted in Figure \ref{fig:F-ROM method}(b), is to build a reduced order model from this decomposition. Because we order the singular values such that $\sigma_1 > \sigma_2 > \sigma_3 ...$, the rate at which the gains in $\Sigma$ decay provides a natural way to truncate the SVD and only retain a few I/O pairs (columns of $U_k$, $V_k$) at each wavenumber. If the gains decay sufficiently fast with respect to some error measure, then only a few singular values and singular vectors are needed to accurately reconstruct the physics. We can define the truncated SVD as 
\begin{equation}
\tilde{y}(x,r,m_k) \approx \check{U}_k \check{\Sigma}_k \check{V}_k^H \hat{f}(\psi) = \check{H}_k \hat{f}(\psi).
\label{svd_trunc}
\end{equation}

In Step 3, we can reconstruct the full 3D I/O analysis by vertically concatenating the $\check{V}_k$ blocks, creating block-diagonal matrices using the $\check{U}_k$ and $\check{\Sigma}_k$ blocks, and applying the inverse Fourier transform.  The reconstruction, depicted in Figure \ref{fig:F-ROM method}(c) can be expressed as 
\begin{equation}
\hat{y}(x,r,\theta) = \mathcal{F}^{-1}
    \check{U} \check{\Sigma} \check{V} ^H,
\label{eq:F-ROM}
\end{equation}
where
\begin{equation}
\check{U} = 
    \begin{pmatrix}
    \check{U}_1  & 0 & \cdots &  & \\
    0 & \check{U}_2 & 0 & \cdots &  \\
    \vdots  &  0 & \ddots & 0 & \cdots \\
     & \vdots & 0 & \check{U}_{N - 1} & 0\\
    &  & \vdots & 0 & \check{U}_{N} 
    \end{pmatrix}, 
\label{eq:ucheck}
\end{equation}
\begin{equation}
\check{\Sigma} = 
    \begin{pmatrix}
    \check{\Sigma}_1  & 0 & \cdots &  & \\
    0 & \check{\Sigma}_2 & 0 & \cdots &  \\
    \vdots  &  0 & \ddots & 0 & \cdots \\
     & \vdots & 0 & \check{\Sigma}_{N - 1} & 0\\
    &  & \vdots & 0 & \check{\Sigma}_{N} 
    \end{pmatrix}, 
\label{eq:sigcheck}
\end{equation}
and 
\begin{equation}
\check{V}^H = 
    \begin{pmatrix}
    \check{V}_1^H  \\
    \check{V}_2^H \\
    \vdots \\
    \check{V}_{N-1}^H\\
     \check{V}_{N}^H
    \end{pmatrix}.
\label{eq:vcheck}
\end{equation}
Here the inverse DFT is again the unitary case where
\begin{equation}
\mathcal{F}\{\cdot\}^{-1}_k = \frac{1}{\sqrt{N}}\sum_{k=0}^{N-1} \tilde{y}(x,r,m_k) \cdot e^{\frac{2 \pi i}{N} k n}. 
\label{idft_def}
\end{equation}

Step 4, depicted in Figure \ref{fig:F-ROM method}(d) is to perform the optimization again, this time using the reconstructed 3D transfer function from the previous step. This final I/O analysis of Eq. \ref{eq:F-ROM} is the heart of the hierarchical I/O (H-IO) method. It allows us to efficiently and quickly three-dimensionalize the I/O analysis via a Fourier decomposition and reconstruction with respect to the azimuthal direction. The computational cost of the final step largely depends on the input and output dimension of the transfer function, but it is not difficult to restrict the input/output such that re-optimization step is in fact the cheapest of the four steps. This final decomposition is denoted
\begin{equation}
H_r(x,r,\theta) = U_{r} \Sigma_{r} V_{r}^H,
\label{eq:final_from}
\end{equation}
where the basis of input directions (columns of $V_r$), correspond to physically realizable free-stream forcing amplitudes and the output directions (columns of $U_r$), correspond to the 3D post-shock response of the boundary layer. 

Finally, the optional Step 5 is to use the input basis from the final I/O analysis to compute the full state for analysis. This step is optional because in some cases we are only interested in examining the output, which is immediately available after step four. However, it is useful to reconstruct the global state to examine receptivity mechanisms and understand how the flow amplifies the input forcing. The direct response $\hat{q}(x,y,\theta)$ can be computed from a modification of Eq. \ref{eq:modified_io_form_D} by neglecting the left multiplication by the output matrix. The direct response at each wavenumber to the $d^{th}$ input direction is found by computing
\begin{equation}
\tilde{q}_d(x,r,m_k) = (-i\omega I - D A)^{-1} (-i \omega) B (V_r)_d,
\label{eq:direct_response_m}
\end{equation}
where the subscript $d$ denotes the $d^{th}$ column of the matrix $V_r$. Once the response is computed at each wavenumber, we do one final Fourier inversion to find the 3D state:
\begin{equation}
\hat{q}_d(x,r,\theta) = \mathcal{F}^{-1} \tilde{q}_d(x,r,m).
\label{eq:state_direct_response}
\end{equation}

The full computational cost of the H-IO method is largely determined by the size of the 2-D/axisymmetric resolvent inversion at each wavenumber. The initial I/O step requires $N$ independent (and thus parallel) resolvent inversions, while Steps 2-4 require minimal cost with respect to Step 1. If Step 5 is performed, another $N$ independent resolvent inversions are required. The computational advantage of this method is that it leverages the Fourier decomposition such that it keeps each resolvent computation in the regime where fast direct algorithms work well, while providing a means by which to return to the full three-dimensional global linear physics to understand the nature of complex receptivity processes.
\begin{table*}[hbt!]
\caption{\label{tab:complexity_comparison}%
Computational complexity compared to published high degree-of-freedom computations}
\begin{ruledtabular}
\begin{tabular}{lccccc}
\textrm{Case}&
\textrm{$N_\omega$}&
\textrm{DOF}&
\textrm{CPU time (hrs) } &
\textrm{RAM (GB)}\\
\colrule
Supersonic flat plate ($M = 4.5$) \cite{Bugeat2019}
  & 1 & 0.65 M & $4.4$ & $1.3\times10^1$ \\
Flow over parabolic body (incompressible) \cite{Martini2021}
  & 96 & 4.5 M & $3.0\times10^1$ & $1.6\times10^1$ \\
Round turbulent jet ($M = 0.4$) \cite{Farghadan2023}
  & 21 & 39 M & $1.8\times10^4$ & $7.4\times10^2$ \\
Hypersonic conical flow ($M = 5.8$) & 10 & 288 M & $1.3\times10^4$ & $7.3\times10^3 $
\end{tabular}
\end{ruledtabular}
\end{table*}

\subsubsection{Receptivity coefficients and N-factors}
It is useful to employ a receptivity coefficient to quantitatively connect the free-stream to the boundary layer. If the forced boundary layer instability is modal in nature, then it is possible to fit the response of the boundary layer to the predicted growth from a linear stability analysis such that the amplitude $A = A_0 e^{N}$, where N is the spatially integrated growth rate, and $A_0$ is the fitting parameter (initial amplitude). This is related to the classic $e^N$ method for transition prediction~\cite{VanIngen2008}. In the case where the forcing is acoustic, the initial amplitude can be defined in terms of pressure, and the receptivity coefficient is simply the pressure at the neutral point normalized by the acoustic forcing pressure amplitude \cite{Balakumar2015}. It is also possible to correlate receptivity in terms of the nose bluntness radius itself \cite{Marineau2017}, but \jwnresolved{in this paper we choose to focus on understanding receptivity mechanisms in detail for one particular representative bluntness radius,} and so do not employ the more general correlation. \jwnresolved{We do, however,} generalize the pressure amplitude method to account for different types of forcing waves by using the Chu energy amplitude as the starting place for defining the receptivity coefficient. The generalized amplitude receptivity coefficient is defined as 
\begin{equation}
    C_{a} = \frac{A_{0}}{A_{fs}},
     \label{eq:amplitude_receptivity}
\end{equation}
where $A_0$ is the Chu amplitude (the square root of the Chu energy) in the boundary layer at the upstream neutral point of the modal instability, and $A_{fs}$ is the peak Chu amplitude in the forcing wave outside of the shock. In cases where we force with one type of wave at a time, this provides a simple connection between the forcing wave amplitude and the initial modal amplitude. 

    \label{eq:efficiency_receptivity}

\dac{
It is worth emphasizing here that for flows with tip bluntness, the modal grow starts downstream of the blunt tip and so modal-based transition prediction lacks sufficient treatment of the receptivity mechanisms. Initial amplitudes obtained empirically also neglect a physical treatment of non-modal growth via entropy layer instability and shock-perturbation interaction.  For these reasons, the $e^N$ method is not predictive from a first principles perspective.} \jwn{I/O analysis, however, enables us} \dac{to define a receptivity related N-factor from the flow responses to the optimal input forcing in order to connect the worst-case spatial growth of flow disturbances to first-principles mechanisms. Instead of an N-factor which is relative to an arbitrary initial amplitude, we directly obtain the initial amplitude from the peak of the forcing wave outside the shock. This N-factor can be considered the true upper bound N-factor with respect to the optimal inputs as constrained in the input-output formulation. In other words, this absolute N-factor represents the integration of downstream instability with free-stream receptivity.} \jwn{We define} \dac{this receptivity N-factor as 
\begin{equation}
    N_r = \log\left(\frac{A_C(\xi)}{A_{fs}}\right), 
\end{equation}
where $\xi$ is the streamwise coordinate direction. 

}

\subsection{\label{sec:goveqs}Numerical methods}

We solve the governing equations using a parallel, structured, in-house solver that employs the SKBC-equipped finite volume method \cite{Cook2022}. We approximate the inviscid fluxes using a 3\ts{rd} order MUSCL scheme \cite{vanLeer1997} with quadratic reconstruction. The viscous fluxes are computed using a 2\ts{nd} order least squares reconstruction. Once the solution is obtained on an initial mesh, the mesh is iteratively refined to fit to the stationary shock surface, and elliptically smoothed such that the mesh is orthogonal at both the wall and the shock surface. Once the mesh is tailored to the shock, we apply a shock fitting routine based on the pre-shock and post-shock mean state such that the shock surface is no longer discretized, but can be defined by a single streamwise grid line. Once the shock fitting is complete, we numerically extract the global linear dynamics via complex step differentiation to retain full double precision in the Jacobian. 

We also use a custom parallel solver to build and implement the free-stream receptivity matrices as well as perform the I/O analysis step. The resolvent factorization at each wavenumber is accomplished via the parallel sparse direct solver MUMPS \cite{Amestoy2001,Amestoy2019}. The singular value decomposition is computed using the sparse eigenvalue package ARPACK \cite{ARPACK}. 

The flow domain was discretized using 1920 points in the streamwise direction and 300 points in the wall-normal direction. In the mean flow computation, the grid points were clustered near the wall such that the boundary layer was well resolved ($y^+ < 1$). Additionally, the streamwise discretization ensured that slow acoustic waves up to $f = 100 \unit{\kilo\hertz}$ were discretized with no less than 10 grid points per wavelength. For use in the H-IO analysis, the Fourier transform was computed by discretizing the $\theta$ direction with 128 points, which is sufficient to resolve flow features with wavenumber components up to $m = 64$. While it may require this many (or more) Fourier coefficients to rebuild high frequency planar waves in the free-stream with good accuracy, above a certain threshold, high wavenumber physics are not amplified in the downstream flow. Figure \ref{fig:gain_vs_m} shows the I/O gain vs. wavenumber for three cases considered in this paper. For each of these computations, the 3D results were reconstructed using wavenumbers from $m = -48\text{--}49$, because for $|m| > 50$ the I/O gain was found to be less than 10\% of the highest gain in the least conservative case. The chosen spatial discretization, five state variables, and Fourier coefficients, leads to a problem with 288 million degrees of freedom, which to the authors knowledge, is the largest input-output analysis successfully performed.

\dac{The performance of the H-IO approach for this problem is summarized alongside recently published three-dimensional computations in Table \ref{tab:complexity_comparison}. We measure and report CPU time by $T_w N_p N_{\omega}$, where $T_w$ is the wall time, $N_p$ is the number of CPUs used, and $N_{\omega}$ is the number of frequencies computed. The memory usage is reported as the total amount of memory required for the sub-level resolvent inversions. The computational study used to report the timings computed 25 leading I/O directions (on the sub-levels) for the sharp cone flow and was run in parallel on 48 AMD EPYC 7451 processors, each with a clock speed of  2.3 GHz. We compare our H-IO results to resolvent analyses of a supersonic flat plate \cite{Bugeat2019}, incompressible flow over a parabolic body \cite{Martini2021}, and a round turbulent jet flow \cite{Farghadan2023}. Because each of the computational studies to which we compare is performed potentially on different processors and computer architectures, it is difficult to obtain a one-to-one comparison. Each of these studies to which we compare, however, were performed within the last four years, and so we assume that our comparisons accurately reflect algorithmic performance. Details of the underlying computations can be found in the provided references.

As is shown in Table \ref{tab:complexity_comparison}, the scale of the present computation has a similar computational cost in terms of CPU time than next largest DOF case, the round turbulent jet, while containing over seven times the total number of degrees of freedom. The H-IO analysis does, however, exceed the memory requirements of the round turbulent jet computation by an order of magnitude. This amount is manageable in our case, because each of the sub-level computations only requires $1/N_m$ times the total memory at a time, which is not excessive for modern high performance computers, where $N_m$ is the number of Fourier coefficients in the azimuthal discretization. The hierarchical approach allows the computation to efficiently utilize the available memory on the sub-level in order to decrease the total amount of CPU time required in the direct and adjoint resolvent inversions. }

\dac{
\subsection{Error analysis of the hierarchical input-output approach}
One of the main ideas enabling hierarchical input-output (H-IO) analysis is the use of a low-rank compression of a flow decomposition such that the re-composition into full three-dimensional space is more computationally affordable. It is important to consider the effect of this compression on the amount of error present in the reconstructed three-dimensional flow features. Two sources of error are present in the Fourier H-IO analysis. First, the truncation of the Fourier decomposition could leave higher wavenumber components of the forcing waves and response unresolved or under-resolved. Secondly, the truncation of the transfer function at each wavenumber could propagate error to the reconstruction.
\begin{figure}[t!]
    \centering
    \includegraphics[trim=4 4 4 4, clip,width = 0.50\textwidth]{ 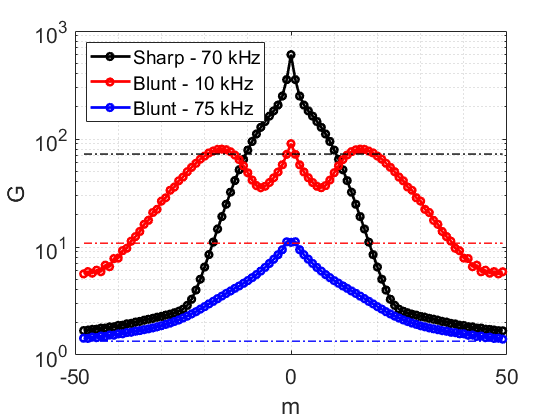} 
    \caption{H-IO gain versus wavenumber for several sharp and blunt cone analyses. For each case shown, the gain at $m = 50$ is less than 10\% of the maximum gain. }
    \label{fig:gain_vs_m}
\end{figure}

We can begin to address the first type of error by examining the gain from the wavenumber specific I/O analyses in the first step of the H-IO process. The gain from I/O analyses across wavenumber are shown in Figure \ref{fig:gain_vs_m} for several full-scale test cases and frequencies. After around 50 coefficients, the gain associated with each successive wavenumber contributes less than 10\% of the gain at the maximum amplified wavenumber. This gives a sense that there is a wavenumber threshold past which the flow physics do not amplify the inputs.

To quantify the effect of truncating the Fourier decomposition, we examine the relative error in the H-IO analysis as we include an increasing number of coefficients. As an example, consider two H-IO analyses for the sharp cone at $f = 10 \unit{\kilo\hertz}$ and $f = 70 \unit{\kilo\hertz}$. We compute the H-IO analysis for each of these flows using three different truncation points in the Fourier decomposition and compare the relative error of final I/O directions and gains to the maximum truncation case. In order for the approximation to be acceptable, the relative error should diminish as we include more Fourier coefficients such that the results are not expected to change if more coefficients were to be added. The relative error between any input amplitude distribution $a_{m_t}$ and the amplitude distribution resulting from the truncation with $m_t = 50$ is
\begin{equation}
e_{a} = \frac{|a_{m_{50}}| - |a_{m_t}|}{|a_{m_{50}}|}.
\end{equation}

\begin{figure*}[t!]
    \begin{tabular}{lll}
       (a) & (b) & (c)  \\
        \includegraphics[trim=4 4 4 4, clip,width = 0.32\textwidth]{ 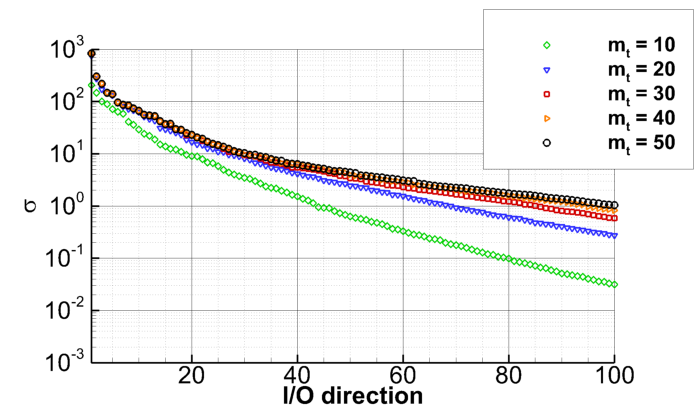} & \includegraphics[trim=4 4 4 4, clip,width = 0.32\textwidth]{ 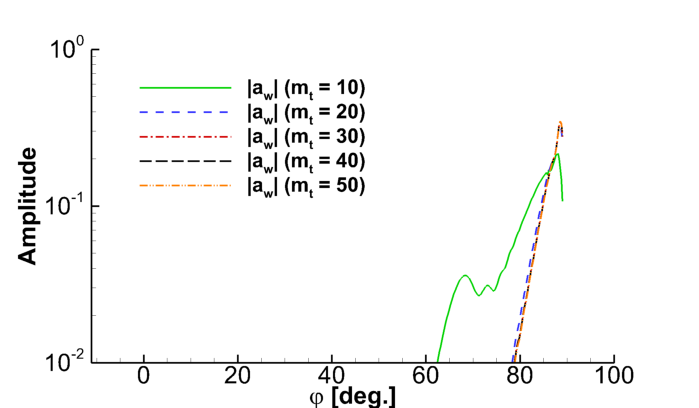} & \includegraphics[trim=4 4 4 4, clip,width = 0.32\textwidth]{ 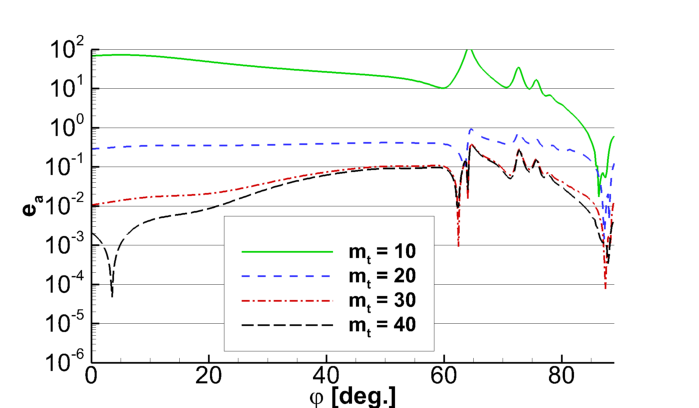} \\
        (d) & (e) & (f) \\
        \includegraphics[trim=4 4 4 4, clip,width = 0.32\textwidth]{ 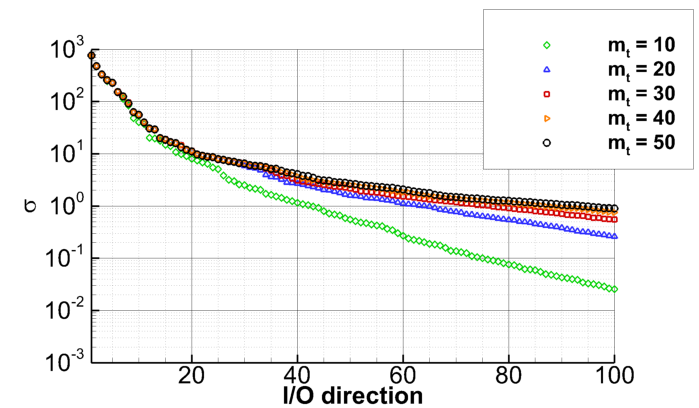} & \includegraphics[trim=4 4 4 4, clip,width = 0.32\textwidth]{ 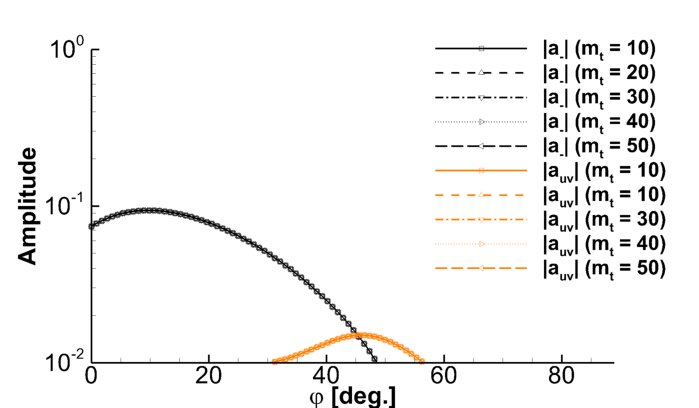} &\includegraphics[trim=4 4 4 4, clip,width = 0.32\textwidth]{ 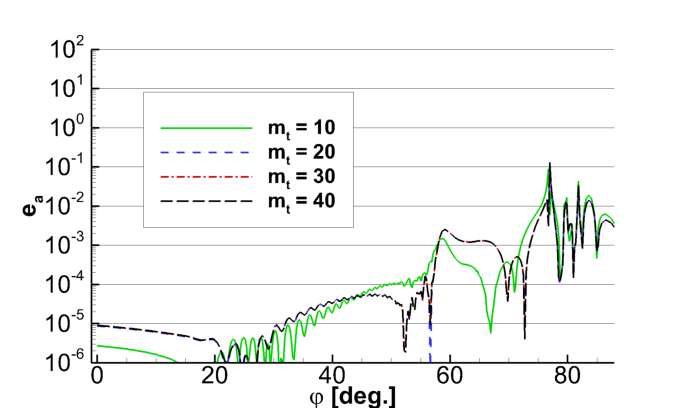} \\
        (g) & (h) & (i) \\
        \includegraphics[trim=4 4 4 4, clip,width = 0.32\textwidth]{ 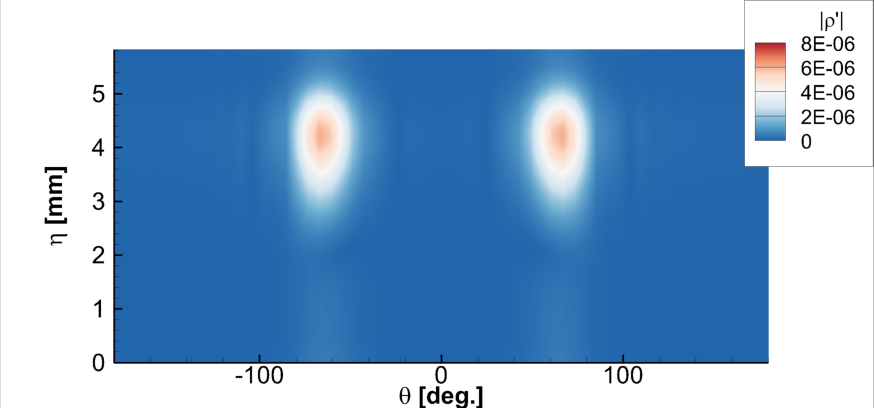} &
        \includegraphics[trim=4 4 4 4, clip,width = 0.32\textwidth]{ 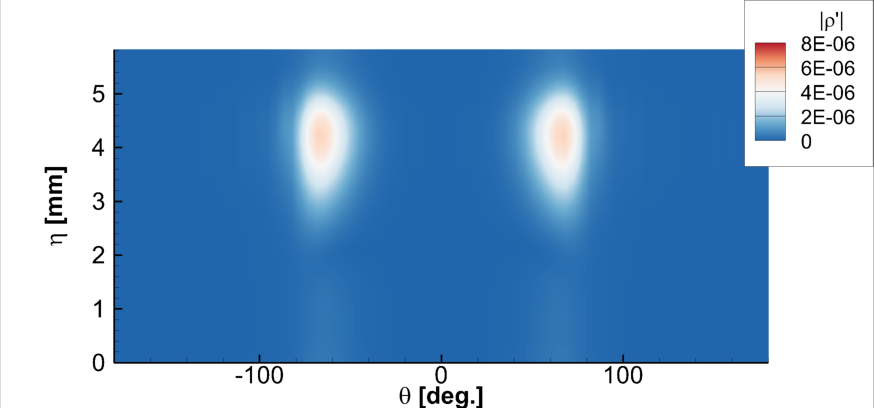} &
        \includegraphics[trim=4 4 4 4, clip,width = 0.32\textwidth]{ 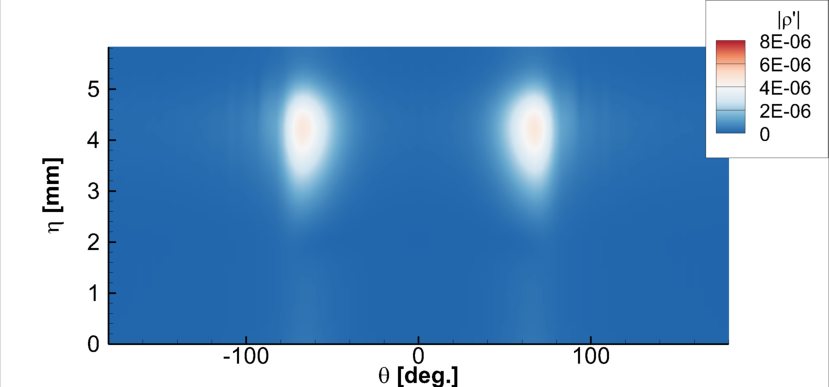}
    \end{tabular}
    \caption{Gain and input error quantification for H-IO analyses of the sharp cone at (a)--(c) 10 kHz and (d)--(f) 70 kHz. Error is quantified by varying the number of included Fourier coefficients and then showing (a), (d) the leading gains, (b), (c) the input distributions, and (c), (f) log-scale relative input error with respect to $m_t = 50$. Also shown are $D_1$ outputs for sharp cone H-IO analyses at 10 kHz and for (g) $m_t = 30$, (h) $m_t = 40$, and (i) $m_t = 50$. As more wavenumbers are included, the output physics converge to a single physical mechanism.}
    \label{fig:m_error}
\end{figure*}

Figures \ref{fig:m_error}(a) and (d) show the gains for sharp cone H-IO analyses at 10 kHz and 70 kHz as a function of $m_t$. As the number of Fourier coefficients are increased, the gains become increasingly similar. It is worth noting that the error in the gain reduces more quickly for the larger gains. Both frequencies show the $D_1$ gain error reduces more quickly than the other directions as $m_t$ is increased. This is another indicator that the dominant physical mechanisms are sufficiently resolved with $m_t = 50$.

Figure \ref{fig:m_error}(b) shows the effect of varying the Fourier truncation on the $D_1$ input distributions for $m_t = 10\text{--}50$ at a frequency of 10 kHz. A significant difference in the input distribution at $m_t = 10$ is visible in the input distribution, while the differences between the optimal input distributions for $m_t > 20$ become much smaller. The relative error with respect to $m_t = 50$ is shown in Figure \ref{fig:m_error}(c). The error in the $D_1$ input distribution is largest using $m_t = 10$, but reduces as the number of included wavenumbers increase. For the case where $m_t = 40$, the maximum relative error is around $10^{-1}$. Note also that the error is smallest, around $10^{-3}$ where the peak of the input distribution occurs, indicating that the dominant receptivity mechanisms are well-resolved using $m_t = 50$. 

A comparison at high frequency shows a similar trend. The $D_1$ input distributions for the H-IO analysis of the sharp cone at 70 kHz are shown in Figure \ref{fig:m_error}(e), and are virtually indistinguishable from one another. The relative error is again shown in Figure \ref{fig:m_error}(f), which shows a much lower error for all values of $m_t$. This indicates that the physical mechanisms captured by the H-IO analysis at 70 kHz are more axisymmetric in nature and therefore require fewer wavenumbers to resolve, indicating that $m_t = 50$ is more than sufficient. 

The effect of varying the Fourier truncation on the leading output directions can also be quantified. Figure \ref{fig:m_error}(g)--(i) shows the absolute value of the density fluctuations in the output region as functions of the wall-normal and azimuthal coordinate directions. As $m_t$ is increased, the output direction shapes converge onto a single shape function, although the amplitudes slightly reduce as energy is distributed to the higher wavenumber components. The largest shape changes are visible for the low frequency oblique structures due to the higher number of Fourier coefficients needed to resolve them. All of this suggests that $m_t = 50$ is sufficient for these cases to capture the dominant physics with an H-IO analysis. 
\begin{figure*}[t!]
    \begin{tabular}{lll}
       (a) & (b) & (c)\\
        \includegraphics[trim=4 4 4 4, clip,width = 0.30\textwidth]{ 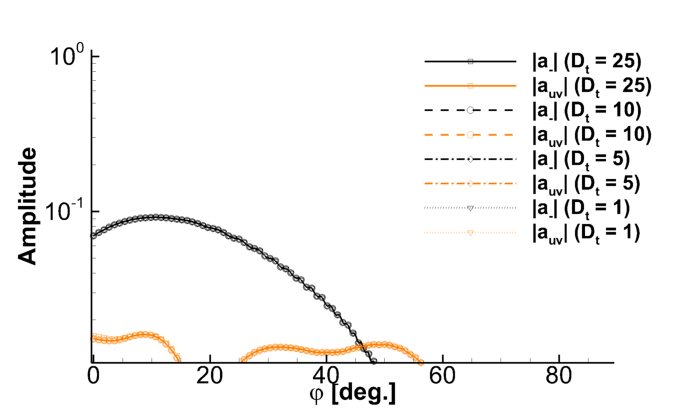} & \includegraphics[trim=4 4 4 4, clip,width = 0.30\textwidth]{ 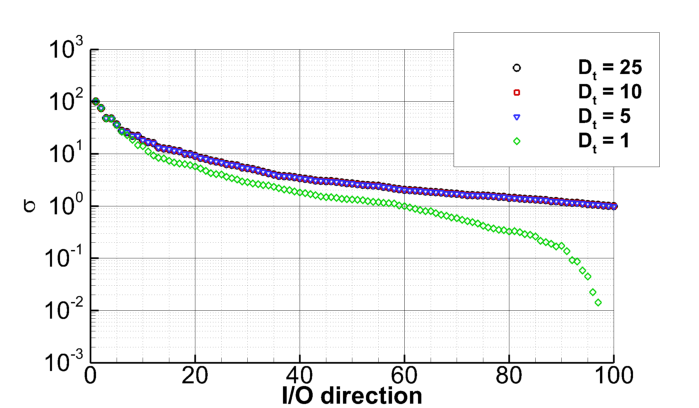} & \includegraphics[trim=4 4 4 4, clip,width = 0.37\textwidth]{ 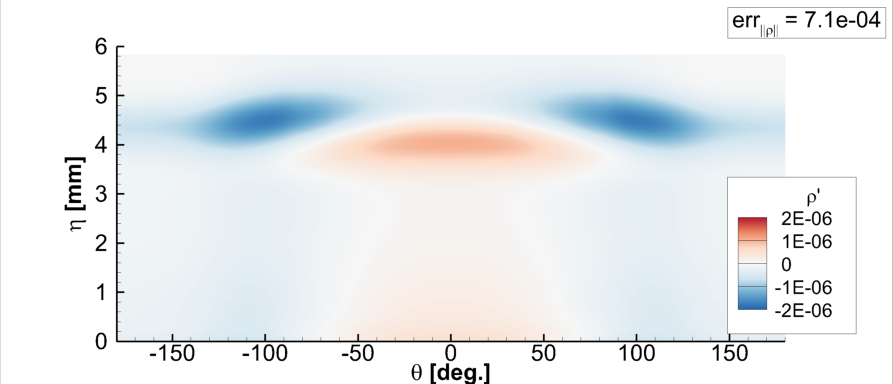}\\
        (d) & (e) & (f)\\
        \includegraphics[trim=4 4 4 4, clip,width = 0.30\textwidth]{ 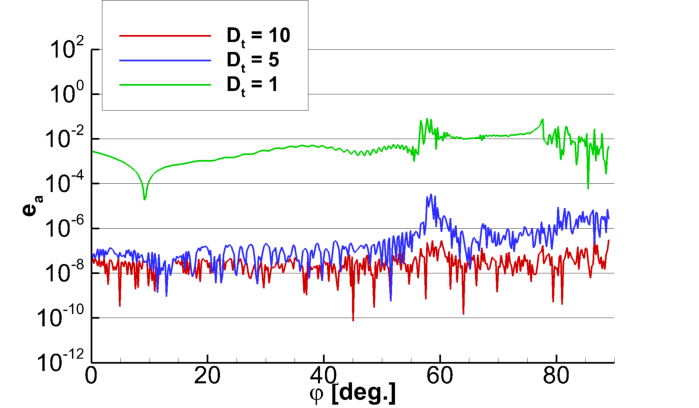} & \includegraphics[trim=4 4 4 4, clip,width = 0.30\textwidth]{ 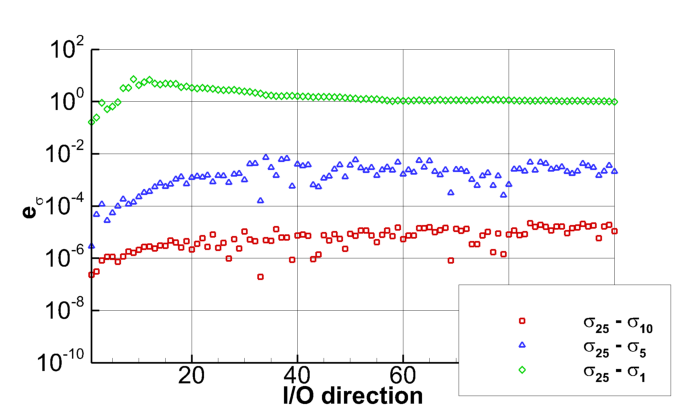} & \includegraphics[trim=4 4 4 4, clip,width = 0.37\textwidth]{ 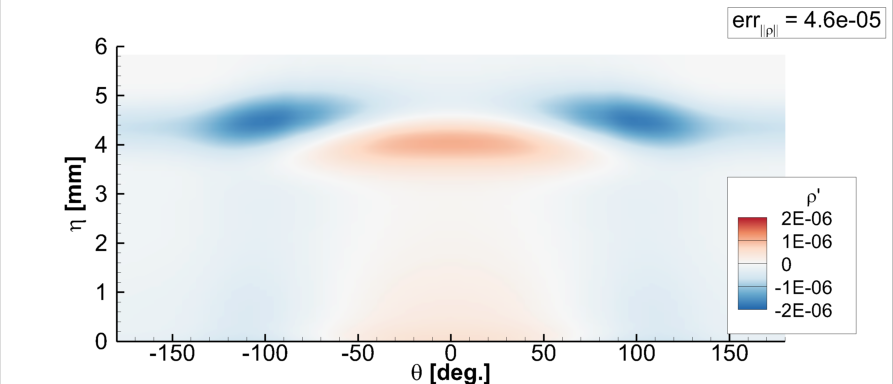} 
    \end{tabular}
    
    \caption{Low-rank truncation error are quantified for sharp cone H-IO analysis at 60 kHz in terms of (a), (d) the input error, (b), (e) the gain error, and (c), (f) the output error. For several truncation numbers, (a) shows good agreement between the $D_1$ forcing distributions as a function of $\psi$, and (d) shows log-scale relative input error with respect to $D_t = 25$. For several truncation numbers, (b) shows good agreement between the first 100 gains for $D_t>1$, and (e) shows log-scale relative gain error with respect to $D_t = 25$. Output is shown for two truncation numbers: (c) $D_t = 1$ and (f) $D_t = 10$. Relative density norm error between (c), (f) and $D_t = 25$ are shown in (c),(f).}
    \label{fig:trunc_error}
\end{figure*}

We now address the second error source: low-rank approximation of the wavenumber specific transfer functions. Again, this quantification is performed in terms of a relative error measure. The relative error of the input distribution can be defined as the relative difference between an arbitrary input distribution and the input distribution in the most accurate test case performed. As the number of I/O directions included in the transfer function truncation is increased, if the error becomes vanishingly small, then we have high confidence that the error in the approximation is low. The input error is defined by 
\begin{equation}
e_{a} = \frac{|a_{D_{25}}| - |a_{test}|}{|a_{D_{25}}|},
\end{equation}
where $a_{test}$ is the input amplitude distribution and $a_{max}$ is the same amplitude distribution for the most accurate test case computed. This provides a relative way to evaluate how many directions are necessary in the transfer function reconstruction. The relative error in the gain is similarly defined as 
\begin{equation}
e_{\sigma} = \frac{|\sigma_{D_{25}} - \sigma_{test}|}{|\sigma_{D_{25}}|}. 
\end{equation}
The relative error in the output is assessed by computing the two-norm of output density fields for various truncation thresholds. This error is given by 
\begin{equation}
e_{out} = ||\frac{\rho_{D_{25}} - \rho_{test}}{\rho_{D_{25}}}||_2. 
\end{equation}

For a test case, we consider an H-IO analysis of the sharp cone boundary layer at $f = 60 \unit{\kilo\hertz}$. The relative $D_1$ input error quantification is shown in Figure \ref{fig:trunc_error}(a). There is no discernible difference in the selected optimal input distributions, even when only the leading I/O direction is retained in the model reduction step of the H-IO analysis. Figure \ref{fig:trunc_error}(d) shows the log of the relative error for truncation values from $D_t = 1\text{--}10$, relative to the max case where $D_t = 25$. Retaining a single I/O direction is sufficient to bring the relative error below a 10\% threshold, whereas including ten directions in the rank-reduction is sufficient for a relative error on the order of $10^{-7}$. The error of the leading input distribution when using $D_t = 25$ is expected to be even lower. 

The relative gain error quantification is shown in Figure \ref{fig:trunc_error}(b) and (e). The only case for which there is visible error in the first 100 H-IO directions is the case where only a single direction is retained in the low-rank step, whereas the gains from the other analyses are identical. Even in the case where a single direction is retained, the error in the leading H-IO direction is the lowest and is on the order of 10\%. Again, a log plot of the error reveals that the error in the gain is sufficiently low when $D_t = 10$. 

The relative $D_1$ output error is shown in Figure \ref{fig:trunc_error}(c) and (f) in terms of the density fluctuation in the output region as a function of the wall-normal coordinate $\eta$ and the azimuthal coordinate $\theta$. No discernible difference is visible between the output signatures of density. For the cases where the relative error was computed, it is given in the upper right corner of the figure. The error in each of the cases is very low. 

To ensure that this particular test case was not abnormal, the $D_1$ error gain as a function of frequency for the sharp cone is shown in Figure \ref{fig:gain_error_v_freq}, verifying that this trend is consistent across frequencies. The relative error for the $D_t = 10$ case is on the order of $10^{-6}$. All of this suggests that using $D_t = 25$ is more than sufficient to ensure that the error associated with the model truncation is negligible. 
\begin{figure}
    \centering
     \includegraphics[trim=4 4 4 4, clip,width = 0.45\textwidth]{ 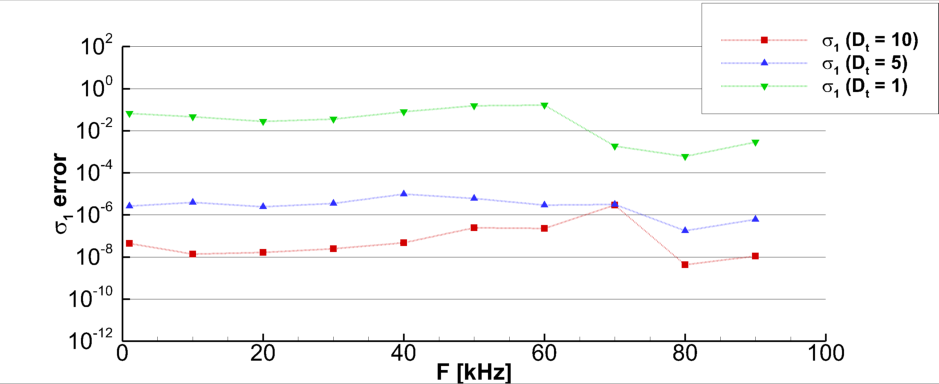}
   
    \caption[$D_1$ gain error quantification for H-IO analysis of the sharp cone across several frequencies and truncation values.]{$D_1$ gain error quantification for H-IO analysis of the sharp cone across several frequencies and truncation values. Error is relative to case with $D_t = 25$. }
    \label{fig:gain_error_v_freq}
\end{figure}

Having performed a comprehensive characterization of the possible error sources---truncation of the Fourier series and the low-rank approximation---we conclude that the error associated with the hierarchical input-output analysis approach is very low. 
}
\section{\label{sec:RESULTS}Results}
\dac{In this section, we present hierarchical input-output (H-IO) analyses of three $M = 5.8$ flows over sharp and blunt cones. H-IO analyses were performed at frequencies from 10 \unit{\kilo\hertz} to 90 \unit{\kilo\hertz}. The optimal gains as functions of frequency are shown in Figure \ref{fig:gain_vs_frequency_b0p2_b3p6} for the sharp cone, 3.6 mm and 7.2 mm blunt tipped cones. The gains from the sharp cone analysis contain a low-frequency peak at 10 kHz and a high-frequency peak at 70 kHz. The gains for the blunt cones are lower than those from the sharp cone analysis and decrease monotonically with frequency above 10 kHz. Both of the strong peaks present in the sharp cone gains are absent from the blunt cone gains. Of the three cones, the H-IO analysis of the 7.2 mm blunt cone has the lowest gain at 10 kHz, but then the trend reverses, and the gains from the 7.2 mm cone analyses exceed those of the 3.6 mm cone analysis at frequencies above 20 kHz. The largest difference in the blunt cone gains occurs at 40 kHz. The gray rectangles in Figure \ref{fig:gain_vs_frequency_b0p2_b3p6} highlight several cases selected for more detailed analysis. The gains at 10 kHz and 70 kHz demonstrate an overall stabilizing effect from the addition of nose-tip bluntness. The gains at 40 kHz cases show a stabilizing effect with the initial blunting of the sharp cone, but show a destabilizing effect when the bluntness is increased from 3.6 mm to 7.2 mm. Cases are selected from these frequencies in order to understand both the mechanisms and the receptivity underlying these observations.
\begin{figure}[t!]
\centering
 \includegraphics[trim=4 4 4 4 , clip,width = 0.45\textwidth]{ 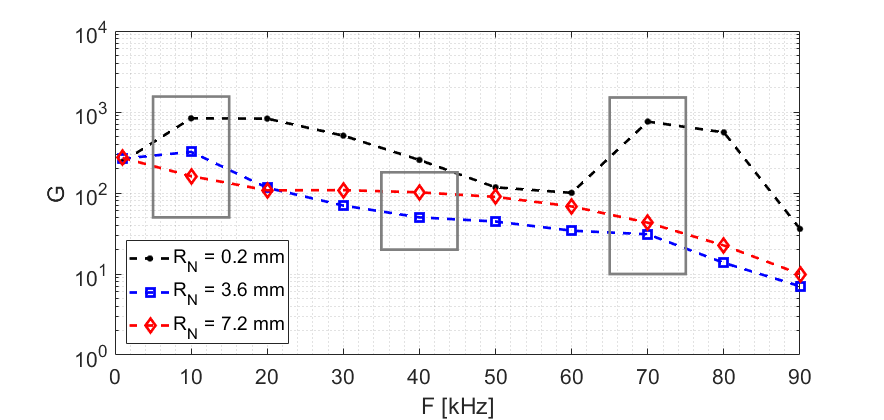} \\
\caption[$D_1$ gain as a function of frequency from H-IO analysis of $M = 5.8$ flows over sharp and blunt cones.]{\label{fig:gain_vs_frequency_b0p2_b3p6} $D_1$ gain as a function of frequency from hierarchical input-output analysis of $M = 5.8$ flows over sharp and blunt cones.}
\end{figure}

The results are organized into five sub-sections. First, in \S \ref{subsec:mean_flow}, we present the mean flow solutions for three 1 \unit{\metre} long, \ang{7} half-angle cones, one with a nominally sharp ($R_N = 0.2 \unit{\milli\metre}$) tip, one with a 3.6 \unit{\milli\metre} tip, and one with a 7.2 \unit{\milli\metre} tip. Next, in \S \ref{subsec:modal_results}, we examine the modal boundary layer mechanisms predicted by linear stability theory. We then proceed in \S \ref{subsec:high_frequency} to perform a verification of H-IO analysis using the sharp cone boundary layer at 70 kHz, where the well-known Mack mode is present and then examine the stabilizing effect of adding nose-tip bluntness. In \S \ref{subsec:low_frequency}, we present results at 10 kHz in order to examine the stabilizing effect of nose-tip bluntness. \S \ref{subsec:mid_frequency} discusses the H-IO at 40 kHz, at which the gain reversal trend was observed. Finally, \S \ref{subsec:summary} examines the receptivity and instability trends across several frequencies and sub-optimal directions and summarizes the results.}

\subsection{\label{subsec:mean_flow} Mean flow}
The mean flow is a Mach 5.8 flow over a 1 \unit{\metre} long, \ang{7} half-angle cone. For verification we consider a nearly sharp cone ($R_N =  0.2 \unit{\milli\meter}$), and then consider \dac{two cones with spherically blunted tips with nose radii $R_N = 3.6 \unit{\milli\meter}$ and $R_N = 7.2 \unit{\milli\meter}$}. The flow conditions are given in Table \ref{tab:flow_conditions}, as well as the isothermal wall temperature. \dac{These flow conditions were initially chosen to correspond to the experiments of Rufer \cite{Rufer2005}. Subsequent modal analysis of these flows was also performed by Robarge \cite{Robarge2005}. However, in order to strike a compromise between computational tractability while still providing physical insights, the Reynolds number has been reduced by a factor of three. This still provides a flow which supports significant modal instability, while keeping the overall, degrees of freedom in a tractable range for methodological development. 
}

\begin{table}[b]
\caption{\label{tab:flow_conditions}%
Mean flow conditions
}
\begin{ruledtabular}
\begin{tabular}{lllll}
\textrm{$Re_u$}&
\textrm{$M$}&
\textrm{$\rho_{\infty}$}&
\textrm{$T_{\infty}$ } &
\textrm{$T_{wall}$ }\\
\colrule
2.4$\times 10^6$ \unit{\per\meter} & 5.8 & 0.01 [\unit{\kilo\gram\per\metre\cubed}]& 56.0 [\unit{\kelvin}]
& 300.0 [\unit{\kelvin}]\\
\end{tabular}
\end{ruledtabular}
\end{table}
\begin{figure*}[thb!]
\begin{tabular}{l c l c l}
    (a) & \hspace{1cm} & (b) & \hspace{1cm} & (c)\\
   \includegraphics[trim=4 4 4 4, clip, width = 0.225\textwidth]{ 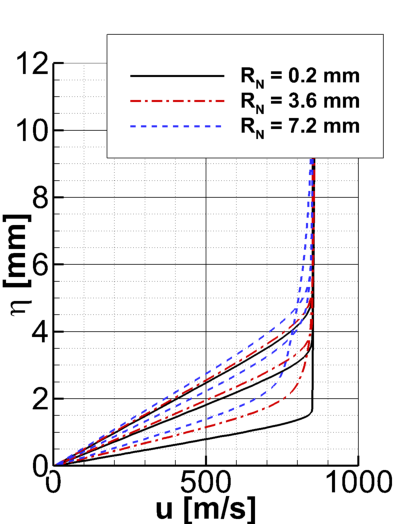}  & &
   \includegraphics[trim=4 4 4 4, clip, width = 0.225\textwidth]{ 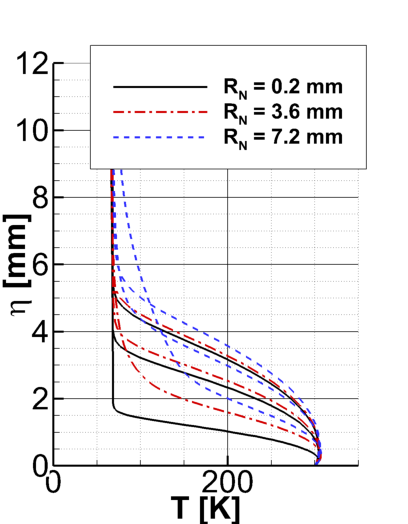} & & 
   \includegraphics[trim=4 4 4 4, clip, width = 0.225\textwidth]{ 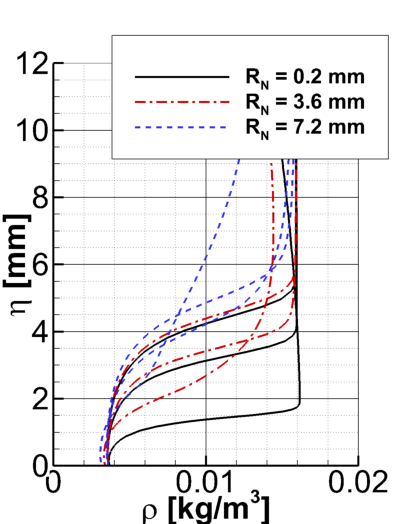} \\
   \multicolumn{5}{l}{(d)} \\
   \multicolumn{5}{c}{ \includegraphics[trim=2 2 2 2, clip, width = 0.75\textwidth]{ 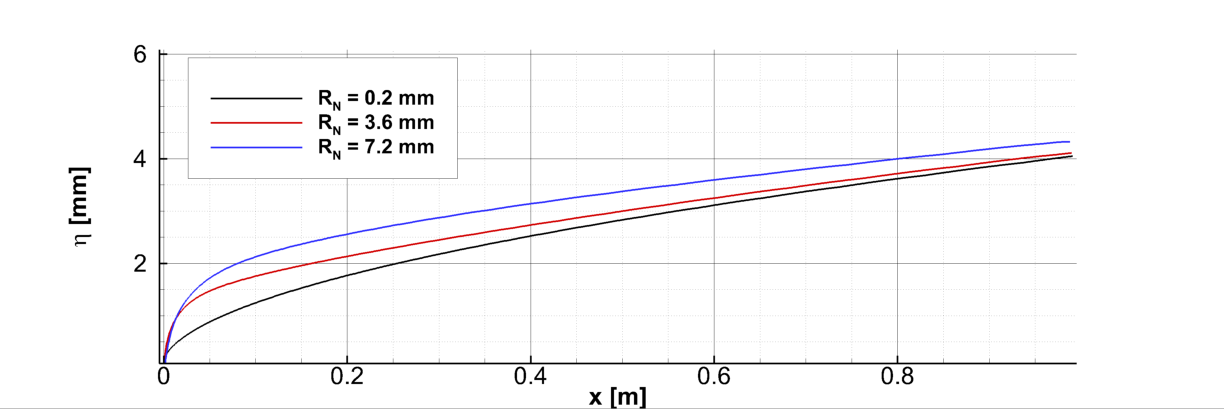}}
\end{tabular}
\caption{\label{fig:mean_bl_profiles} Mean boundary layer profiles of \jwnresolved{(a)} velocity, \jwnresolved{(b)} temperature, and \jwnresolved{(c)} density at several streamwise position\jwnresolved{s} as a function of wall-normal coordinate $\eta$ for both flows. Shown in (d) are mean boundary layer thicknesses based on edge enthalpy $\delta_h$.}
\end{figure*}
In contrast \dac{to a sharp cone,} a blunt tip generates a curved bow shock which creates a high entropy layer near the tip. This entropy layer persists above the boundary layer downstream of the tip before it is slowly absorbed, or swallowed, due to radial expansion of the flow and the growth of the boundary layer. \dac{The swallowing length---theoretically correlated with free-stream parameters and nose-tip bluntness first by Rotta \cite{Rotta1966} and more recently by Zhou et al.\cite{Zhou2021}---is defined as the streamwise location at which the total mass flow through the entropy layer region is equal to that of the boundary layer. The swallowing lengths for the sharp, 3.6 mm, and 7.2 mm cones are $X_{SW} = 0.014 \unit{\m}$, $X_{SW} = 0.67 \unit{\m}$, $X_{SW} = 1.69 \unit{\m}$, respectively. We consider the $R_N = 0.2 \unit{\milli\metre}$ cone to be sharp because the entropy swallowing distance is vanishing with respect to the streamwise extent of the flow. The boundary layer over the $R_N = 3.6 \unit{\milli\metre}$ cone swallows the entropy layer at a position around two-thirds of the streamwise extent of the computational domain, whereas the computational domain for the $R_N = 7.2 \unit{\milli\metre}$ cone does not contain the swallowing point. This provides an informative comparison of the relative effect of entropy layer swallowing on the stability of the boundary layer.} 

\dac{
Mean boundary layer profiles of velocity, temperature, and density along with the mean boundary layer thicknesses are shown in Figure \ref{fig:mean_bl_profiles}. At the earliest streamwise station ($x = 0.1 \unit{\metre}$) shown in Figure \ref{fig:mean_bl_profiles}(a)--(c), the entropy layer is quite pronounced, resulting in shallower shear stresses and gradients. The temperature profiles at this station also show the high temperature and lower density maintained by the entropy layer upstream of its swallowing point. By $x = 0.5 \unit{\metre}$, the $R_N = 3.6 \unit{\milli\metre}$ cone almost fully matches the sharp cone boundary layer, though marginally thicker.  The mean velocity at the end of the flow domain is nearly the same for the sharp and 3.6 mm blunt cone, but the 7.2 mm blunt cone maintains a slightly thicker boundary layer at the end of the domain. The boundary layer thickness can be measured precisely by a total enthalpy criterion where the boundary layer edge $\delta_h$ is defined as the wall-normal location where $h(\delta_h) = 0.995 h_t$. Here, $h$ is the enthalpy and $h_t = C_p \bar{T} + 0.5(\bar{u}^2 + \bar{v}^2 + \bar{w}^2)$ is the total enthalpy, defined by the specific heat $C_p$, mean temperature $\bar{T}$, and mean velocities $\bar{u}$, $\bar{v}$, and $\bar{w}$. The boundary layer thicknesses for each of the flows as a function of streamwise distance is shown in Figure \ref{fig:mean_bl_profiles}(d). Whereas the boundary layer over 3.6 mm cone is nearly the same height as that of the sharp cone by the end of the domain, the 7.2 mm tip causes an overall thickening of the boundary layer which persists to the end of the flow domain. 
}

\subsection{\label{subsec:modal_results}Modal mechanisms}
\begin{figure*}[htb]
\begin{tabular}{l l l}
    (a) & (b) & (c)\\
    \includegraphics[trim=4 4 4 4, clip,width = 0.3\textwidth]{ 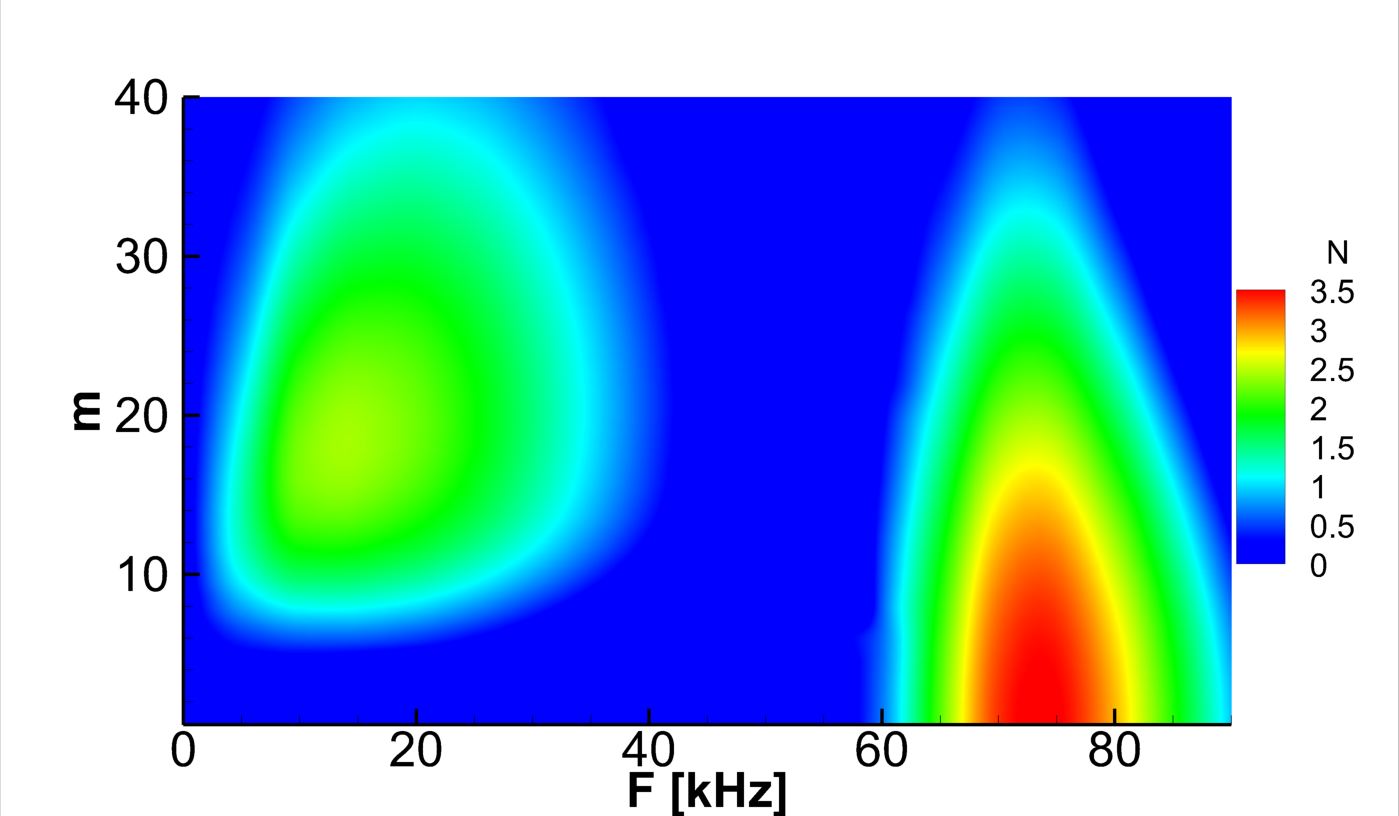} & 
    \includegraphics[trim=2 2 2 2, clip,width = 0.3\textwidth]{ 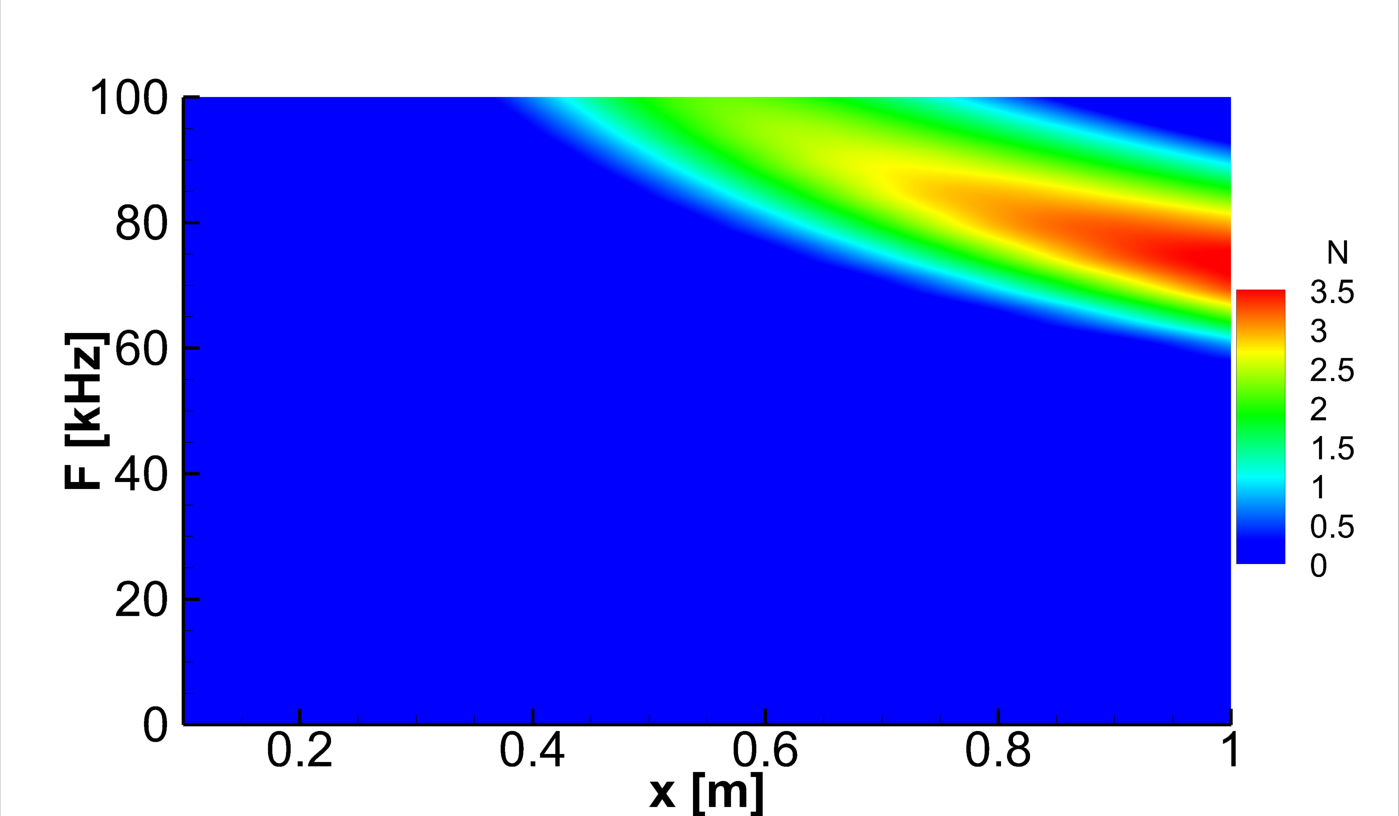}  & 
    \includegraphics[trim=2 2 2 2, clip,width = 0.3\textwidth]{ 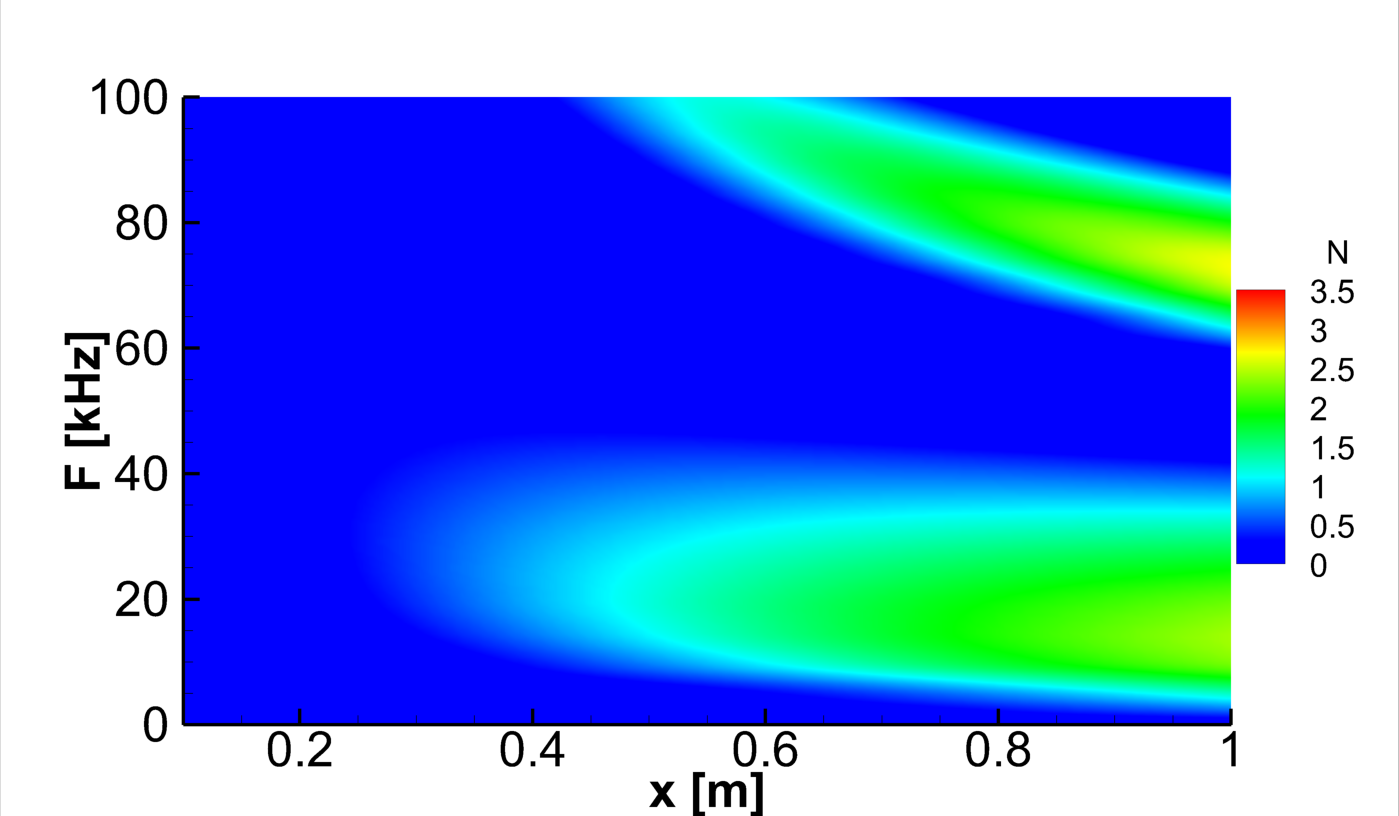} \\
    (d) & (e) & (f)\\      
    \includegraphics[trim=4 4 4 4, clip,width = 0.3\textwidth]{ 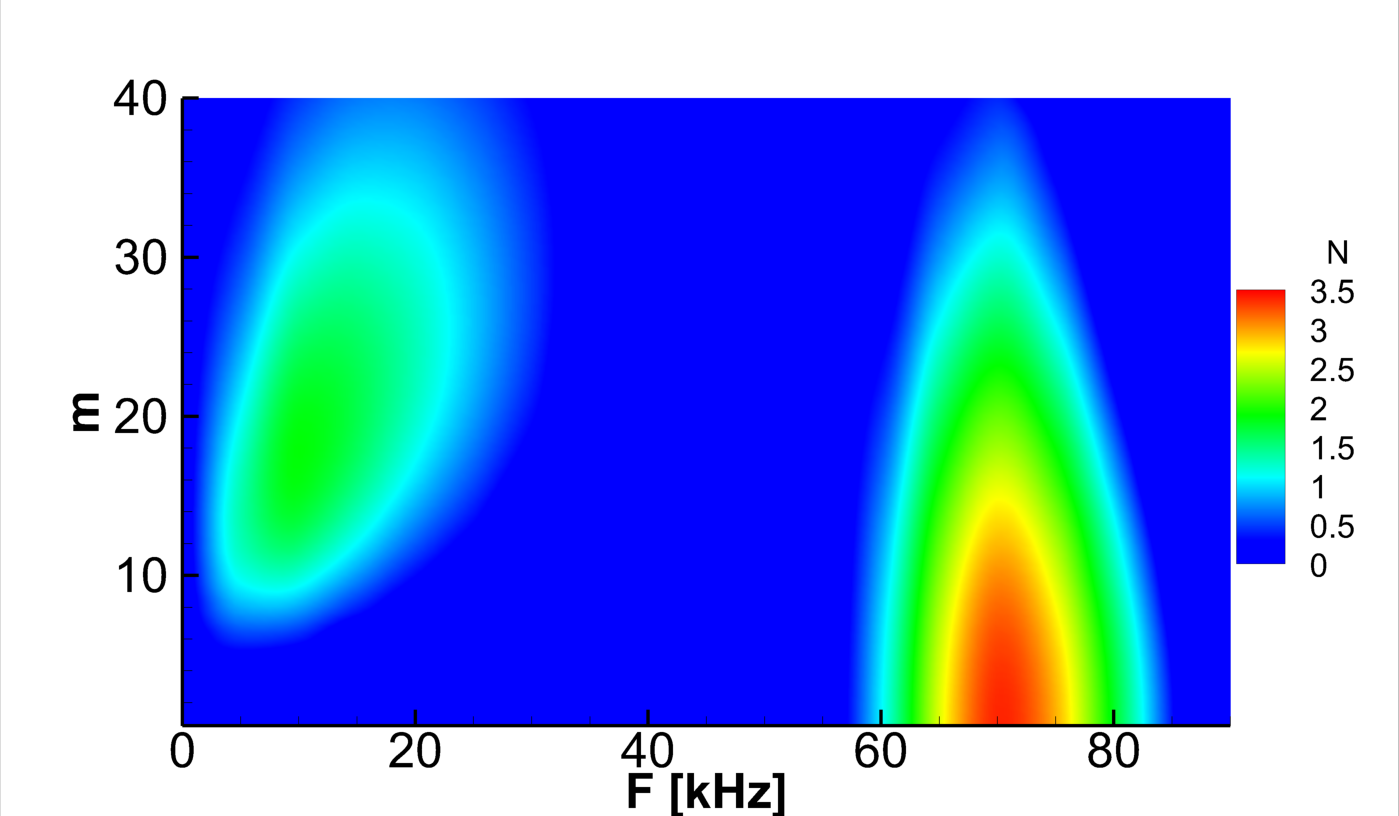} &
    \includegraphics[trim=2 2 2 2, clip,width = 0.3\textwidth]{ 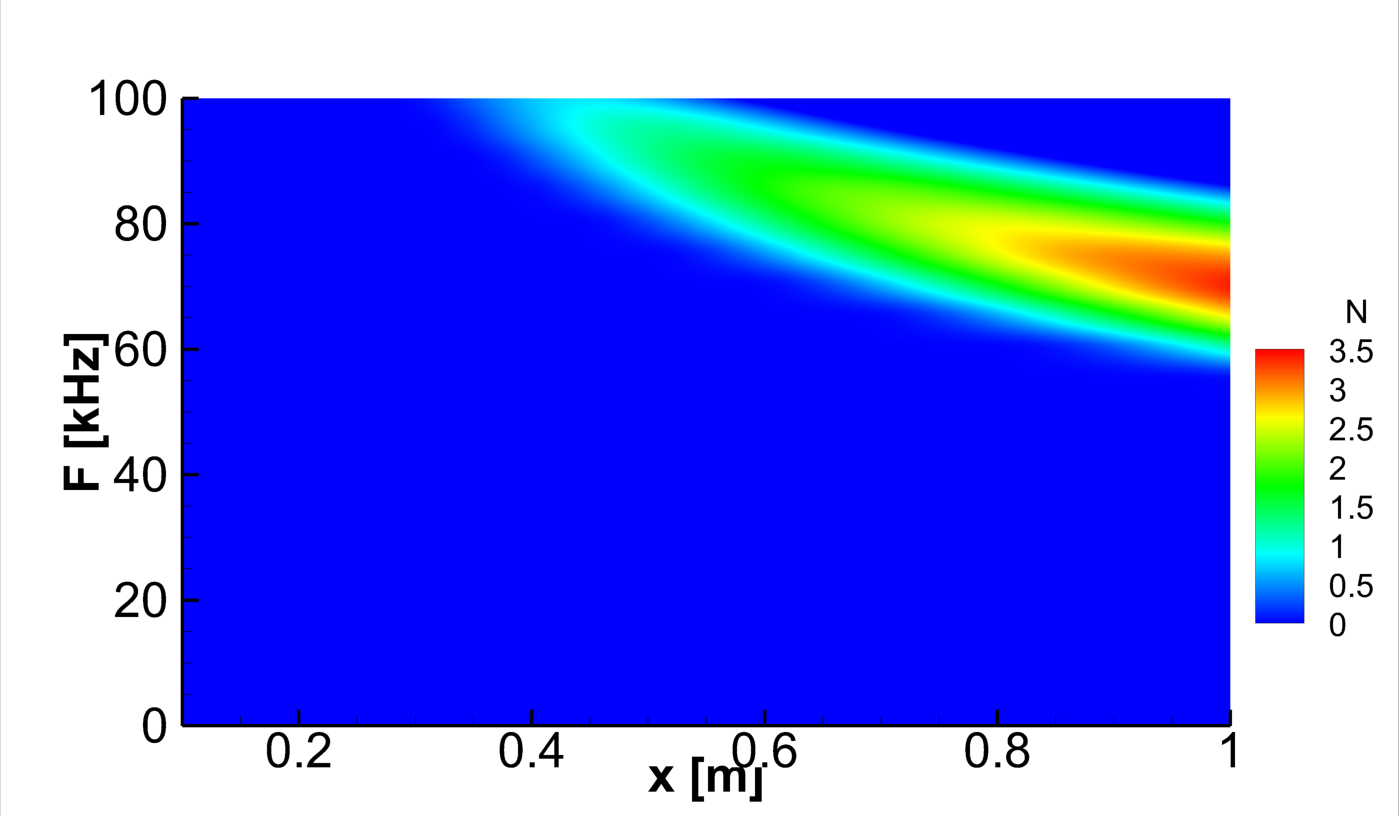}  & 
    \includegraphics[trim=2 2 2 2, clip,width = 0.3\textwidth]{ 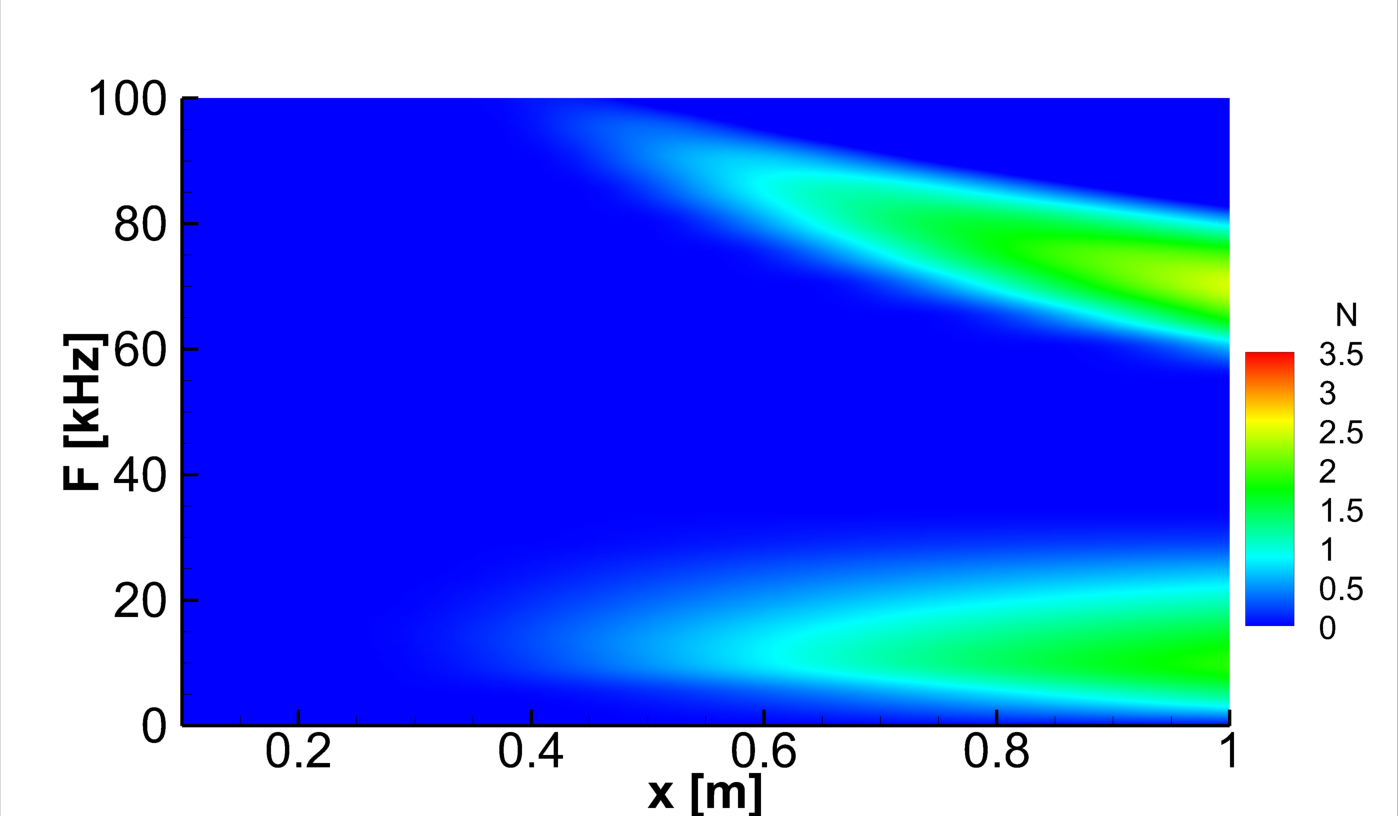} \\
     (g) & (h) & (i)\\      
    \includegraphics[trim=4 4 4 4, clip,width = 0.3\textwidth]{ 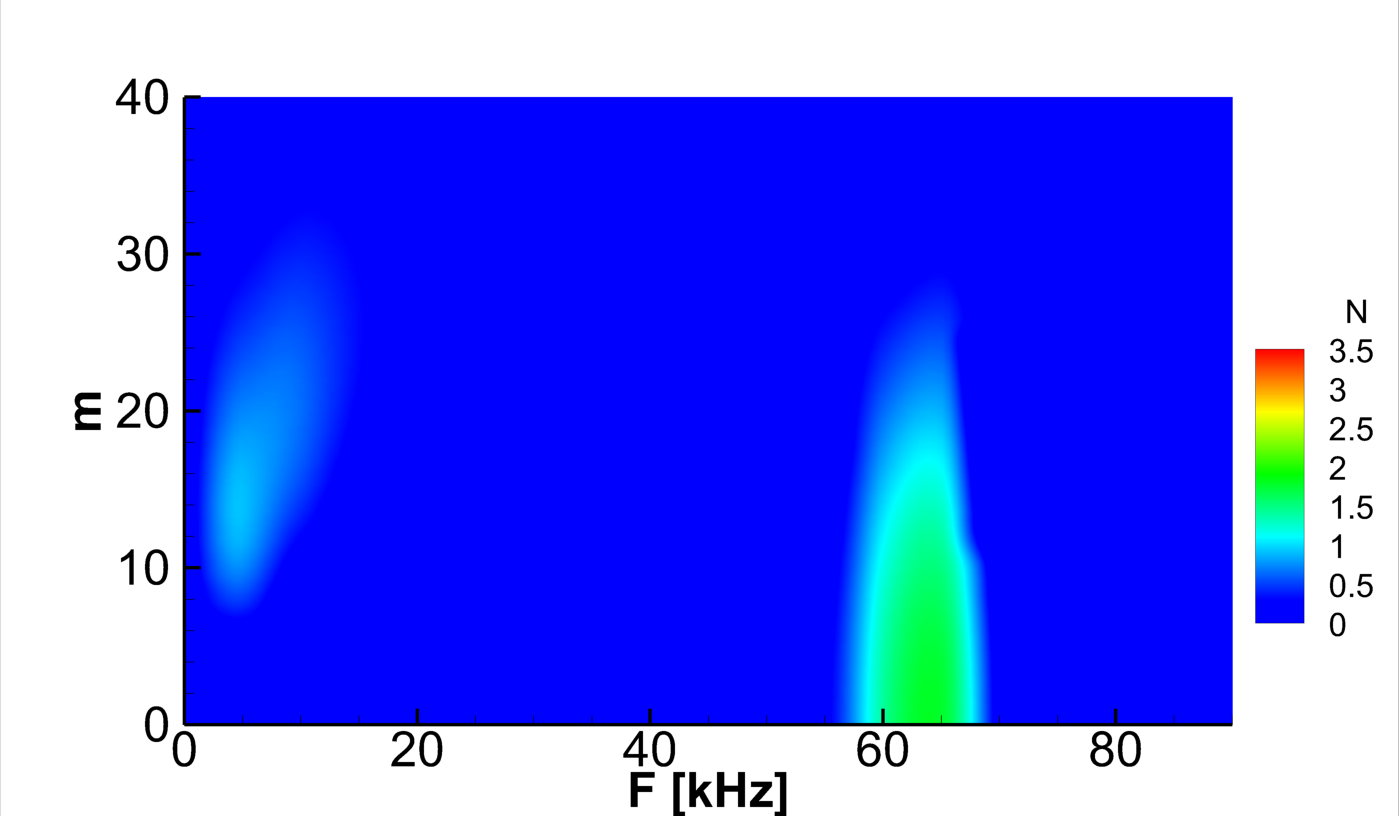} &
    \includegraphics[trim=2 2 2 2, clip,width = 0.3\textwidth]{ 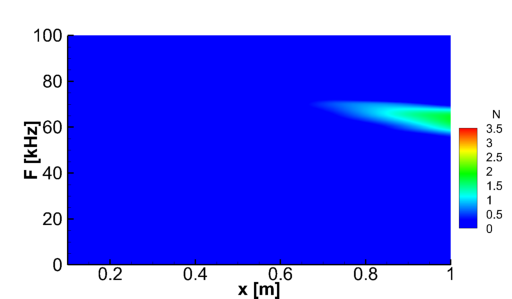}  & 
    \includegraphics[trim=2 2 2 2, clip,width = 0.3\textwidth]{ 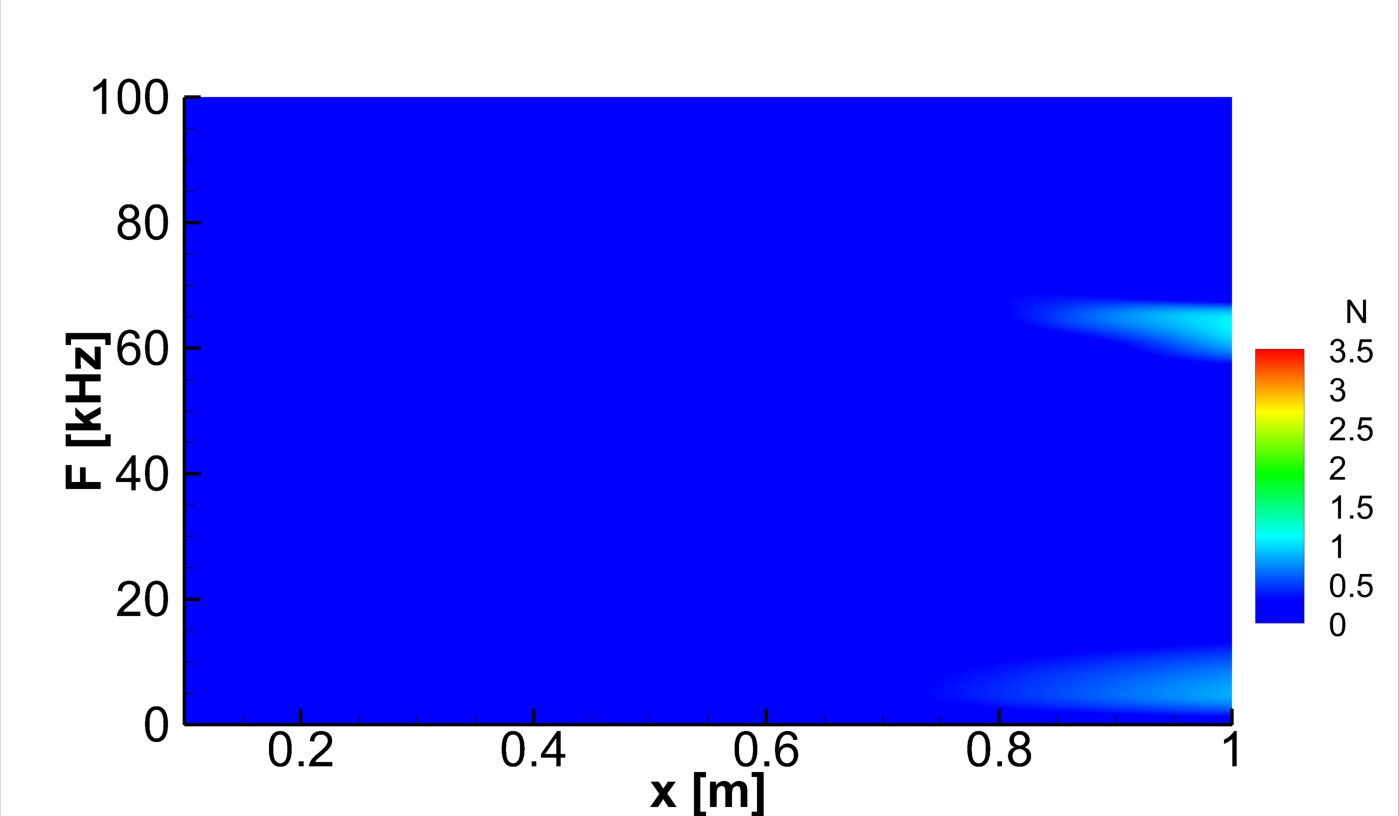}
   \end{tabular}
\caption{\label{fig:lst_results} LST N-factors contours for (a)--(c) the sharp cone, (d)--(f) the $R_N = 3.6 \unit{\milli\metre}$  blunt cone, and (g)--(i) the $R_N = 7.2 \unit{\milli\metre}$ blunt cone. N-factors are shown at $x = 1.0 \unit{\metre}$ as functions of frequency $f$ and azimuthal wavenumber $m$ in (a),(d),(g). N-factors are shown as a functions of $x$ and $f$ at $m = 0$ in (b),(e),(h), and $m = 17$ in (c), (f), (i). Panels (a),(d) both contain a peak at low frequency and high wavenumber corresponding to oblique Mack first mode instability, and a second peak at high frequency but low wavenumber corresponding to Mack second mode instability.  Comparing (a)--(c), (d)--(f) and (g)--(i), bluntness reduces the N-factors associated with both modes of instability, and thus should delay the onset of laminar to turbulent transition.}
\end{figure*}
\dac{Linear stability theory (LST) analyses were performed for all three boundary layers for frequencies $f = 0\text{--}100 \unit{\kilo\hertz}$ and for azimuthal wavenumbers $m = 0\text{--}40$. The least stable N-factors were computed by integrating the growth rate of unstable boundary layer modes in the streamwise direction, starting from the upstream neutral point in each case. Contours of N-factor amplification at $x = 1.0 \unit{\metre}$ are shown in Figure \ref{fig:lst_results}(a), (d), and (g) as functions of frequency and azimuthal wavenumber. The sharp cone boundary layer stability, shown in \ref{fig:lst_results}(a), supports an unstable modal lobe with a peak at 75 \unit{\kilo\hertz}, which is predominantly axisymmetric. This lobe is the well-known Mack second mode instability\cite{Mack1963,Mack1984,Malik1990,Malik1990b}. The less amplified lobe is oblique and is most unstable at 10 \unit{\kilo\hertz} and $m = 17$. This lobe is the oblique Mack first mode instability.  The first mode is unstable at frequencies from 5 kHz to 35 kHz and wavenumbers from 10 to 40, reaching a maximum N-factor around $N = 2$. The second mode is unstable from 60 kHz to 85 kHz and wavenumber from 0 to 30, reaching a maximum N-factor around $N = 3.5$. As the nose radius is increased to 3.6 mm, both the first mode and second mode lobes, shown in Figure \ref{fig:lst_results}(d), are slightly stabilized. The most amplified frequencies and wavenumbers are approximately the same as those of the sharp cone boundary layer, but the addition of bluntness reduces the frequency bandwidth and wavenumber range of both instabilities. The first mode is unstable in a frequency range from 5 kHz to 20 kHz up to $m = 30$. The effect of nose bluntness on the second mode is similar, supporting instabilities from 60 kHz to 80 kHz up to $m = 30$. Like the sharp cone, the 3.6 mm blunt cone contains modal amplification up to $N = 3.5$ for the second mode and $N = 2$ for the first mode at $x = 1.0 \unit{\metre}$. The increase in nose-tip bluntness from 3.6 mm to 7.2 mm has a much more drastic effect on the boundary layer stability, as visible in Figure \ref{fig:lst_results}(g). The most amplified frequency of the second mode at the end of the cone is closer to 60 kHz, and results in a much smaller N-factor around $N = 2$. The first mode instability is almost completely absent in the 7.2 mm cone boundary layer. Note that a maximum N-factor of 3.5 is somewhat modest. The maximum N-factor is very dependent on the Reynolds number, and we chose the Reynolds number as a compromise between simulating significant modal effects while at the same time keeping the grid resolution manageable as we developed new methods.}

\dac{Contours of N-factor as a function of streamwise distance are shown in Figure \ref{fig:lst_results}(b) and (c) for the sharp cone, (e) and (f) for the 3.6 mm blunt cone, and (h) and (i) for the 7.2 mm blunt cone. The streamwise amplification of the second mode for the sharp cone (Figure \ref{fig:lst_results}(b)) demonstrates that the instabilities upstream are tuned to higher frequencies. This occurs because the thinner boundary layer upstream supports the exponential amplification of trapped acoustic waves at higher frequencies. An increase in obliquity (going from Figure \ref{fig:lst_results}(b), where $m = 0$, to Figure \ref{fig:lst_results}(c), where $m = 17$) stabilizes the second mode and destabilizes the first mode. Similarly, N-factor contours are shown in Figure \ref{fig:lst_results}(e) and \ref{fig:lst_results}(f) for the 3.6 mm blunt cone. The neutral point for both the first and second modes shift downstream with the addition of nose bluntness, and the frequency tuning effect remains very similar, but the most unstable frequencies are lower for this blunt cone. The contours for the 7.2 mm tipped cone are shown in \ref{fig:lst_results}(h) and \ref{fig:lst_results}(i). The neutral point is significantly pushed downstream and the N-factors are much smaller than either of the other two. This is consistent with the well-documented prediction of the transition delay phenomenon \cite{Jewell2017}. Modal analysis predicts that an increase in nose bluntness has a monotonically stabilizing effect on the boundary layer.}

\subsection{Hierarchical input-output analyses at 70 kHz\label{subsec:high_frequency}}
\begin{figure*}[htb]    
\begin{tabular}{l l}
(a) & (b)\\
\includegraphics[trim=4 4 4 4, clip,width = 0.5\textwidth]{ 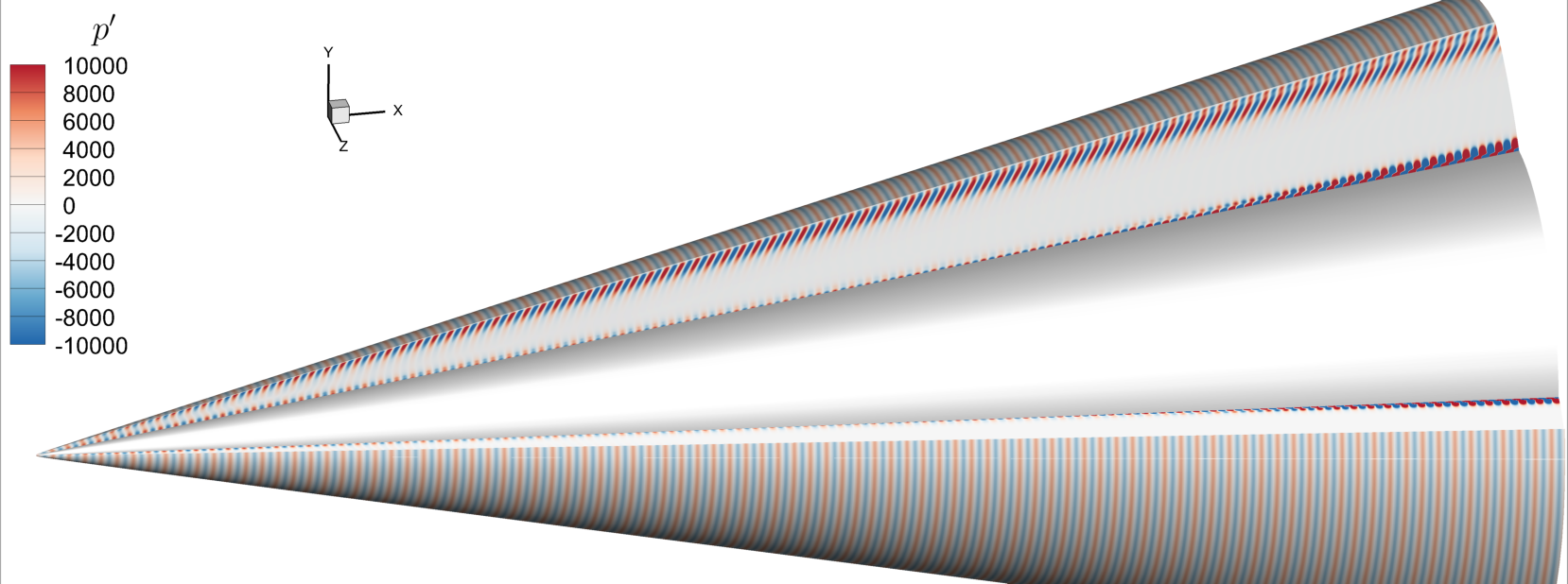} &
\includegraphics[trim=4 4 4 4, clip,width = 0.5\textwidth]{ 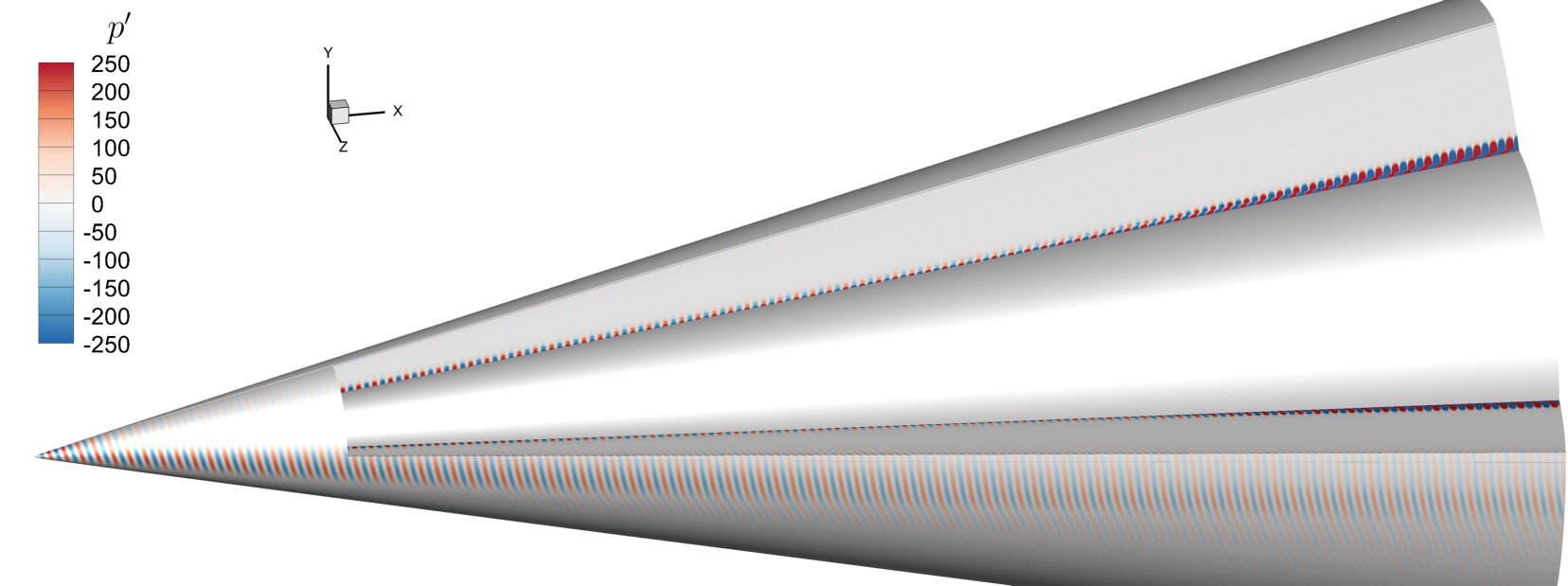}
\end{tabular}
\caption{\label{fig:sharp_cone_responses} Global linear response of the sharp cone to (a) a free-stream slow acoustic wave at $f = 70 \unit{\kilo\hertz}$ and $\psi = \ang{0}$, and (b) the leading input direction ($D_1$) at $f = 70 \unit{\kilo\hertz}$. Pressure contours on the outermost surface are shown in the free-stream. Pressure contours on the inner surfaces show the boundary layer response.}
\end{figure*}

We verify our approach by computing the global linear response in three dimensions to each type of free-stream wave (slow acoustic, fast acoustic, entropic, and vortical) at zero incidence angle ($\psi = \ang{0}$). We also compute the global linear response in three-dimensions to the slow acoustic wave at $\psi = \ang{0}$, $\psi = \ang{10}$, and $\psi = \ang{20}$. \dac{Instead of performing H-IO, we start by forcing a single free-stream wave at a time. In doing so, we still solve the resolvent system in much the same way as is done in the optional fifth step of H-IO. Note that this is different from performing a time-domain simulation; the resolvent system is not subject to the CFL condition and its accompanying numerical dissipation requirements. Single wave forcing is useful as a first study in order to create a clear comparison to existing numerical studies.} Figure \ref{fig:sharp_cone_responses}(a) shows pressure contours of the global three-dimensional response to a free-stream slow acoustic wave at 70 \unit{\kilo\hertz} and $\psi = \ang{0}$. The outermost surface shows contours of pressure in the free-stream, where the slow acoustic wave is visible. The inner slices along $\theta = \ang{0}$ and $\theta = \ang{90}$ show the amplification of post-shock disturbances in the boundary layer near the end of the cone. 

\begin{figure*}[htb]    
\begin{tabular}{c c c l}
\multicolumn{3}{l}{(a)} & (b) \\
(i) & (ii) & (iii) & 
  \multirow{2}{*}{ 
  \includegraphics[trim=4 4 4 4, clip,width = 0.4\textwidth]{ 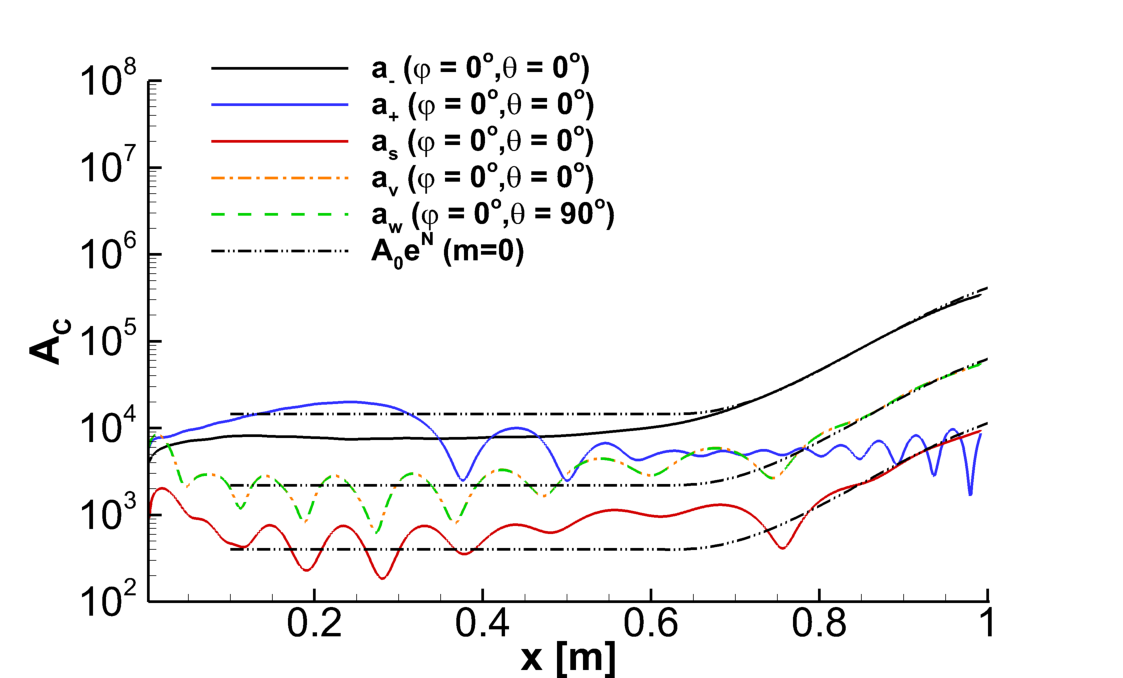}} \\
 \includegraphics[trim=2 2 2 2, clip,width = 0.15\textwidth]{ 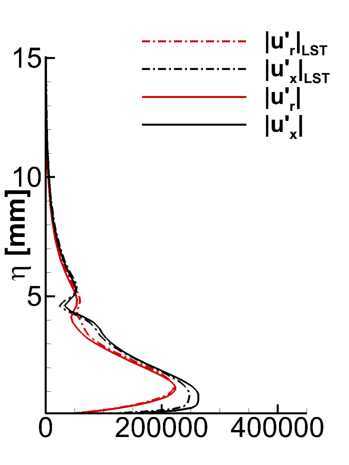} & 
 \includegraphics[trim=2 2 2 2, clip,width = 0.15\textwidth]{ 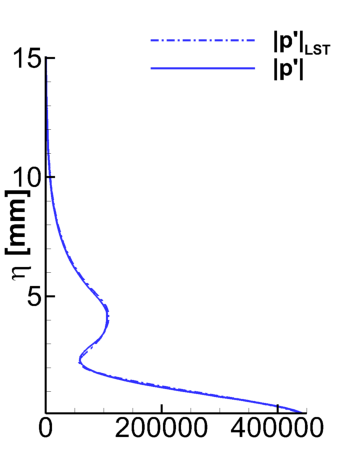} & 
 \includegraphics[trim=2 2 2 2, clip,width = 0.15\textwidth]{ 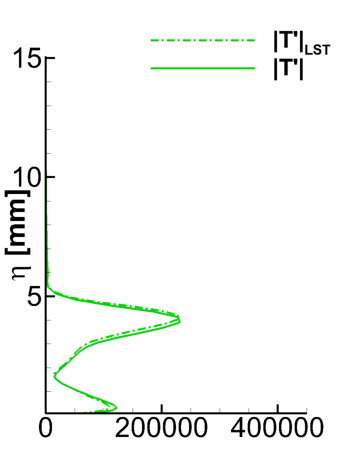} \\
\multicolumn{3}{l}{(c)} & (d) \\
\multicolumn{3}{l}{
\includegraphics[trim=4 4 4 4, clip,width = 0.45\textwidth]{ 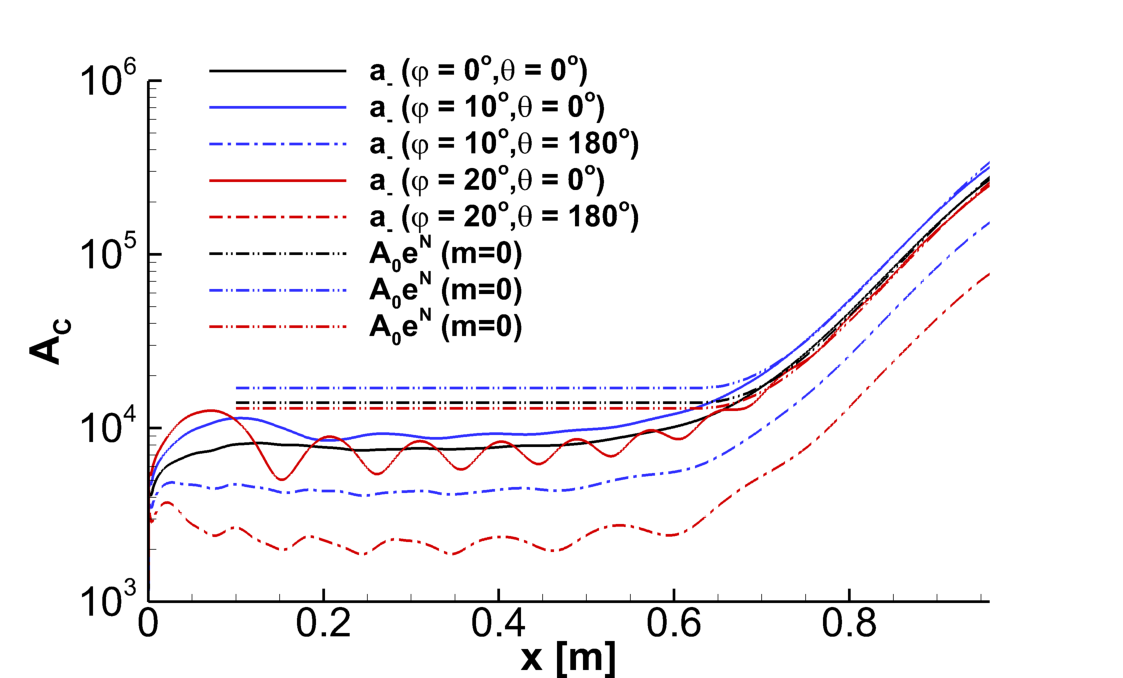}} &
\includegraphics[trim=4 4 4 4, clip,width = 0.45\textwidth]{ 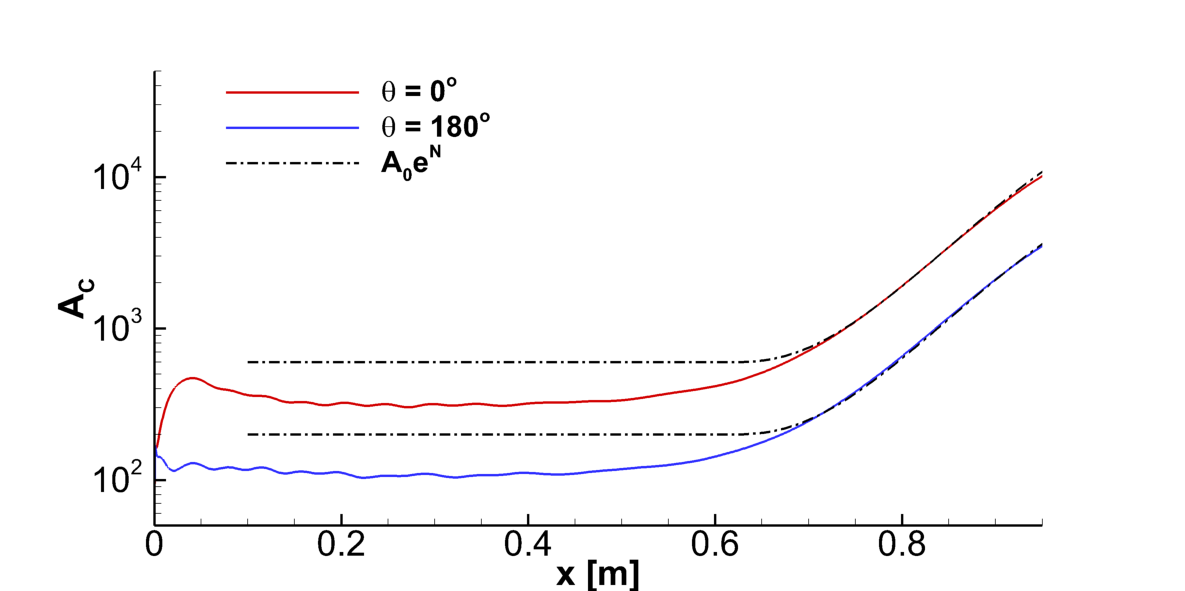}
\end{tabular}
\caption{\label{fig:sharp_data} \jwnresolved{(a) Wall-normal profiles of fluctuating (i) velocity, (ii) pressure, and (iii) temperature at $x = 1.0\unit{\metre}$ and $\theta = \ang{0}$ for the sharp cone boundary layer. The profiles from the forced response (solid lines) are compared to LST eigenfunctions (dash-dotted lines) corresponding to axisymmetric Mack 2\ts{nd} mode instability.  Wall-parallel profiles along the sharp cone of Chu energy amplitude in response to (b) five different types of free-stream waves at $\psi =\ang{0}$ angle of incidence, (c) slow acoustic waves at three different incidence angles, and (d) H-IO $D_1$ forcing.  In panels (b-d), the wall profiles are plotted with fitted $A_0 e^N$ functions corresponding to LST N-factors.}}
\end{figure*}

Comparisons between wall-normal profiles through the instability at the end of the domain ($x = 1.0 \unit{\metre}$, $\theta = \ang{0}$) and the Mack 2\ts{nd} mode eigenfunctions from the LST at $f = 70 \unit {\kilo\hertz}$ are shown in Figure \jwnresolved{\ref{fig:sharp_data}(a)}, verifying that the instability is the axisymmetric Mack 2\ts{nd} mode. The agreement of the velocity, pressure, and temperature profiles between the LST and the globally forced response is nearly perfect. 

\dac{Previous studies of the receptivity of sharp cone boundary layers to zero incidence waves have concluded that the Mack second mode is most receptive to slow acoustic waves \cite{Balakumar2015,Balakumar2011}.} Vortical waves also activate the mode, although less efficiently. Figure \jwnresolved{\ref{fig:sharp_data}(b)} shows the streamwise growth of the Chu energy amplitude at the wall in response to free-stream forcing of each of the five wave types. \dac{The slow acoustic wave ($a_{-}$) produces the strongest response, followed by the vortical waves ($a_{uv}$ and $a_w$), and lastly the fast acoustic ($a_+$) and entropic ($a_s$) waves.} The two different types of vortical waves produce identical responses, but the azimuthal locations of the maximum growth are $\ang{90}$ out of phase with each other due to the orthogonality of the vortical wave vectors to each other. The fast acoustic wave produces an initial amplification of a fast boundary layer mode upstream of $x = 0.3 \unit{m}$, but is not able to activate the Mack mode downstream at this frequency. These results agree well with previous receptivity studies which show that slow acoustic waves produce the strongest response, because of their resonance with the Mack mode~\cite{Ma2005}. \dac{This study also showed that} the fast acoustic wave initially excites an upstream boundary layer mode, which is then damped prior to a delayed onset of the Mack 2\ts{nd} mode. At this Reynolds number, we have insufficient streamwise extent of the domain to capture this effect. The vortical/entropic wave successfully activate the modal growth, but produce lower amplitudes than the slow acoustic wave. The receptivity of the slow acoustic wave is the highest with $C_a = 5.54$, which is over six times higher than that of the vortical waves and thirty times higher than for entropy waves. This is similar to results for acoustic forcing over sharp cones in which the receptivity to slow waves is higher for flows with adiabatic wall conditions. Although the mean flow was computed using an isothermal wall, it is very near the adiabatic wall temperature for the mean flow conditions, and so the receptivity coefficients do agree \jwnresolved{well} with what has been previously observed \cite{Balakumar2015}.

The effect of incidence angle on boundary layer receptivity has also been studied, which showed higher receptivity on the leeward side of the wave incidence than on the windward side \cite{Balakumar2007,Balakumar2011,Balakumar2015}. Figure \ref{fig:sharp_data}(c) shows the wall energy amplitude in response to a forced slow acoustic wave at three incidence angles, showing higher amplification on the leeward side (with respect to the wave incidence) than on the windward side for waves with non-zero incidence, which agrees with previous studies. The highest receptivity occurs at $\psi = \ang{10}$ with a receptivity coefficient of $C_a = 6.75$. This confirms that the local receptivity on one side of the cone is higher for waves at non-zero incidence angles. \dac{Furthermore, it identifies $\psi = \ang{10}$ as the optimal angle for slow acoustic receptivity.}

\begin{figure*}[thb!]
\begin{tabular}{lll}
 (a) & (b) & (c)\\
  \includegraphics[trim=4 4 4 4, clip,width = 0.30\textwidth]{ 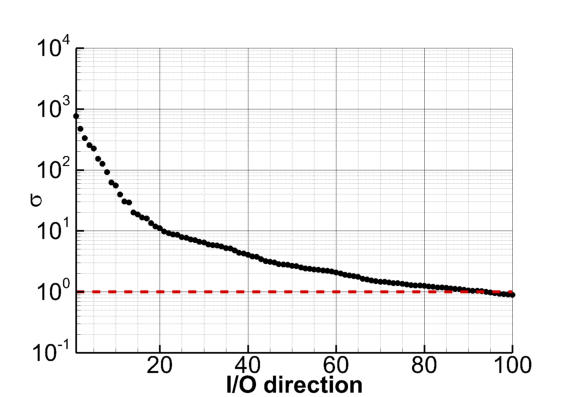}  & \includegraphics[trim=4 1 4 4, clip,width = 0.37\textwidth]{ 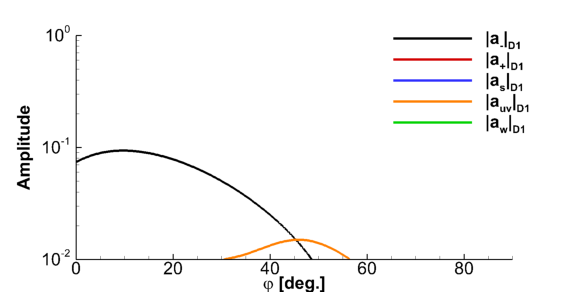}
  & \includegraphics[trim=4 4 4 4, clip,width = 0.32\textwidth]{ 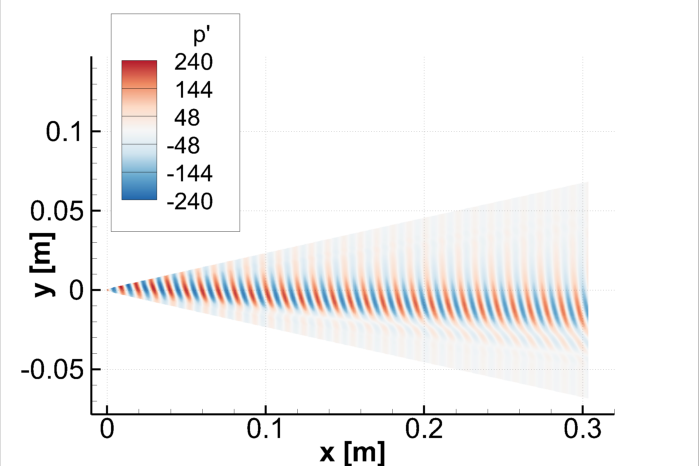}
  \\
 (d) & (e) & (f) \\
 \includegraphics[trim=4 4 4 4, clip,width = 0.30\textwidth]{ 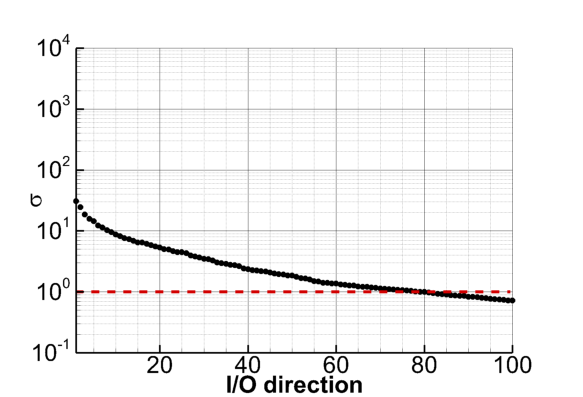} & \includegraphics[trim=4 1 4 4, clip,width = 0.37\textwidth]{ 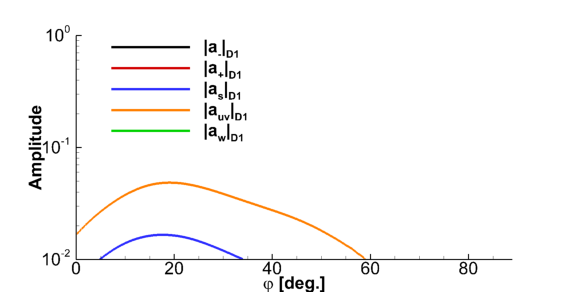} & \includegraphics[trim=4 4 4 4, clip,width = 0.32\textwidth]{ 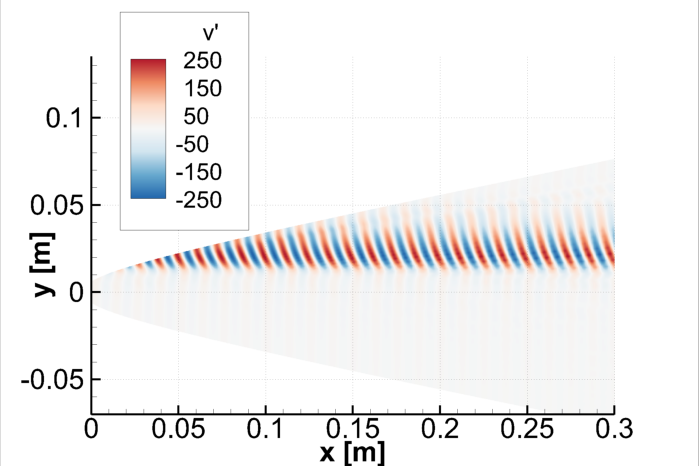}\\
 (g) & (h) & (i)\\
  \includegraphics[trim=4 4 4 4, clip,width = 0.30\textwidth]{ 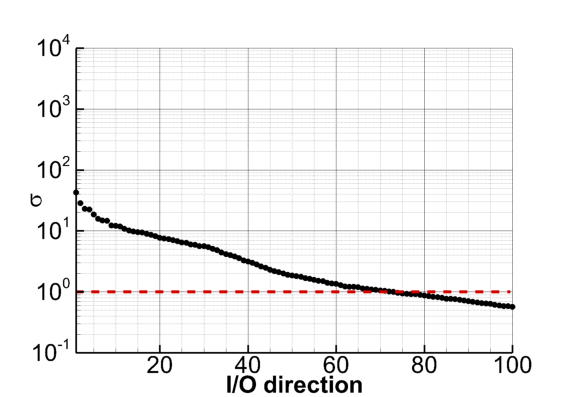} & \includegraphics[trim=4 1 4 4, clip,width = 0.37\textwidth]{ 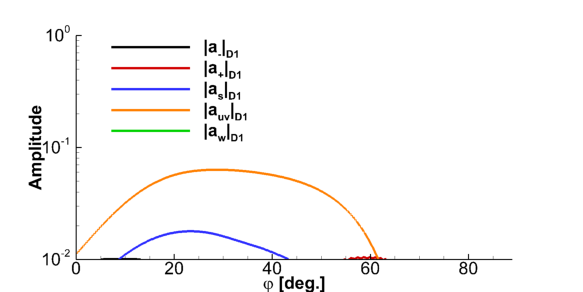} & \includegraphics[trim=4 4 4 4, clip,width = 0.32\textwidth]{ 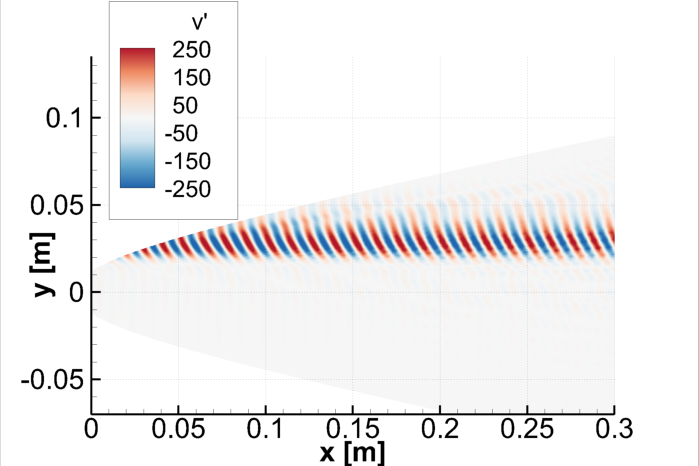}
\end{tabular}
\caption{\label{fig:io_results_f70} H-IO results at 70 kHz for (a)--(c) the sharp cone, (d)--(f) the 3.6 mm blunt cone, and (g)--(i) the 7.2 mm blunt cone. Shown are (a), (d), (e) gains versus I/O direction, (b), (e), (h) $D_1$ input directions, and (c), (f), (i) physical realizations of the optimal forcing in the free-stream. }
\end{figure*}

With some understanding of the receptivity mechanisms, we now demonstrate the ability of the H-IO analysis to accurately predict the global receptivity to free-stream waves. The gains from an H-IO analysis at $f = 70 \unit{kHz}$ are shown in Figure \ref{fig:io_results_f70}(a). We term the leading I/O direction (the direction with the highest gain) $D_1$, the second direction $D_2$, and so on. The gain associated with $D_1$ is around 760. The magnitude of the gains trail off very rapidly, and by $D_{100}$, the gain drops below 1, indicating that directions past this threshold do not produce amplified physical responses in the output region. The amplitude distribution and its physical realization in the free-stream for $D_1$ are shown in Figure \ref{fig:io_results_f70}(b) and \ref{fig:io_results_f70}(c). Two distributions are visible, the first and largest of which contains slow acoustic waves with a peak amplitude occurring at $\psi = \ang{10}$. The second smaller peak occurs around $\psi = \ang{45}$ and is of the first type of vortical waves, which corresponds to fluctuations in $u$ and $v$ velocity.

This agrees very well with the predictions from computing the response to direct forcing. The boundary layer is receptive first to slow acoustic waves, followed by vortical waves. Fast acoustic and entropy waves are not selected for $D_1$, but occur in sub-optimal forcing directions. Furthermore, the optimal angle for slow acoustic waves in the free-stream is a distribution of waves from $\psi = 0 \text{--} \ang{45}$ with a peak at $\psi = \ang{10}$, which is exactly the worst case angle observed in the single wave forcing results. This verifies that I/O analysis is capturing relevant and understood receptivity mechanisms: this boundary layer is most receptive to slow acoustic waves at angle of incidence $\psi = \ang{10}$. 

The direct response corresponding the $D_1$ is shown in Figure \jwnresolved{\ref{fig:sharp_cone_responses}(b).} The outermost surface shows contours of free-stream pressure corresponding the the physical realization of the $D_1$ input distribution. The slices downstream show the activation and growth of the Mack 2\ts{nd} mode instability. Profiles of Chu energy along the cone surface as a function of streamwise distance are shown in Figure \jwnresolved{\ref{fig:sharp_data}(d),} along with a fitted N-factor from the LST. Profiles are taken along the top and bottom of the cone at $\theta = \ang{0}$ and $\theta = \ang{180}$. The response to the optimal forcing is larger along the top of the cone (the leeward side), reaching a maximum amplitude triple that than to profile along the bottom (the windward side). \dac{Receptivity coefficients for the $D_1$ forcing are $C_a = 4.96$ on the leeward side and $C_1 = 1.65$ on the windward side. These coefficients are similar, but smaller than those from the single wave forcing case. This is due to the localized nature of the $D_1$ forcing packet in the free-stream. While the peak forcing amplitude of the $D_1$ free-stream wave packet produces a lower peak neutral point amplitude than the single slow acoustic wave, it is a much more efficient means by which to generate a similar downstream response.

The overall receptivity process predicted by the H-IO follows the known trends for sharp cone boundary layers\cite{Balakumar2011,Balakumar2015}. First, slow acoustic waves impinge on the shock and transmit through it, amplifying as they do so. Because the attached shock is in the close vicinity of the boundary layer, the slow acoustic mode is directly activated upstream and persists in the boundary layer until it reaches the upstream neutral point and becomes the Mack second mode instability.}

\dac{We now consider H-IO analyses at the same frequency but for the two cones with blunted tips. The first 100 gains from an H-IO analysis of the $R_N = 3.6 \unit{\milli\metre}$ blunt cone at 70 \unit{\kilo\hertz} are shown in Figure \ref{fig:io_results_f70}(d). Overall, the gains are much lower than those for the sharp cone by more than a factor of ten, indicating far less energy amplification between the free-stream and boundary layer. The maximum gain only reaches around 30, and trails off below to less than one by $D_{60}$. Similarly, the first 100 gains for the $R_N = 7.2 \unit{\milli\metre}$ blunt cone are shown in \ref{fig:io_results_f70}(g). The leading gain is slightly higher than the other blunt cone, but only marginally so. Overall, the addition of nose bluntness reduces the global gain.
\begin{figure}[t!]
\begin{tabular}{l}
 (a) \\
 \includegraphics[trim=2 2 2 2, clip,width = 0.45\textwidth]{ 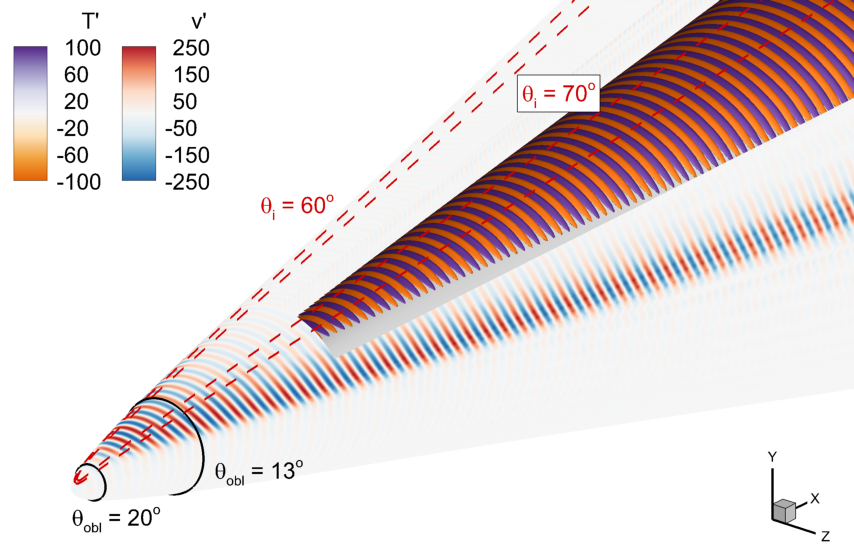} \\
 (b) \\
 \includegraphics[trim=2 2 2 2, clip,width = 0.45\textwidth]{ 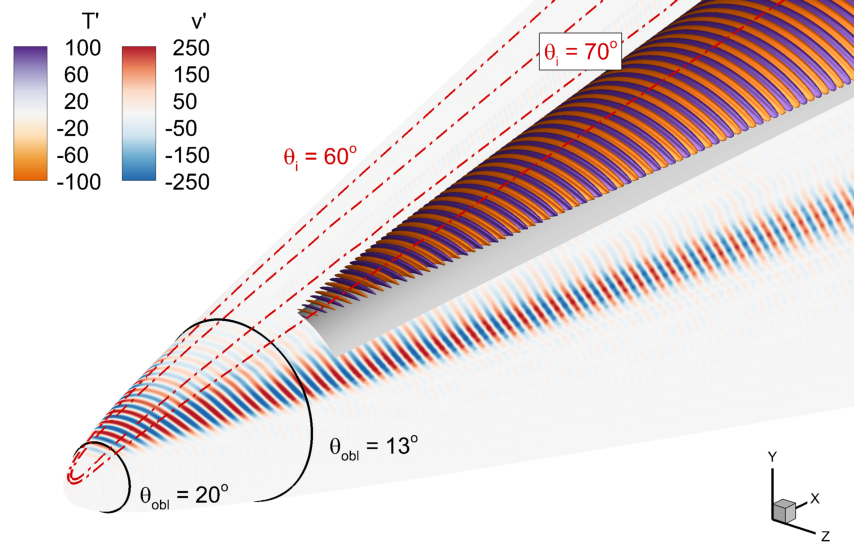} 
\end{tabular}
\caption{\label{fig:IO_f75_b3p6_3D} Global response of (a) the $R_N = 3.6\unit{\milli\metre}$ blunt cone and (b) the $R_N = 7.2\unit{\milli\metre}$ blunt cone to the $D_1$ forcing direction at 70 \unit{\kilo\hertz}. Contours on the outermost surface are in the free-stream, while density iso-surfaces show the downstream response. Reference contours of local shock obliqueness (solid lines) and incidence angles (dashed lines) are included.}
\end{figure}

The $D_1$ input distributions are shown in Figure \ref{fig:io_results_f70}(e) and (h) for each blunt cone, along with their physical realizations in the free-stream in \ref{fig:io_results_f70}(f) and (i). Each of the leading direction wave distributions is comprised of two peaks. For the 3.6 mm blunt cone, the first wave type is a $u$-$v$ vorticity distribution with a peak around $\psi = \ang{20}$. A smaller distribution of entropy waves is also present at the same incidence angle, but with a much smaller amplitude. The 7.2 mm blunt cone contains the same two wave types in the same relative proportions. The vortical distribution, however, is much more broadband, including incidence angles of up to $\psi = \ang{60}$. The physical realizations for both blunt cones are very similar to each other, including a thin band of waves down each side of the bow shock, which wraps around and impinges on the shock directly above the origin of the entropy layer. In contrast to the sharp cone boundary at 70 kHz, the blunt cone boundary layer is most receptive to vortical waves and entropic waves, instead of slow acoustic waves. 

The global responses of the blunt cones to the $D_1$ forcing directions are shown in Figure \ref{fig:IO_f75_b3p6_3D}. Contours of $y$-velocity fluctuations are shown in the free-stream to show the spatial realization of the $u$-$v$ vorticity wave. The wave selected is clustered around the top of the shock near the cone tip, but also extends in the streamwise direction down either side of the shock in two thin bands. The cutaway portion of the shock shows the growth of temperature fluctuations downstream. Azimuthal lines on the exterior of the shock location show contours of the mean shock obliqueness angle ($\theta_{obl}$) with respect to the free-stream mean flow direction. The dashed lines show contours of wave incidence angle ($\theta_i$) with respect to the shock obliqueness angle, computed from measuring the angle between peak free-stream wave vector (e.g., given by peak $\psi$ from Figure \ref{fig:io_results_f70}(e),(h)) and the local shock obliqueness angle. This allows us to map the peak spatial location of the $D_1$ forcing onto predictions from shock theory. 

Acoustic generation from incident vorticity waves is shown in Figure \ref{fig:shock_theory_Gv} as predicted by shock-disturbance theory \cite{McKenzie1968}. Incident vorticity waves impinge on the shock with shock-relative incidence angle $\theta_i$, and produce acoustic waves in three distinct regions. These regions correspond to whether the produced acoustic wave is fast, slow, or damped (with respect to the shock normal direction) \cite{Huang2019}. The arrows on Figure \ref{fig:shock_theory_Gv} show where on this plot the peak $D_1$ forcing occurs. The direction of the arrows indicates moving from upstream to downstream along the shock, following the position of the peak $D_1$ forcing.  The base of the arrow corresponds to the farthest upstream portion of the forcing wave, and the arrowhead follows the thin band of the forcing downstream. The peak of the forcing coincides with the production of a damped wave and a weakly generated fast acoustic wave. However, this incidence configuration is also the incidence for which transmitted vorticity is the highest \cite{Huang2019}, so the optimal forcing selects the vortical wave for which the maximum amount of vorticity is transmitted along with weak production of a fast acoustic wave. 
\begin{figure}[t!]
\centering
\includegraphics[trim=4 4 4 4, clip,width = 0.45\textwidth]{ 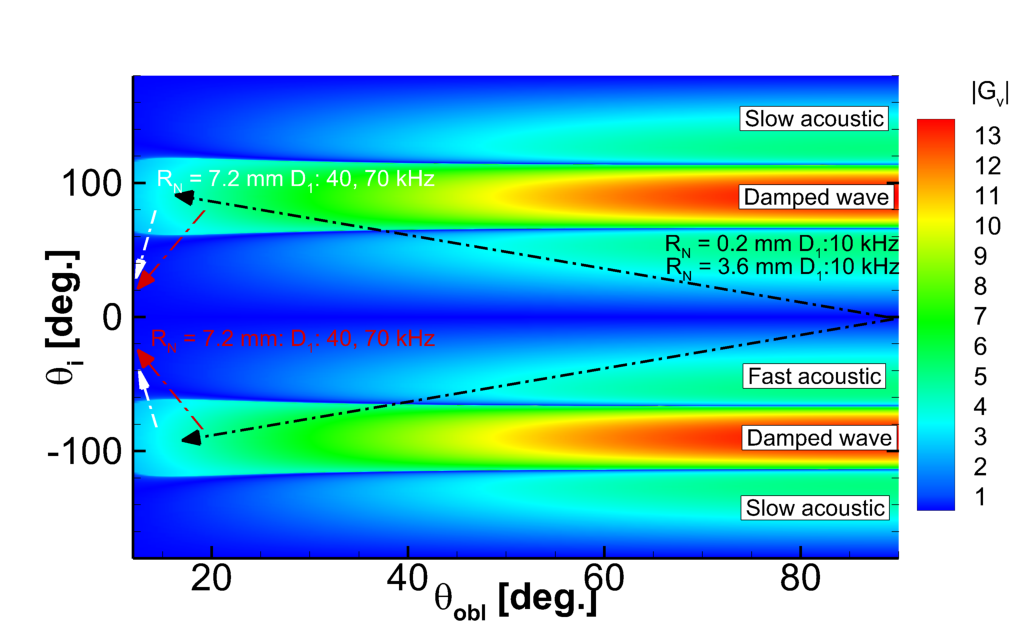}
\caption[Theoretical acoustic generation from vortical waves impinging on oblique shock waves as a function of obliqueness angle and incidence angle.]{\label{fig:shock_theory_Gv} Theoretical acoustic generation from vortical waves impinging on oblique shock waves as a function of obliqueness angle ($\theta_{obl}$) and incidence angle ($\theta_i$). }
\end{figure}
\begin{figure*}[t!]
\centering
\begin{tabular}{ll}
 (a) & (b) \\ 
 \includegraphics[trim=2 2 2 220, clip,width = 0.48\textwidth]{ 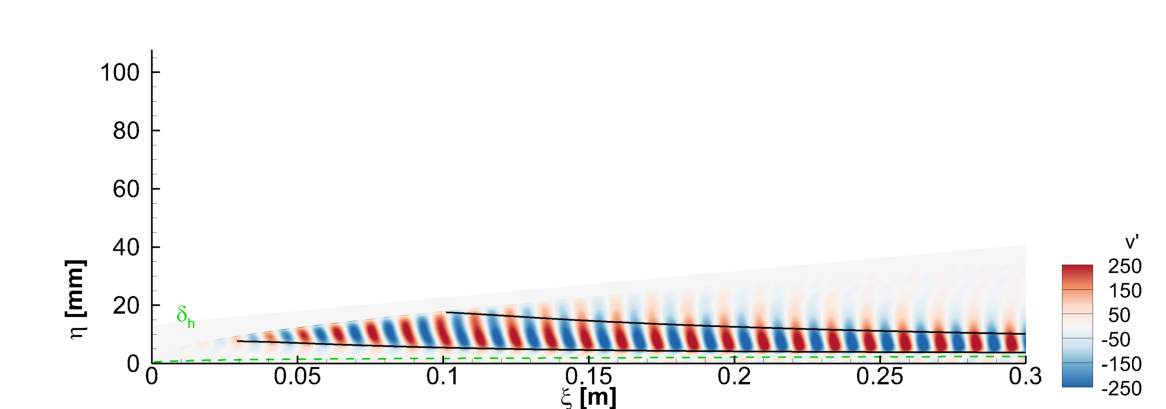} & \includegraphics[trim=2 2 2 220, clip,width = 0.48\textwidth]{ 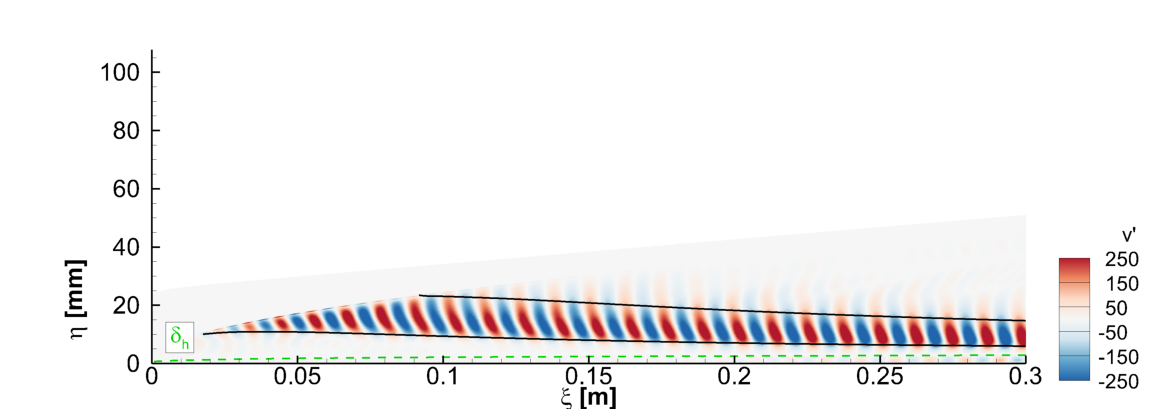} \\
 (c) & (d)\\
 \includegraphics[trim=2 2 2 85, clip,width = 0.48\textwidth]{ 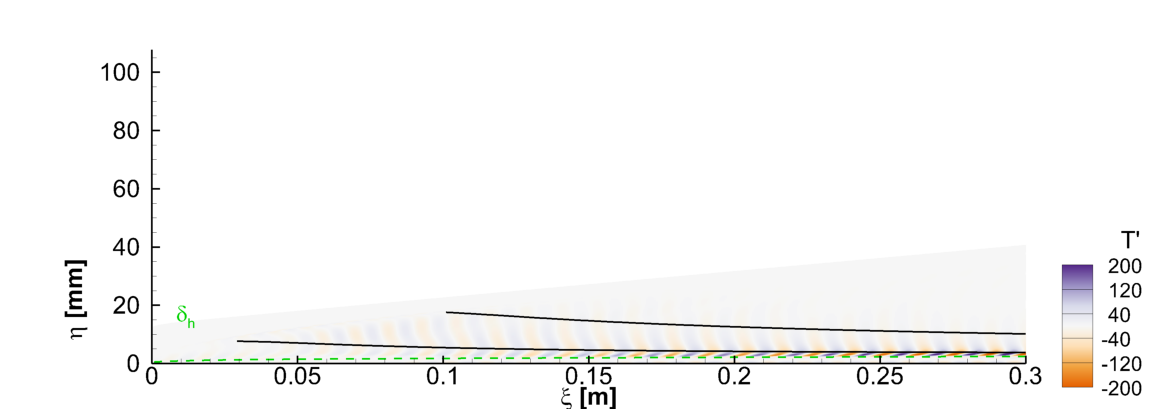} & 
 \includegraphics[trim=2 2 2 220, clip,width = 0.48\textwidth]{ 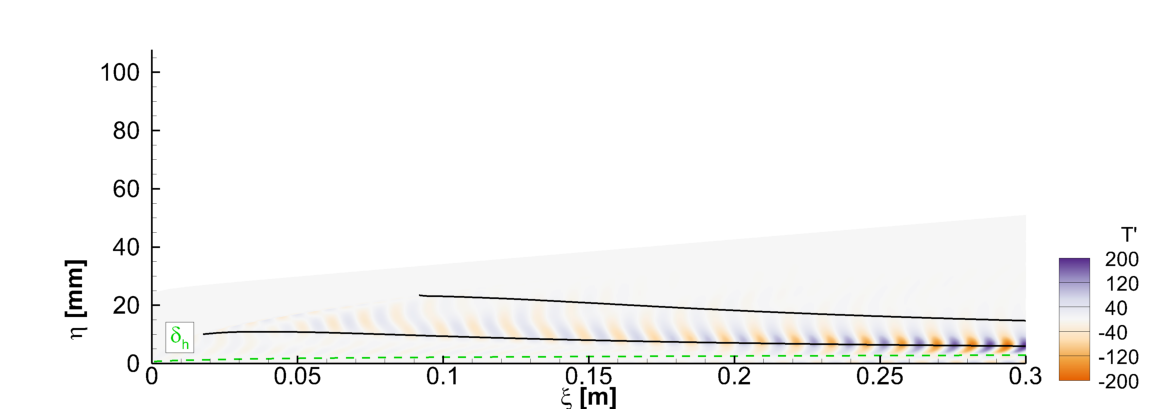}  \\
 (e) & (f)\\
 \includegraphics[trim=2 2 2 220, clip,width = 0.48\textwidth]{ 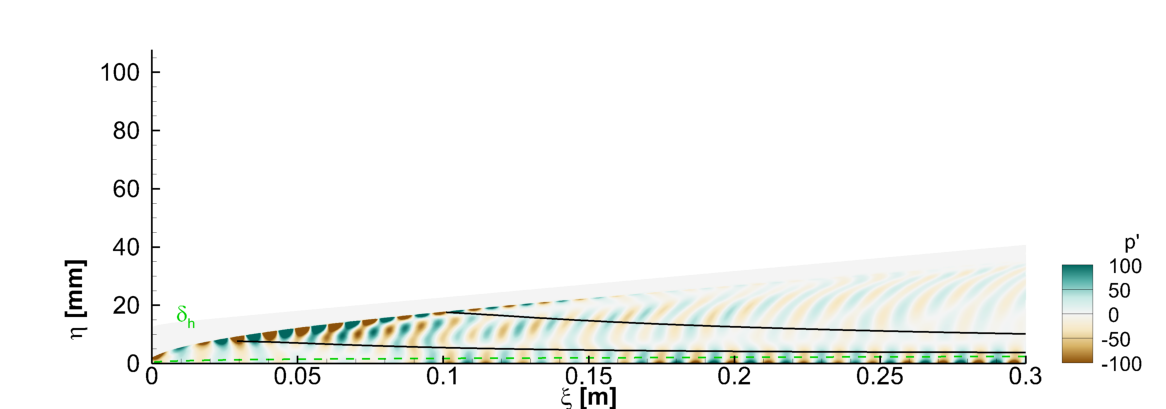} & 
 \includegraphics[trim=2 2 2 220, clip,width = 0.48\textwidth]{ 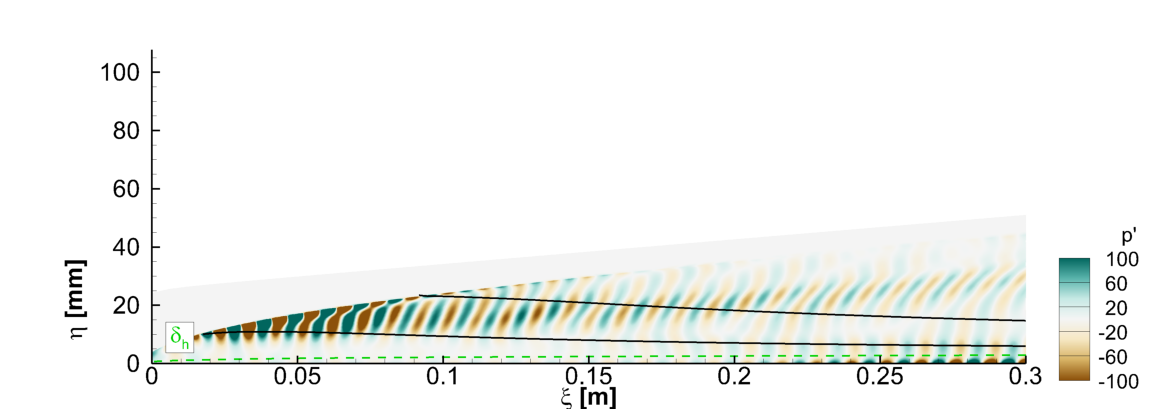}
\end{tabular}
\caption{\label{fig:2d_entropy_layer_f70} Contours of spatially amplifying (a)--(b) velocity, (c)--(d) temperature, and (e)--(f) pressure for the (a), (c), (e) $R_N = 3.6 \unit{\milli\metre}$ blunt cone and (b), (d), (f) $R_N = 7.2 \unit{\milli\metre}$ blunt cone at 70 kHz. The solid streamlines are extracted at the boundaries of the injected velocity packet, and the dashed lines show the boundary layer edge.}
\end{figure*}

Near the tip of the cone, contours spatially amplifying normal velocity, temperature, and pressure fluctuations are shown in Figure \ref{fig:2d_entropy_layer_f70} for both cones, along with a dashed line denoting the edge of the boundary layer. Contours are shown on a rotated grid such that the coordinate system is streamwise ($\xi$) and wall-normal ($\eta$) instead of $x$ and $y$. Also shown are two streamlines placed at the top and bottom edge of the injected velocity packet, downstream of the normal shock. In \ref{fig:2d_entropy_layer_f70}(a), the structure of the injected vorticity over the 3.6 mm cone is tilted upstream at an angle near the incidence angle of the pre-shock forcing wave. Because entropy and vorticity convect with the mean flow, the structure of the instability follows the inviscid entropy layer streamlines. Because the entropy layer is rotational, the injected disturbances rotate as they convent downstream, and the energy amplifies algebraically. This mechanism is related to the Orr mechanism, but it is inviscid, not viscous \cite{Orr1907a,Orr1907b}. As the streamlines move downstream, they also converge, even when the radial expansion of the flow is taken into account. This convergence corresponds to a flow deceleration, which further amplifies the rotating structures, compressing them into the top of the boundary layer. This process also amplifies the temperature fluctuations in a thin layer just outside the edge of the boundary layer, visible in \ref{fig:2d_entropy_layer_f70}(c). As noted earlier, the injection point of the vorticity wave theoretically produces a damped acoustic wave post-shock, and this is precisely what is observed in \ref{fig:2d_entropy_layer_f70}(e). Strong pressure fluctuations are visible where the pre-shock waves impinge, but the frequency is high enough and the distance between the wall and the shock is large enough that these acoustic waves are lensed away from the boundary layer on the underside of the shock. There are also some pressure fluctuations visible inside the boundary layer, starting near $x = 0.2 \unit{\metre}$. This is not related to the acoustic wave trapped beneath the shock, but arises directly from the entropy layer instability itself. 

Streamwise development of the entropy layer instabilities along with the wall energy signature is shown in Figure \ref{fig:streamwise_growth_f70_blunt} for both blunt cones. Because the entropy layer instability is due to the entropy and vorticity wave type, the amplification envelope can be easily captured by extracting several streamlines and plotting the Chu energy amplitudes along these streamlines. Together, these streamlines form an envelope of maximum amplification across the streamwise extent of the flow. Depending on the size of the entropy layer and the Reynolds number, some of the streamlines are swallowed by the entropy layer. Also shown in Figure \ref{fig:streamwise_growth_f70_blunt} are the height of the streamlines above the wall ($\delta_{str}$) and the boundary layer height ($\delta_{bl}$). For the 3.6 mm blunt cone, the entropy layer growth experiences a strong initial rise before it more gently plateaus. This initial rise is due to the combined effects of the rotation of the flow and the deceleration of the flow. Around $x = 0.5 \unit{\metre}$, some of the streamlines begin to be swallowed by the boundary layer, causing the rapid decay of energy as each streamline intersects the boundary layer edge. 
\begin{figure}[t!]
\begin{tabular}{ll}
 (a)  \\
 \includegraphics[trim=2 2 2 2, clip,width = 0.45\textwidth]{ 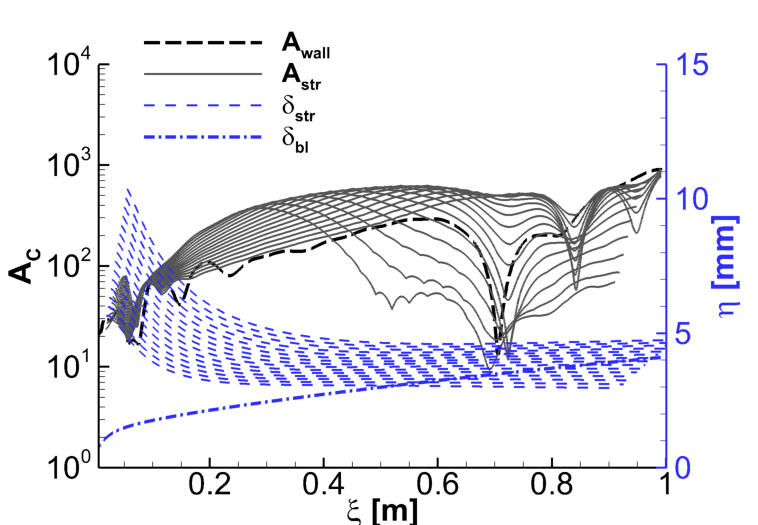} \\
 (b) \\
 \includegraphics[trim=2 2 2 2, clip,width = 0.45\textwidth]{ 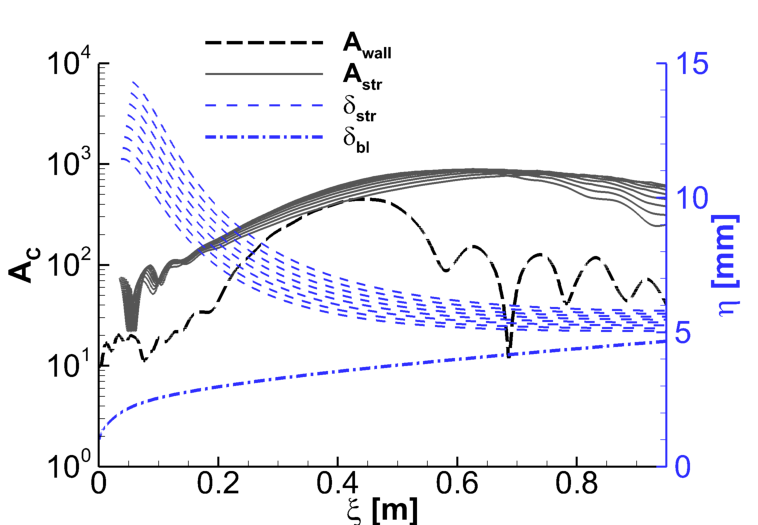} 
\end{tabular}
\caption[Streamwise energy amplification envelope of the entropy layer instability taken along several entropy layer streamlines for the $R_N = 3.6 \unit{\milli\metre}$ cone and $R_N = 7.2 \unit{\milli\metre}$ cone at 70 kHz.]{\label{fig:streamwise_growth_f70_blunt} Streamwise energy amplification envelope of the entropy layer instability taken along several entropy layer streamlines for (a) the $R_N = 3.6 \unit{\milli\metre}$ blunt cone and (b) the $R_N = 7.2 \unit{\milli\metre}$ blunt cone at 70 kHz. Also shown are streamline heights above the wall ($\delta_{str}$) and the boundary layer edge height ($\delta_{bl}$) as a visualization of where the entropy layer interacts with the boundary layer. }
\end{figure}
One feature visible in \ref{fig:streamwise_growth_f70_blunt}(a) is the slow growth of wall energy in response to the entropy layer instability. Note that the shock curvature and distance from the wall prevent the direct injection of acoustic energy into the boundary layer, as shown in \ref{fig:2d_entropy_layer_f70}(e); the downstream pressure is only present in the near shock region immediately downstream of the tip, and is not directly stimulating a boundary layer mode upstream. It is curious, then, that the wall energy is growing significantly upstream of the neutral point where both the F and S discrete modes are stable, and where they are not directly destabilized by the shock. Slices through the boundary layer and entropy layer profiles at $x = 0.5 \unit{\metre}$ shown in Figure \ref{fig:bl_f70_b3p6_b7p2} shed some light on the underlying physical mechanisms. Within the boundary layer, the profiles agree very well with the discrete F mode from the LST, while outside of the boundary layer, the profiles deviate due to the strong entropy layer signature. The amplification of energy at the wall seems to be due to F mode destabilization by the entropy layer itself.
\begin{figure}[t!]
\centering
\begin{tabular}{lll}
 (a) & (b) \\
 \includegraphics[trim=2 2 2 2, clip,width = 0.21\textwidth]{ 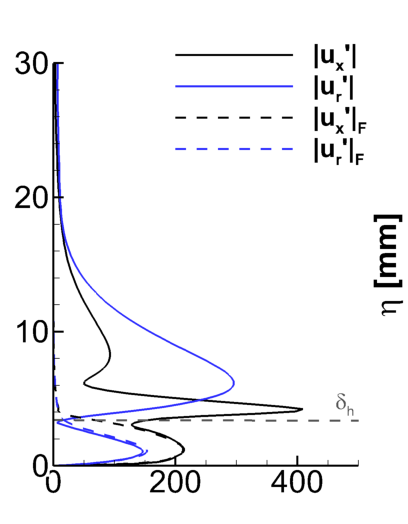} &
 \includegraphics[trim=2 2 2 2, clip,width = 0.21\textwidth]{ 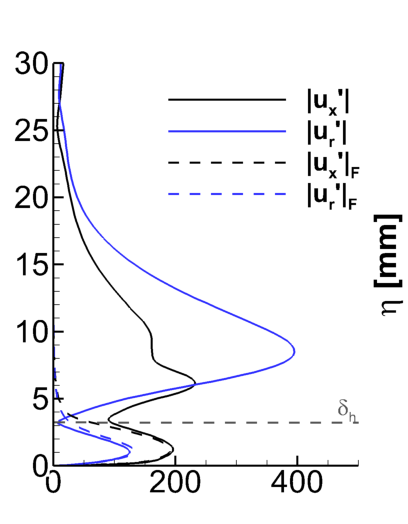} 
\end{tabular}
\caption{\label{fig:bl_f70_b3p6_b7p2} Profiles of absolute value of fluctuating velocity from the direct response to the $D_1$ forcing for the (a) $R_N = 3.6 \unit{\milli\metre}$ blunt cone at and (b) the $R_N = 7.2 \unit{\milli\metre}$ blunt cone. Profiles are shown along with the F modes from the LST at the same streamwise positions at $m = 0$.}
\end{figure}

In Figure \ref{fig:streamwise_growth_f70_blunt}(a), a significant effect occurs at $x = 0.7 \unit{\metre}$, leading to a large dip and subsequent growth of the wall energy. This particular flow contains the overlapping of several key phenomena at a single point. First, the upstream neutral point at this frequency is around 0.6 m, while the theoretical entropy swallowing point is around 0.67 m. At this streamwise position, the swallowing of the entropy layer plays a key role in the stimulation of the Mack second mode. The swallowing process is visible at the end of the flow domain, and is shown in Figure \ref{fig:2d_entropy_swallowing_f70}(a) for the 3.6 mm blunt cone via contours of fluctuating temperature. As the entropy layer undergoes the swallowing process, the disturbances penetrate into the boundary layer edge and decelerate quickly, stimulating and activating the beginning of the Mack second mode. This particular effect has been known to occur in some cases. The entropy layer can stimulate fast acoustic growth, which damps, and then directly enters the boundary layer, activating the Mack mode further downstream~\cite{Wan2020}.
\begin{figure*}[t!]
\centering
\begin{tabular}{ll}
 (a) \\ 
 \includegraphics[trim=2 2 2 220, clip,width = 0.95\textwidth]{ 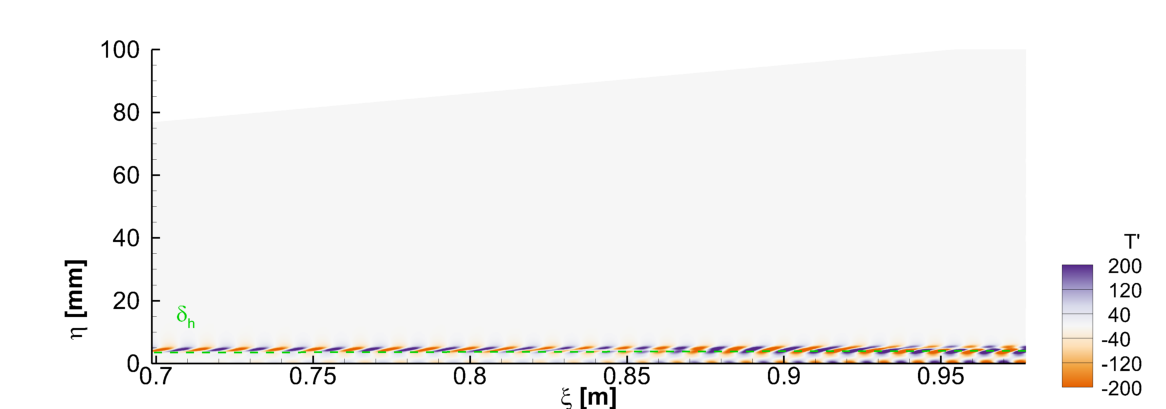} \\
 (b) \\
 \includegraphics[trim=2 2 2 220, clip,width = 0.95\textwidth]{ 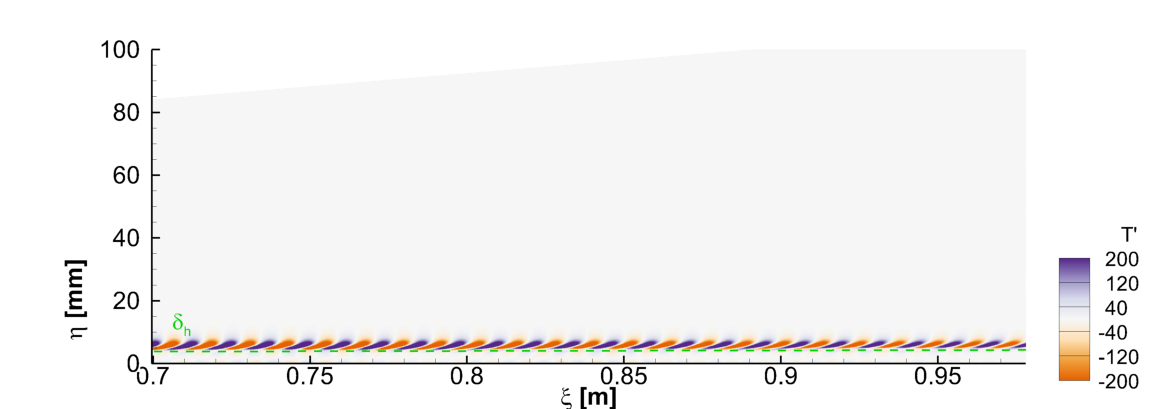} 
\end{tabular}
\caption[Contours of fluctuating temperature near the end of the domain for the $R_N = 3.6 \unit{\milli\metre}$ blunt cone and $R_N = 7.2 \unit{\milli\metre}$ blunt cone, highlighting the effect of entropy layer swallowing.]{\label{fig:2d_entropy_swallowing_f70} Contours of fluctuating temperature near the end of the domain for the (a) $R_N = 3.6 \unit{\milli\metre}$ blunt cone and (b) $R_N = 7.2 \unit{\milli\metre}$ blunt cone. The entropy layer in (a) undergoes swallowing, whereas no swallowing occurs in (b). }
\end{figure*}

The $R_N = 7.2 \unit{\milli\metre}$ cone shares an initial upstream receptivity process with the $R_N = 3.6 \unit{\milli\metre}$ cone. The free-stream optimal wave packet is positioned in the same spatial region with the same dominant effect---maximization of transmitted vorticity through the shock. The comparison with shock theory in Figure \ref{fig:shock_theory_Gv} shows the peak of the spatial distribution to align very similarly to the previous blunt case. The initial phase of the downstream receptivity process also closely mirrors the previous case. Contours of spatially amplifying normal velocity, temperature, and pressure fluctuations for the $R_N = 7.2 \unit{\milli\metre}$ blunt cone are shown in Figure \ref{fig:2d_entropy_layer_f70}(b), (d) and (f). Upstream tilted fluctuating velocity structures tilt and compress as they decelerate downstream, causing amplification in the velocity and temperature signatures in the entropy layer. The pressure contours also show a pressure capturing effect. As observed earlier, the generated acoustic wave is lensed away from the boundary layer and does not immediately inject energy into the boundary layer. 

The spatial growth of the entropy layer envelope streamlines is shown in Figure \ref{fig:streamwise_growth_f70_blunt}(b) for the $R_N = 7.2 \unit{\milli\metre}$ cone. It shares several key features with the 3.6 mm blunt cone. First, the entropy layer grows via the rotation and deceleration mechanism. Furthermore, the deceleration of the entropy layer destabilizes the discrete F mode and causes significant growth in the wall energy amplitude. The growth of the F mode, confirmed by its presence in the boundary layer in Figure \ref{fig:bl_f70_b3p6_b7p2}, is stronger upstream for two reasons. First, the merging of the F mode with the continuous spectra occurs further upstream. The second reason for the stronger boundary layer response is that the larger nose bluntness causes a more aggressive deceleration, so the amount of acoustic energy generated by the entropy layer instability is higher. 

There are also several key differences between the two blunt cones. First, the $R_N = 3.6 \unit{\milli\metre}$ cone entropy layer shows a strong amplification upstream of $x = 0.2 \unit{\metre}$, which then slows before reaching a maximum amplitude around ten times that of the initial injected energy. The growth of the high frequency entropy layer instability for the $R_N = 7.2 \unit{\milli\metre}$ cone is initially slower, but reaches a slightly higher maximum amplitude around the same streamwise position. Another key difference between to two boundary layers is the absence of any modal activation in the 7.2 mm cone boundary layer. This is due not only to the fact that the S mode is significantly more damped, but also to the fact that the entropy layer is not swallowed at the end of the domain. This is apparent in Figure \ref{fig:2d_entropy_swallowing_f70}(b), in which the entropy layer instability is clearly above the entropy layer with no boundary layer instabilities visible. 

The overall receptivity process for blunt cones at high frequency activates a combined non-modal and modal interaction downstream. The optimal receptivity begins when vorticity waves at shallow incidence angles around $\psi = \ang{20}$ impinge on the shock. Then, they transmit and amplify, injecting vorticity downstream. This injected vorticity has an upstream tilted spatial structure, which amplifies via a combined rotation and deceleration mechanism, as the inviscid rotational streamlines converge downstream above the boundary layer. The deceleration produces fast acoustics, which destabilize a discrete F mode, upstream of its merging with the vortical and entropic continuous branch. In the absence of entropy swallowing, as in the 7.2 mm blunt cone, the entropy layer instability simply convects downstream, above the boundary layer, slowly fading due to continued rotation and its interaction with the boundary layer. When entropy swallowing is present, as in the 3.6 mm blunt cone, the disturbances can enter the boundary layer and activate the modal growth if any unstable boundary layer modes exist.}

\subsection{Hierarchical input-output analyses at 10 kHz}
\label{subsec:low_frequency}
\begin{figure*}[thb!]
\centering
\begin{tabular}{lll}
 (a) & (b) & (c) \\
 \includegraphics[trim=4 4 4 4, clip,width = 0.30\textwidth]{ 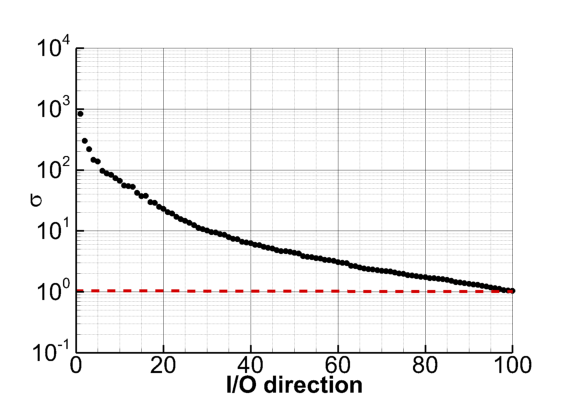}  & \includegraphics[trim=4 1 4 4, clip,width = 0.37\textwidth]{ 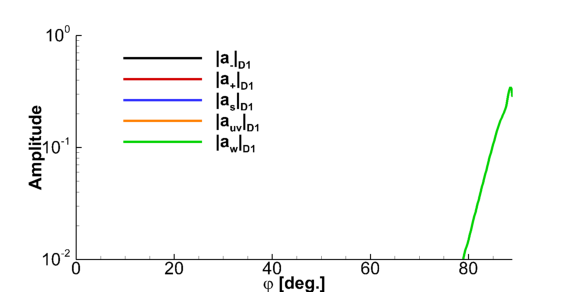} & \includegraphics[trim=4 4 4 4, clip,width = 0.32\textwidth]{ 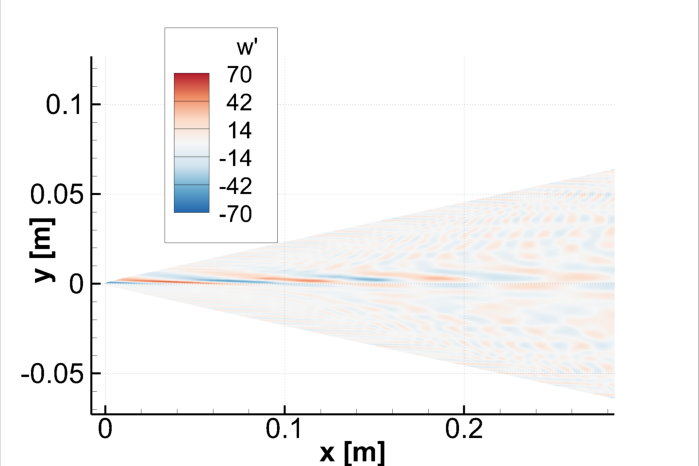} \\
 (d) & (e) & (f) \\
 \includegraphics[trim=4 4 4 4, clip,width = 0.30\textwidth]{ 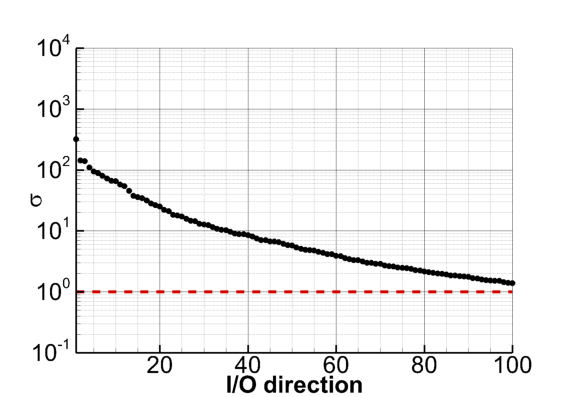} & \includegraphics[trim=4 1 4 4, clip,width = 0.37\textwidth]{ 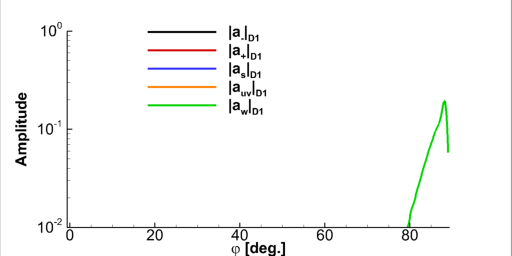} & \includegraphics[trim=4 4 4 4, clip,width = 0.32\textwidth]{ 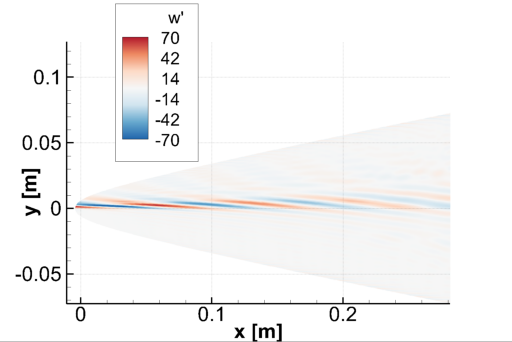}
\end{tabular}
\caption{\label{fig:io_results_f10} H-IO results at 10 kHz for (a)--(c) the sharp cone, (d)--(f) the 3.6 mm blunt cone. Shown are (a), (d) gains versus I/O direction, (b), (e) $D_1$ input directions, and (c), (f) physical realizations of the optimal forcing in the free-stream.}
\end{figure*}
\dac{The first 100 gains from an H-IO analysis of the sharp cone boundary layer are shown in Figure \ref{fig:io_results_f10}(a). The largest gain is around 800, with a sharp trail-off leading to a gain below 100 after $D_5$. By $D_{100}$, the gain is below unity, indicating that 100 directions are sufficient to capture amplifying flow features between the free-stream and the boundary layer. Modal analysis predicts stronger amplification from the Mack second mode than from the Mack first mode (see Figure \ref{fig:lst_results}). The leading gain from the H-IO, however, is actually slightly higher than that of the sharp cone at high frequency, which would not be expected from the modal analysis alone.

The leading input direction wave distribution is shown in Figure \ref{fig:io_results_f10}(b), containing a very dominant peak at $\psi = \ang{88}$ of the $w$-vorticity wave type. None of the other wave distributions are present in large magnitudes in the $D_1$ input distribution. These highly oblique waves create a very thin wave packet, clustered near the $x\text{--}z$ plane and across the sharp shock tip, as shown in Figure \ref{fig:io_results_f10}(c). Because the input distributions only select for wave angles in the $x$-$y$ plane, the $w$-vorticity fluctuations are unique in that the resultant velocity fluctuation amplitude is constant with respect to $\psi$. Unlike its effect on $u$-$v$ vorticity waves,  the effect of $\psi$ on $w$-vorticity is only to determine the spatial structure and placement of the free-stream wave packet, but does not not influence a relative amplitude of the $w$-velocity fluctuation. 

The effect of $\psi$ on both types of vorticity waves is illustrated in Figure \ref{fig:velocity_component_schematic}, where the shock depicted is aligned with the $z$--$y$ plane. The velocity components are denoted with respect to the shock coordinate system for generality, where $u_n$ is fluctuating velocity normal to the shock, and $u_t$ and $u_p$ are the two fluctuating velocity components tangent to the shock. The $u_n$ and $u_t$ components of velocity comprise the first type of vortical wave. At $\theta_i = \ang{0}$, the velocity fluctuations are parallel to the shock. As $\theta_i$ is increased to \ang{90}, the velocity fluctuations rotate such that the velocity fluctuation is in the normal direction only. Notice that the $u_p$ velocity fluctuation, which corresponds to the second type of vortical wave, the $w$-vortical wave, is independent of $\theta_i$. Thus, we must map $w$-vorticity onto the shock theory by thinking of it in terms of the fluctuating velocities instead of a vortical wave with a dependence on $\psi$.
\begin{figure}[b!]
\centering
\includegraphics[trim=4 4 4 4, clip,width = 0.45\textwidth]{ 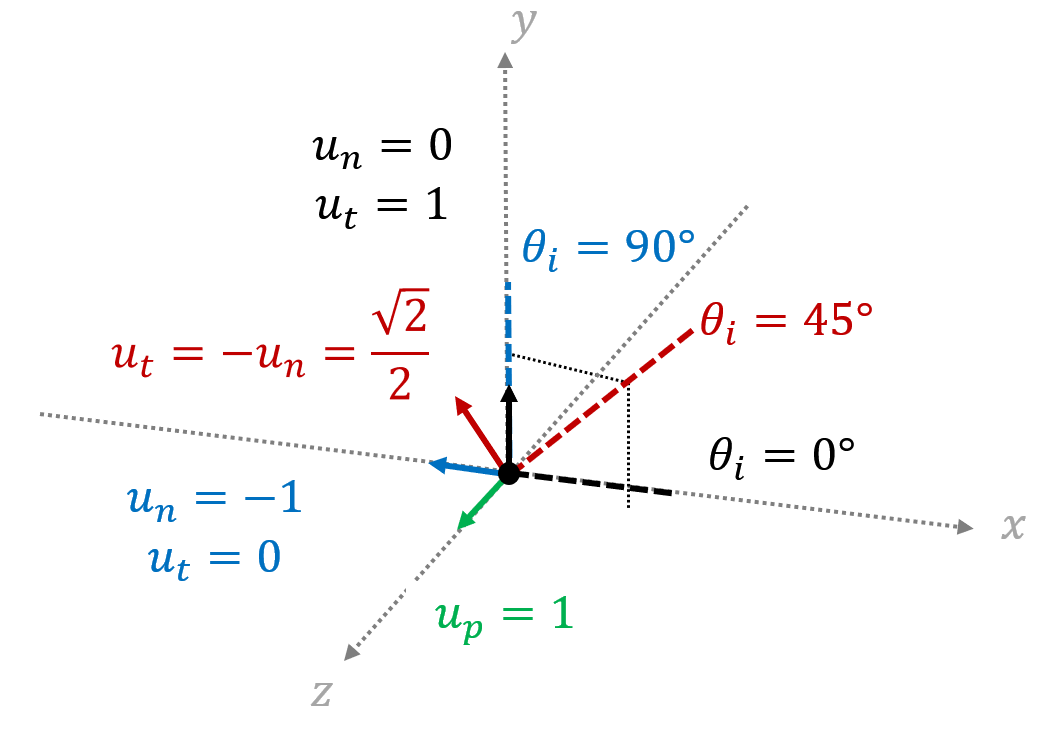} 
\caption[Schematic showing the dependence of fluctuating velocity on incidence angle for different types of vortical waves.]{\label{fig:velocity_component_schematic} Schematic showing the dependence of fluctuating velocity on incidence angle for different types of vortical waves. Velocities $u_n$ and $u_t$ comprise the first type of vortical wave, while $u_p$ comprises the second type of vortical wave.}
\end{figure}

Because the cone is sharp, the oblique shock angle is a constant function of the Mach number and the cone half angle; the oblique shock angle is around \ang{12.5}. We can map the physical realization of the $D_1$ forcing by taking the angle between the fluctuating velocity component and the oblique shock angle. From the $x$-$z$ plane moving upward in the $y$-direction, the peak of the forcing wave-packet occurs at $\theta = \pm \ang{75}\text{--}\ang{90}$, and corresponds to a incidence angle of $\theta_i = \ang{62.5}\text{--}\ang{77.5}$ in the shock theory formulation. A comparison with the theoretical curve shown in Figure \ref{fig:shock_theory_Gv} places this wave near the critical incidence angle for acoustic generation. This range of angles also extends into the damped wave regime, where there is a maximum point in the theoretical vorticity generation. Thus, the selection of the $w$-vorticity at this location maximizes both transmitted vorticity and acoustic waves at an angle very oblique to the free-stream mean flow direction. 
\begin{figure*}[t!]
\begin{tabular}{l l l}
(a) & \hspace{1cm} & (b)  \\
\includegraphics[trim=4 4 4 4, clip,width = 0.42\textwidth]{ 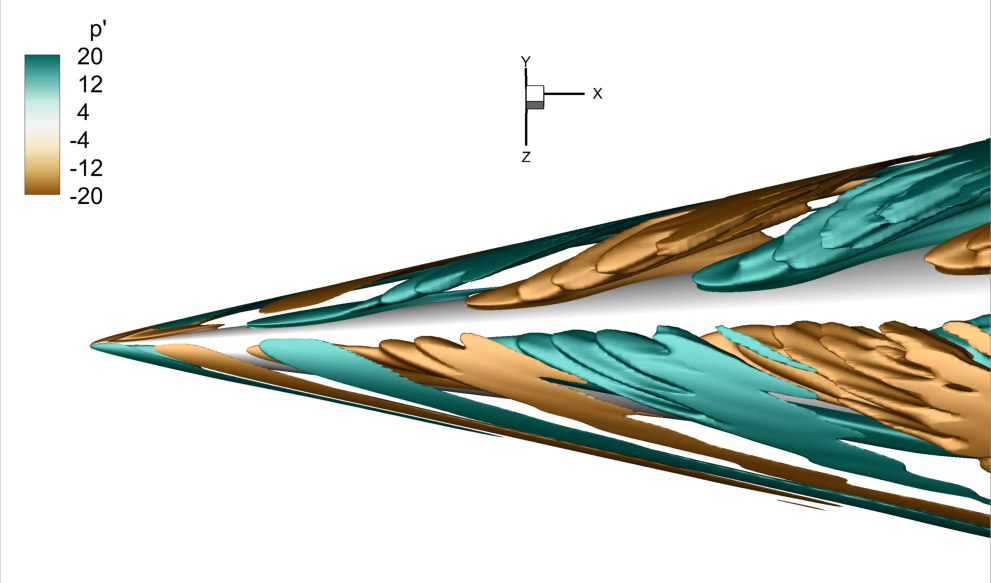} & &
\includegraphics[trim=4 4 4 4, clip,width = 0.42\textwidth]{ 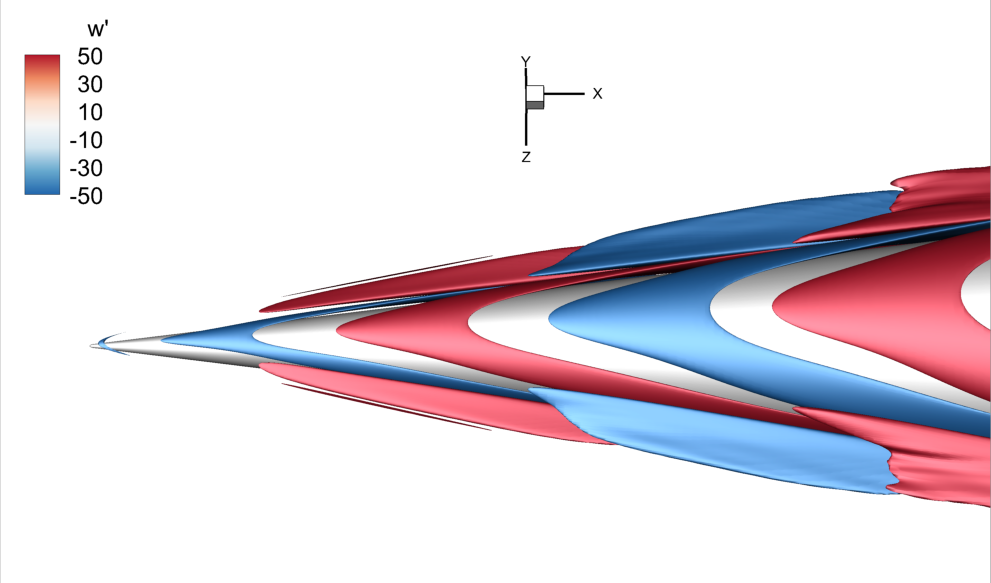} \\
(c) & \hspace{1cm} & (d) \\
\includegraphics[trim=4 4 4 4, clip,width = 0.42\textwidth]{ 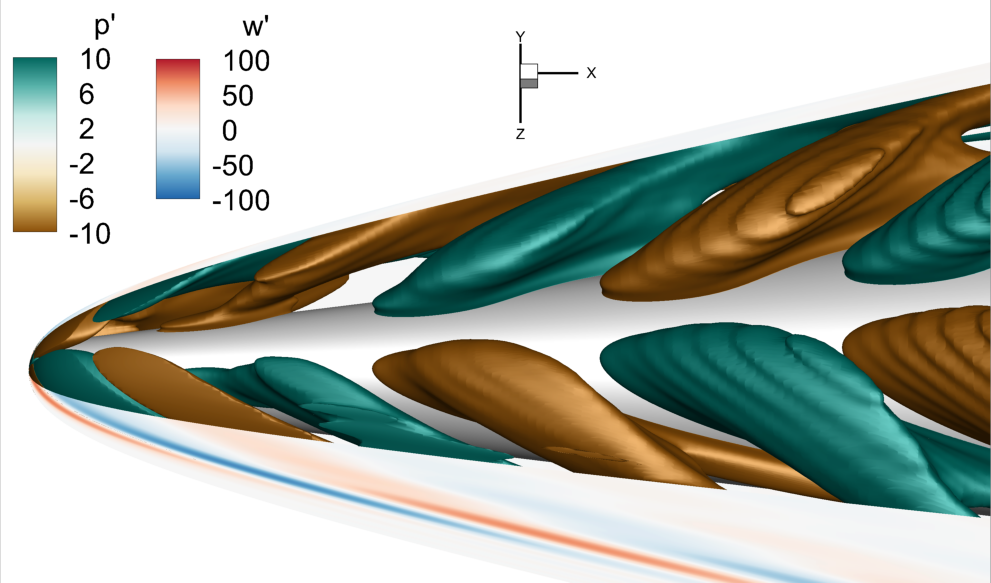} & &
\includegraphics[trim=4 4 4 4, clip,width = 0.42\textwidth]{ 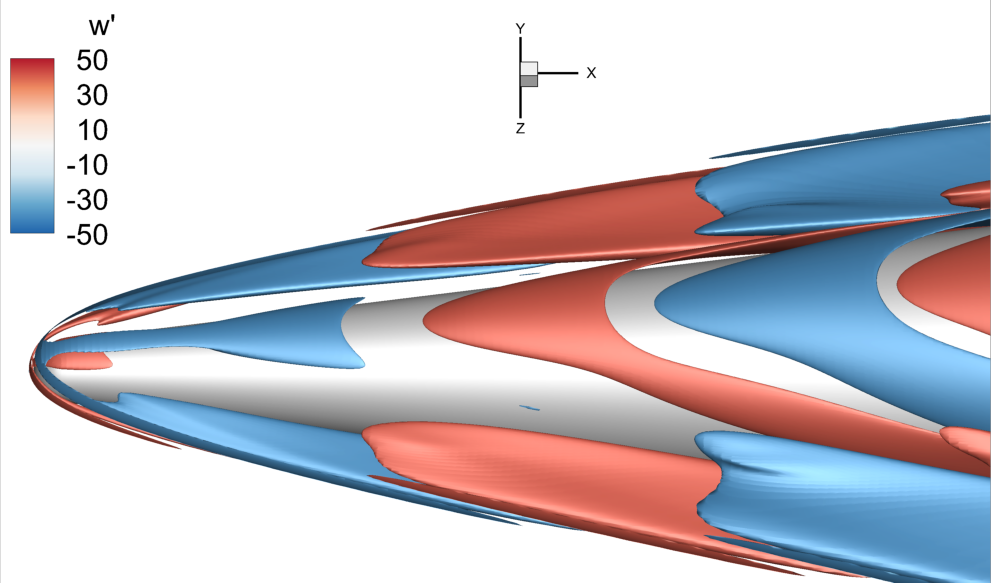} 
\end{tabular}
\caption{\label{fig:tip_receptivity} Iso-surfaces of (a),(c) pressure and (b),(d) $w$-velocity generated by incident $D_1$ vortical waves in the near-tip region of the (a)--(b) sharp cone and (c)--(d) 3.6 mm blunt cone at 10 kHz.}
\end{figure*}

Figures \ref{fig:tip_receptivity}(a) and (b) show iso-surfaces of perturbation pressure and velocity in the near-tip region of the flow over the sharp cone. The free-stream wave packet impinges on the shock and generates strong pressure waves in the sharp tip region post-shock. These waves enter the boundary layer in the vicinity of the sharp tip. The vorticity wave also transmits through the shock and amplifies into highly oblique structures in the near-tip region, visualized by the velocity iso-surfaces. These lobes of injected velocity extend from the shock to the wall but are also visible on the top of the cone in the boundary layer. 

The maximum amplification of the wall pressure occurs in very localized azimuthal positions. Figure \ref{fig:azimuthal_positions}(a) shows the absolute value of wall pressure as a function of the azimuth for several streamwise positions. The peak amplitude occurs around $\theta = \ang{60}$. A slice through the boundary layer instability at $\theta = \ang{60}$ and $x = 1.0 \unit{\metre}$ confirms that the growing instability is the Mack first mode. Wall-normal profiles of fluctuating velocity are shown in Figure \ref{fig:azimuthal_positions}(c), and the agreement between the profiles and the Mack first mode shapes is excellent.
\begin{figure}[t!]
\centering
\begin{tabular}{ll}
\multicolumn{2}{l}{(a)} \\
\multicolumn{2}{l}{\includegraphics[trim=4 4 4 4, clip,width = 0.45\textwidth]{ 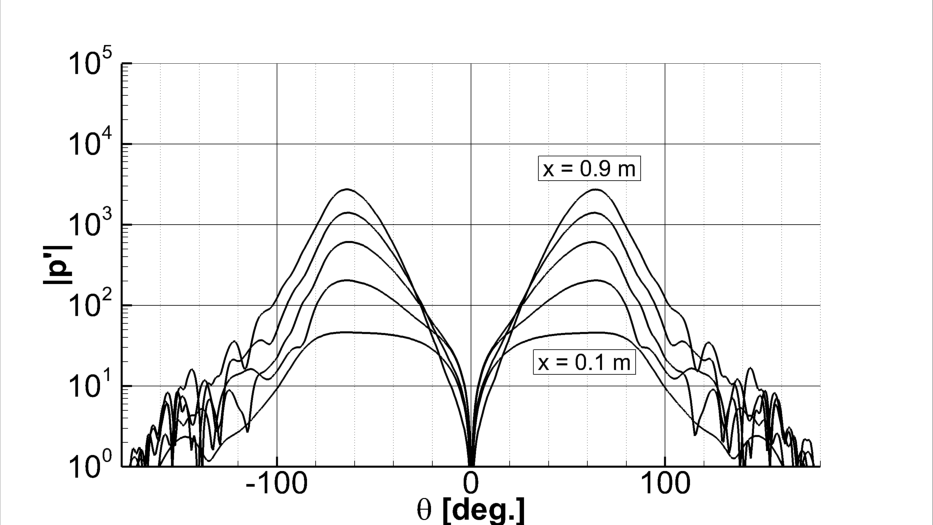}} \\
\multicolumn{2}{l}{(b)} \\
\multicolumn{2}{l}{\includegraphics[trim=4 4 4 4, clip,width = 0.45\textwidth]{ 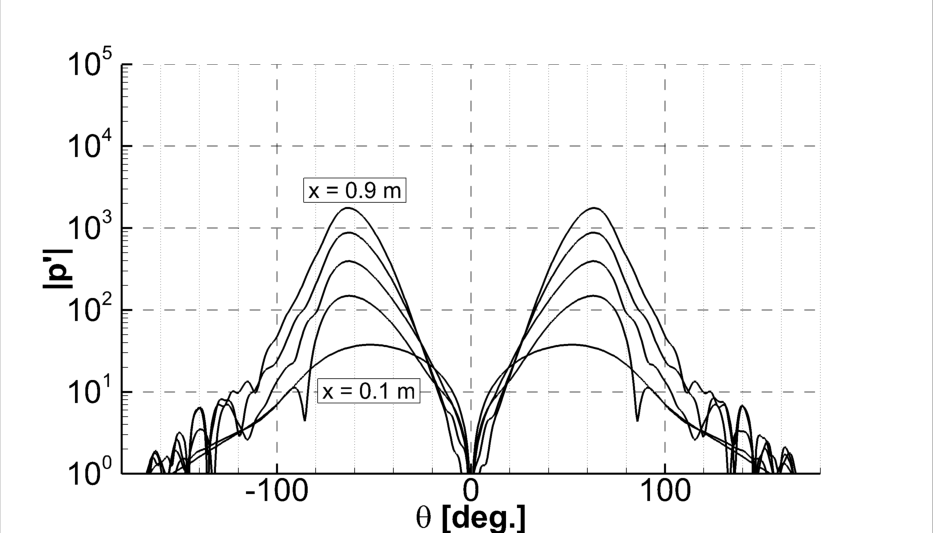}} \\
(c) & (d) \\
\includegraphics[trim=4 4 4 4, clip,width = 0.24\textwidth]{ 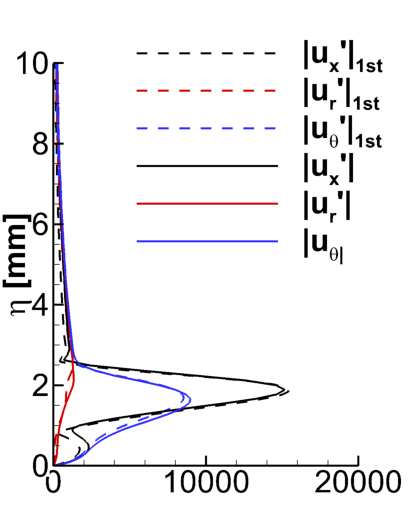} & 
 \includegraphics[trim=4 4 4 4, clip,width = 0.24\textwidth]{ 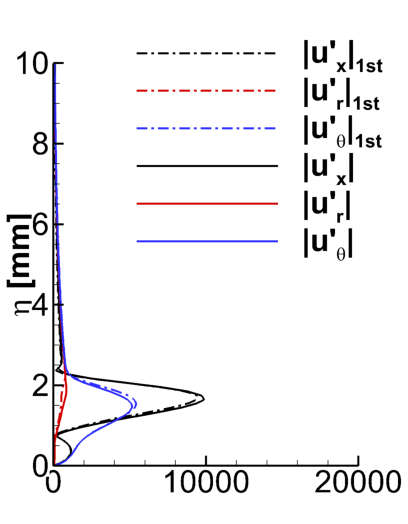} 

\end{tabular}
\caption{\label{fig:azimuthal_positions} Absolute value of fluctuating wall pressure as a function of the azimuth for several streamwise positions as a result of $D_1$ forcing of the (a) sharp cone and (b) 3.6 mm blunt cone at 10 kHz. The disturbances enter the boundary layer upstream and amplify along the $\theta = \ang{60}$ azimuth.}
\end{figure}

Profiles of fluctuating wall pressure and absolute value of the Chu energy amplitude are shown in Figure \ref{fig:f10_wall_amplification}(a) and (c), respectively. Each of the profiles show a best fit with corresponding amplification curves from an LST at several oblique wavenumbers at a frequency of 10 kHz. Because the dominant instability is modal in nature, the wall energy profiles may be used to estimate the initial amplitude using a best fit. In turn, this may be used to compute the receptivity coefficients. The receptivity coefficient for the sharp cone at 10 kHz is $C_a = 346$, which is two orders of magnitude higher than the Mack second  mode receptivity for the sharp cone at 70 kHz. This is largely due to the strong spatial transient growth, which occurs upstream of the predicted neutral point of the first mode. 
\begin{figure*}[t!]
\centering
\begin{tabular}{l l}
(a) & (b) \\
\includegraphics[trim=4 4 4 4, clip,width = 0.45\textwidth]{ 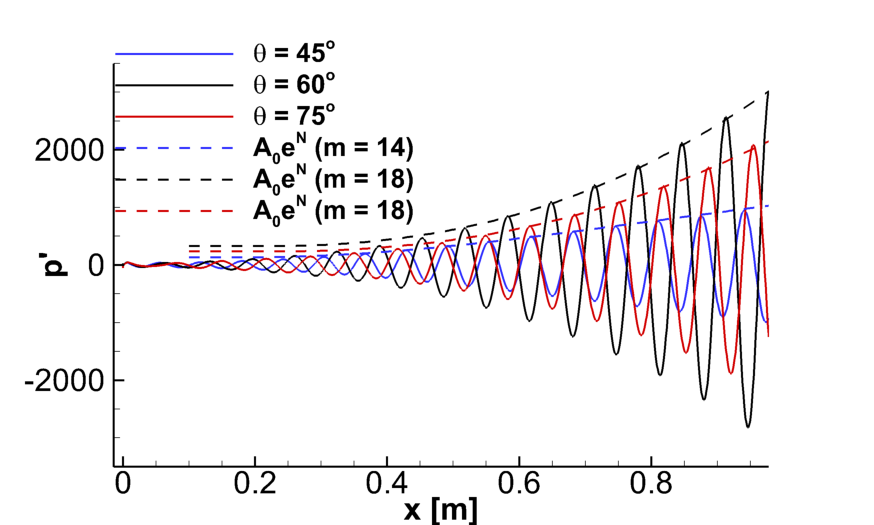} &
\includegraphics[trim=4 4 4 4, clip,width = 0.45\textwidth]{ 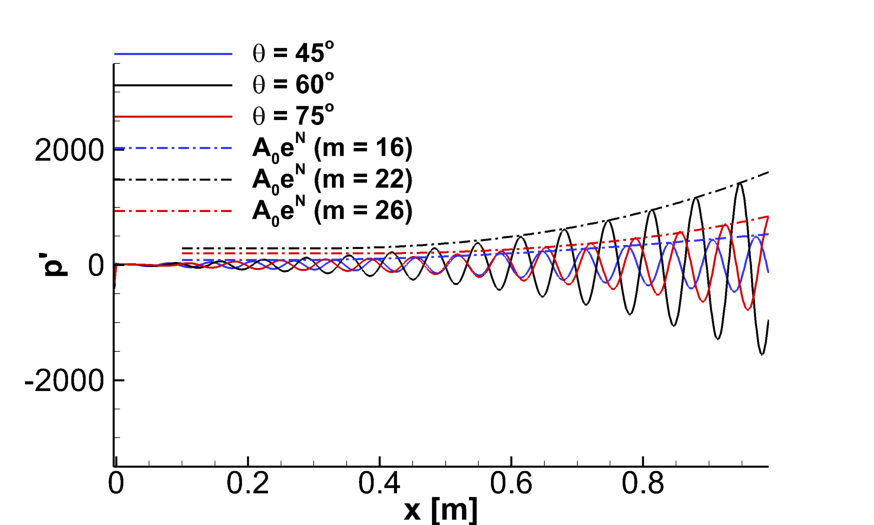} \\
(c) & (d) \\
\includegraphics[trim=4 4 4 4, clip,width = 0.45\textwidth]{ 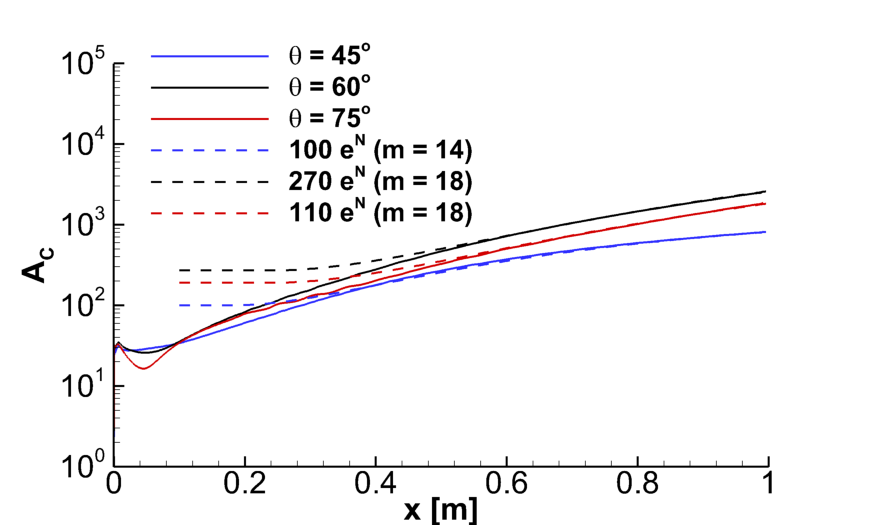} &
\includegraphics[trim=4 4 4 4, clip,width = 0.45\textwidth]{ 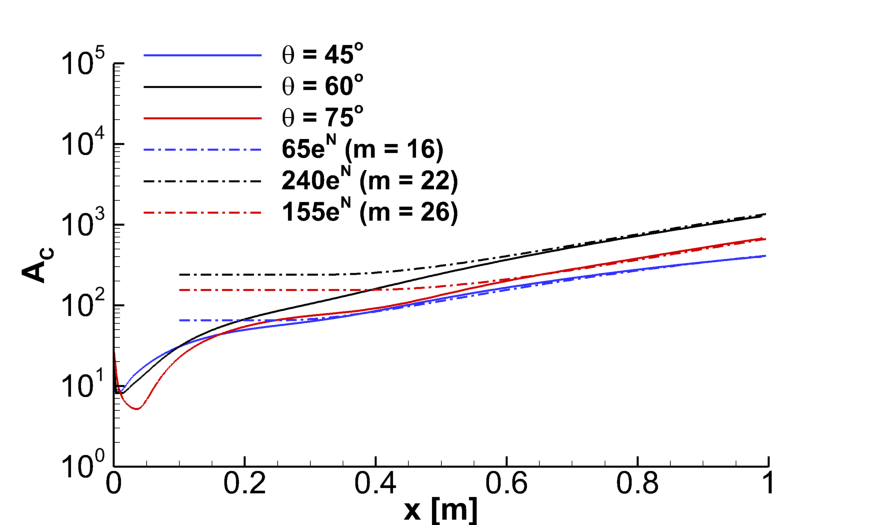} 
\end{tabular}
\caption[Spatial amplification of pressure and Chu energy amplitude at the wall at several azimuthal positions for the sharp and 3.6 mm blunt cones.]{\label{fig:f10_wall_amplification} Spatial amplification of (a)--(b) pressure and (c)--(d) Chu energy amplitude at the wall at several azimuthal positions. Profiles are shown for (a),(c) the sharp cone and (b),(d) the 3.6 mm blunt cone along with the best fit with the LST N-factor at appropriate wavenumbers. Significant amplification occurs upstream of the first mode neutral point.}
\end{figure*}

We now turn to an I/O analysis of the 3.6 mm blunt cone at the same low frequency. The gains from the H-IO analysis of the blunt cone at 10 \unit{\kilo\hertz} is shown in Figure \ref{fig:io_results_f10}(d) alongside the gain from the sharp cone at 10 kHz. The largest gain is around 320 with a decay that approaches one as the number of directions approaches 100. Approximately 20 directions capture 90\% of the amplified physical mechanisms in the flow. The second leading direction has a gain less than half that of the first. This gain is around three times smaller than the sharp cone at 10 kHz. 

The input forcing distribution for $D_1$ is shown in Figure \ref{fig:io_results_f10}(e), along with its physical realization in Figure \ref{fig:io_results_f10}(f). The leading input distribution contains purely a $w$-vortical wave with a very strong, narrow peak around $\psi = \ang{88}$ in the same manner as the sharp cone. The strong peak at a high angle causes a thin band of highly angled waves clustered near the $x$-$z$ plane and the cone tip. The contours in Figure \ref{fig:tip_receptivity}(c) show strong streamwise growth of pressure disturbances in the boundary layer. Figure \ref{fig:tip_receptivity}(d) also shows the strong transmission of vorticity through the shock from the incident wave in a similar manner as before.

The streamwise growth of the pressure disturbance at the wall is plotted for several azimuthal locations in Figure \ref{fig:f10_wall_amplification}(b), showing the exponential amplification of pressure along the wall. Additionally, the streamwise growth of the Chu energy amplitude at the wall is shown in Figure \ref{fig:f10_wall_amplification}(d). Amplification reaches its maximum at an azimuthal position of around $\theta \approx \ang{60}$, which can be seen by plotting the pressure amplitude at the wall as a function of the azimuth for several streamwise positions. These profiles are shown in Figure \ref{fig:azimuthal_positions}(b). There is an injection of pressure into the boundary layer upstream at shallower angles, which then amplifies and spreads along the $\theta = \ang{60}$ azimuth, in the same manner as the sharp cone.

Extracting a slice through the disturbance at $x = 1.0 \unit{m}$ and $\theta = \ang{64}$ and comparing to the corresponding eigenfunctions from the LST clearly indicates that the amplifying mechanism is the oblique first mode instability, as in the sharp cone boundary layer. This comparison is made in Figures \ref{fig:azimuthal_positions}(d) for the 3.6 mm cone. Because the mechanism is modal, we can use the receptivity coefficients to quantify the connection between the free-stream forcing and upstream neutral point. The receptivity coefficient for the $D_1$ forcing is 285, which is slightly less than that of the sharp cone.

In a similar manner to the sharp cone, we can use the contours of mean shock obliqueness and relative incidence to map the peak forcing wave packet onto the theoretical interaction between a vorticity wave with the shock. In this case, however, the bow shock has a variable obliqueness angle with respect to the free-stream, so we take the local obliqueness angle as our reference for the vorticity wave incidence angle. The long dashed line in Figure \ref{fig:shock_theory_Gv} shows a line through the peak forcing packet, where the direction of the arrow indicates moving from upstream to downstream. At the very tip of the shock, the mean shock angle is nearly normal and the incidence angle of vorticity is nearly zero. Farther downstream along the shock, the band cuts through the fast acoustic transmission lobe and terminates in the transmission of a damped wave . The shock also transmits much of this vorticity. Thus, the shock takes this thin band of vortical waves, transmits and amplifies it, as well as generating fast acoustic waves and a strong damped wave in the near tip region. All of these are non-modal effects, which occur upstream of the neutral point of the Mack first mode. 

In summary, at 10 \unit{\kilo\hertz}, the sharp and 3.6 mm blunt cones are most receptive to $w$-vorticity in a thin, highly oblique band along the shock in a manner similar to the sharp cone. This band efficiently transmits oblique vorticity and creates an oblique acoustic wave which non-modally amplifies upstream and stimulates the Mack first mode instability. The first mode then grows according to the modal theory, and is slightly stabilized with the addition of small nose-tip bluntness. The receptivity coefficients for the sharp and blunt cases are 346 and 285, respectively. The modal growth of the 3.6 mm blunt cone boundary layer is also weaker that that of the sharp cone. These two factors are the reason for the reduction in gain from the sharp to the blunt case. The blunt cone boundary layer is less receptive and experiences less first mode growth. While low frequency receptivity of the sharp cone and the 3.6 mm blunt cone are similar, increasing the bluntness even further leads to a completely different mechanism, as we will see in the next section.}

\dac{
\subsection{Hierarchical input-output analyses at 40 kHz}
\label{subsec:mid_frequency}
We now examine the H-IO analysis of both blunt cones at 40 kHz, based on the observed trend in Figure \ref{fig:gain_vs_frequency_b0p2_b3p6}. At this frequency, increasing the nose bluntness slightly leads to a decrease in gain, while increasing the nose bluntness further leads to an increase in gain. We consider the two blunt cones together in this section, since they share a physical mechanism and receptivity process.

The gains from the first 100 H-IO directions are shown in Figure \ref{fig:io_results_f40}(a) for the $R_N = 3.6\unit{\milli\meter}$ cone. The $D_1$ gain is around 50, with a trail off that approaches unity as the number of directions approaches 100. This is approximately one-third as high as the sharp cone at the same frequency. The gains from an H-IO analysis of the $R_N = 7.2 \unit{\milli\metre}$ cone is shown in Figure \ref{fig:io_results_f40}(d), and the largest gain is twice that of the $R_N = 3.6 \unit{\milli\metre}$ cone. The trail-off behavior of the gains from the $R_N = 7.2 \unit{\milli\metre}$ cone analysis is similar to that of the $R_N = 3.6 \unit{\milli\metre}$ cone analysis, with the gain approaching unity as the number of directions approaches 100. 
\begin{figure*}[t!]
\centering
\begin{tabular}{lll}
 (a)  & (b) & (c) \\
 \includegraphics[trim=4 4 4 4, clip,width = 0.30\textwidth]{ 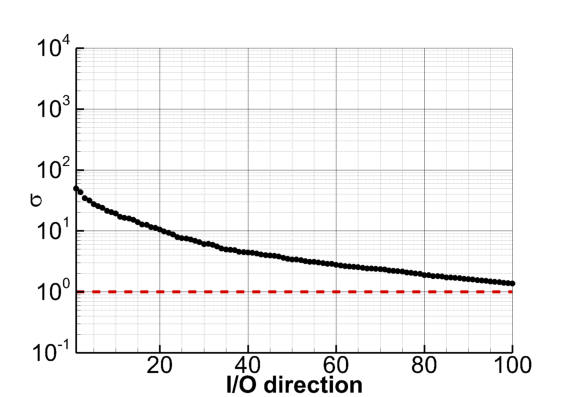} &  \includegraphics[trim=4 1 4 4, clip,width = 0.37\textwidth]{ 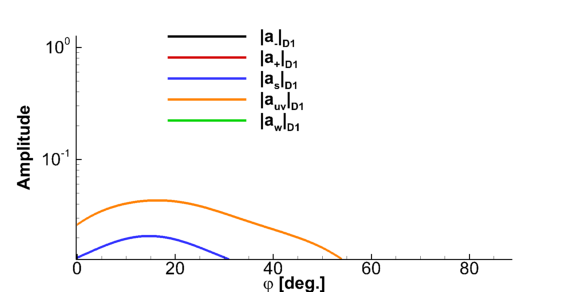} & \includegraphics[trim=4 4 4 4, clip,width = 0.32\textwidth]{ 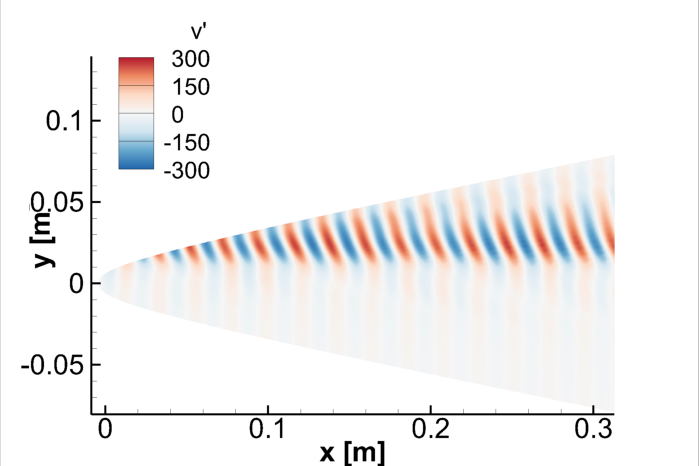} \\
 (d) & (e) & (f) \\
 \includegraphics[trim=4 4 4 4, clip,width = 0.30\textwidth]{ 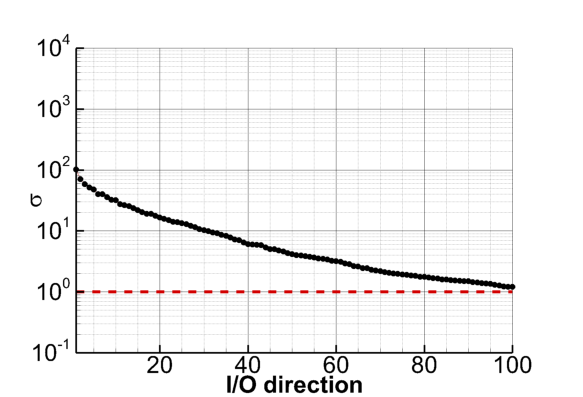} & \includegraphics[trim=4 1 4 4, clip,width = 0.37\textwidth]{ 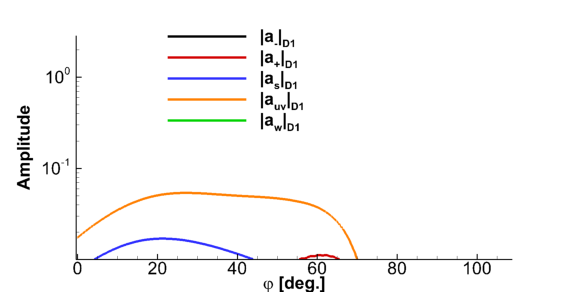} & \includegraphics[trim=4 4 4 4, clip,width = 0.32\textwidth]{ 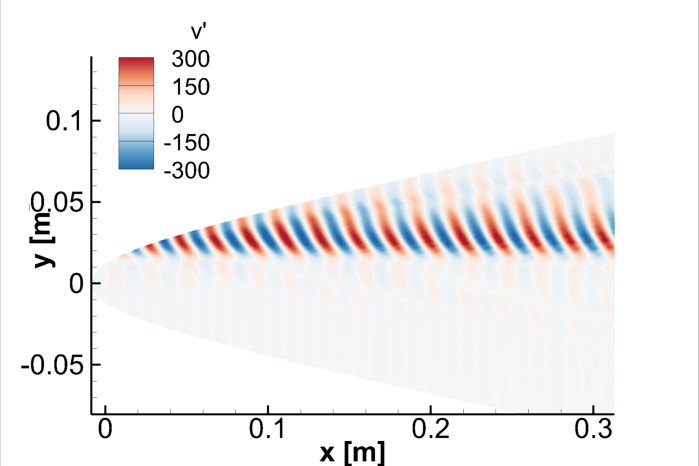}
\end{tabular}
\caption{\label{fig:io_results_f40} H-IO results at 40 kHz for (a)--(c) the 3.6 mm blunt cone, and (d)--(f) the 3.6 mm blunt cone. Shown are (a), (d) gains versus I/O direction, (b), (e) $D_1$ input directions, and (c), (f) physical realizations of the optimal forcing in the free-stream.}
\end{figure*}

The $D_1$ input distribution for the 3.6 mm blunt cone is shown in Figure \ref{fig:io_results_f40}(b), next to its physical realization in the free-stream in \ref{fig:io_results_f40}(c). The dominant wave distribution is of the $u$-$v$ vorticity type in a range of incidence angles up to $\psi = \ang{50}$ with a peak just below $\psi = \ang{20}$. A smaller distribution of entropy waves is also present with the same peak angle. The other wave types are not present in significant amounts in the $D_1$ distribution. The physical realization of this forcing distribution is a band of waves extending down either side of the shock with a peak $y$-location around 0.25 m. This band of fluctuating vorticity also impinges on the top of the shock in the strongly curved region just above the entropy layer. 

The $D_1$ input distribution for the 7.2 mm blunt cone is shown in Figure \ref{fig:io_results_f40}(e). The input distribution contains the same wave types as the previous case---$u$-$v$ vorticity waves first, with a secondary entropy wave distribution. While there is still a peak around $\psi = \ang{20}$, the $D_1$ distribution for the 7.2 mm cone is more broadband and includes waves up to an incidence of $\psi = \ang{70}$. The free-stream physical realization of the forcing distribution is shown in Figure \ref{fig:io_results_f40}(f), which contains a band of waves in nearly the same position as the 3.6 mm blunt cone. 

The forcing wave-packet in three dimensions is shown in Figure \ref{fig:IO_f40_b3p6_b7p2}(a) for the 3.6 mm cone, along with the response of the entropy layer downstream. The contours of fluctuating $y$-velocity on the outermost surface are in the pre-shock region, and the response of the entropy layer is visualized by downstream iso-surfaces of fluctuating temperature. The free-stream waves impinge on the top of the shock above the entropy layer, injecting vorticity and entropy into the downstream flow. The entropy layer only weakly amplifies disturbances injected at this frequency. Figure \ref{fig:IO_f40_b3p6_b7p2}(b) shows similar contours for the 7.2 mm cone. Whereas the free-stream forcing appears very similar to the previous cone, the entropy layer supports much stronger growth of the entropy layer instability, and the temperature iso-surfaces show strong streamwise amplification. The contour levels between the two plots in Figure \ref{fig:IO_f40_b3p6_b7p2} are identical for a direct visual comparison between the two cases. 
\begin{figure}[t!]
\begin{tabular}{l}
 (a) \\
 \includegraphics[trim=4 4 4 4, clip,width = 0.45\textwidth]{ 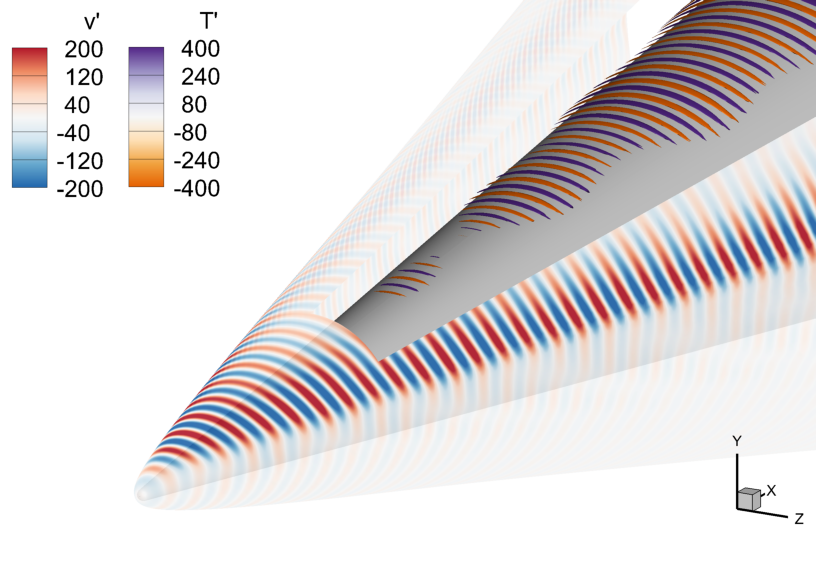} \\
 (b) \\
 \includegraphics[trim=4 4 4 4, clip,width = 0.45\textwidth]{ 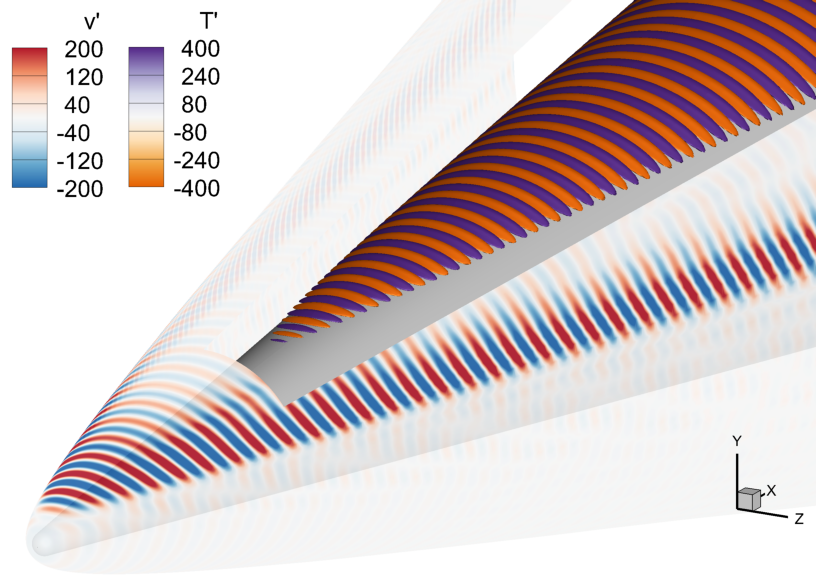} 
\end{tabular}
\caption[Global response of the 3.6 mm blunt cone and the 7.2 mm blunt cone to $D_1$ forcing at 40 \unit{\kilo\hertz}.]{\label{fig:IO_f40_b3p6_b7p2} Global response of (a) the 3.6 mm blunt cone and (b) the 7.2 mm blunt cone to $D_1$ forcing directions at 40 \unit{\kilo\hertz}. Contours on the outermost surface are in the free-stream, while temperature iso-surfaces show the downstream response.}
\end{figure}

The streamwise growth of the Chu energy amplitude extracted from several entropy layer streamlines is shown in Figure \ref{fig:f40_streamwise_growth}(a) for the 3.6 mm cone. Energy profiles along these streamlines show the maximum amplification envelope of the entropy layer instability. The height of the streamlines above the wall is shown via dashed lines corresponding to the axis labels on the right side of the figure. These streamlines originate in the curved portion of the shock. The thick dashed line below the streamlines shows the edge of the boundary layer. Upstream, the boundary layer is very thin, but as it grows, it begins to swallow some of the entropy layer streamlines, causing a rapid decay in the energy along those streamlines. The same quantities for the 7.2 mm cone are shown alongside the previous case in Figure \ref{fig:f40_streamwise_growth}(b). The entropy layer instability grows much more quickly, reaching a peak around $x = 0.6 \unit{\meter}$. The Chu energy amplitude of this peak is fifteen times larger than the peak amplitude of the free-stream forcing packet outside of the shock. 
\begin{figure}[t!]
\begin{tabular}{l}
 (a) \\
 \includegraphics[trim=4 4 4 4, clip,width = 0.45\textwidth]{ 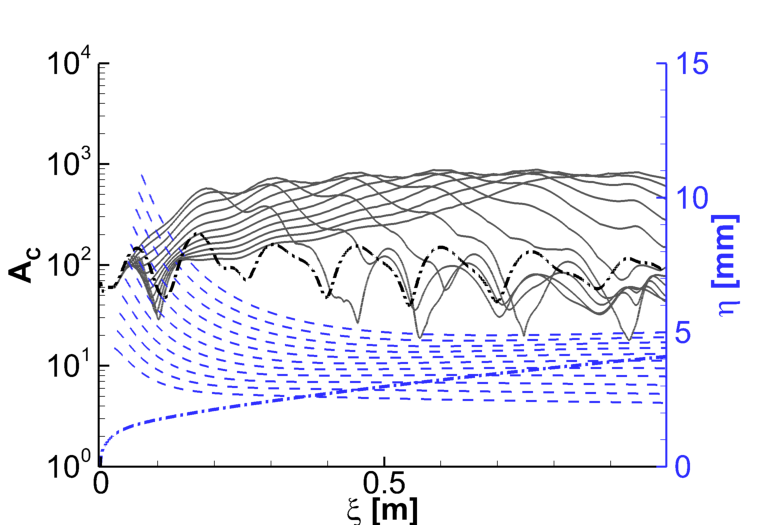} \\
 (b) \\
   \includegraphics[trim=4 4 4 4, clip,width = 0.45\textwidth]{ 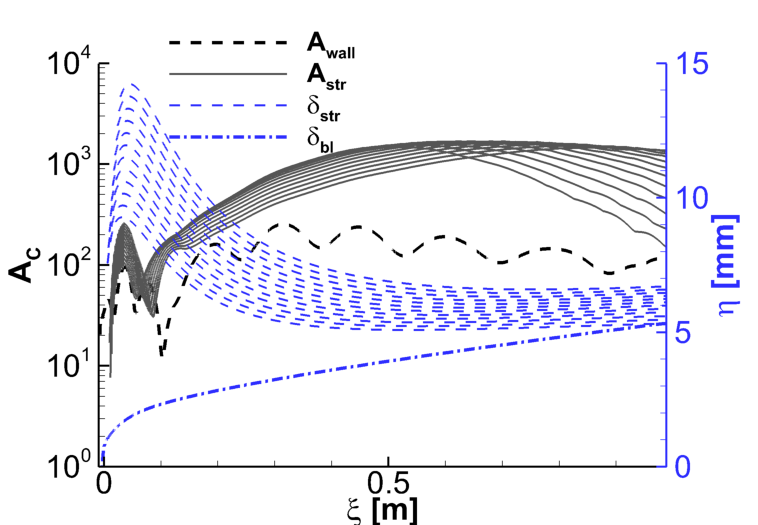} 
\end{tabular}
\caption[Streamwise growth of the Chu energy amplitude in the entropy layer for blunt cones with $R_N = 3.6 \unit{\milli\metre}$ and $R_N = 7.2 \unit{\milli\metre}$.]{\label{fig:f40_streamwise_growth} Streamwise growth of the Chu energy amplitude in the entropy layer for blunt cones with (a) $R_N = 3.6 \unit{\milli\metre}$ and (b)  $R_N = 7.2 \unit{\milli\metre}$. Also shown are streamline heights above the wall ($\delta_{str}$) and the boundary layer edge height ($\delta_{bl}$) as a visualization of where the entropy layer is interacting with the boundary layer.}
\end{figure}
Remember that at 70 kHz, the entropy layer instability destabilized an F mode upstream of its synchronization with the continuous branch. This effect is notably absent here. This is primarily because, at this low frequency, there is not a discrete F mode which the entropy layer can destabilize. Instead of boundary layer activation, the presence of a more large scale beating pattern is visible. Much longer wavelength oscillations are visible in the wall pressure profiles, and the entire streamwise extent of the domain contains several wavelengths of this oscillation pattern. This oscillation pattern is the result of the injection of acoustic disturbances into the boundary layer from the free-stream wave-packet impinging on the shock. Two-dimensional contours of velocity and temperature are shown in Figure \ref{fig:b3p6_f40_entropylayer}(a) and (c) for the 3.6 mm cone. The domain is shown in rotated $\xi$--$\eta$ coordinates for visualization purposes. Also shown are the two streamlines which bound the injected wave packet. The forcing distribution injects vorticity and entropy into the entropy layer above the entropy layer, which is most clearly visible in terms of $y$-velocity fluctuations in \ref{fig:b3p6_f40_entropylayer}(a). The region of the entropy layer bounded by the edge streamlines produces the highest amplification as injected energy is compressed and amplified downstream. The contours of pressure in \ref{fig:b3p6_f40_entropylayer}(e) show that, while the dominant acoustics injected into the downstream flow are damped, the frequency is low enough and the wavelength is long enough that the damped wave reaches the boundary layer and activates an acoustic disturbance. This acoustic disturbance is initially reinforced by internal reflection from the underside of the shock, then travels downstream in the boundary layer. The beating phenomenon is due to the difference in the fast and slow acoustic wavelengths as the pressure signature moves through the boundary layer downstream. 
\begin{figure*}[t!]
\centering
\begin{tabular}{ll}
 (a) & (b) \\
  \includegraphics[trim=4 4 4 220, clip,width = 0.45\textwidth]{ 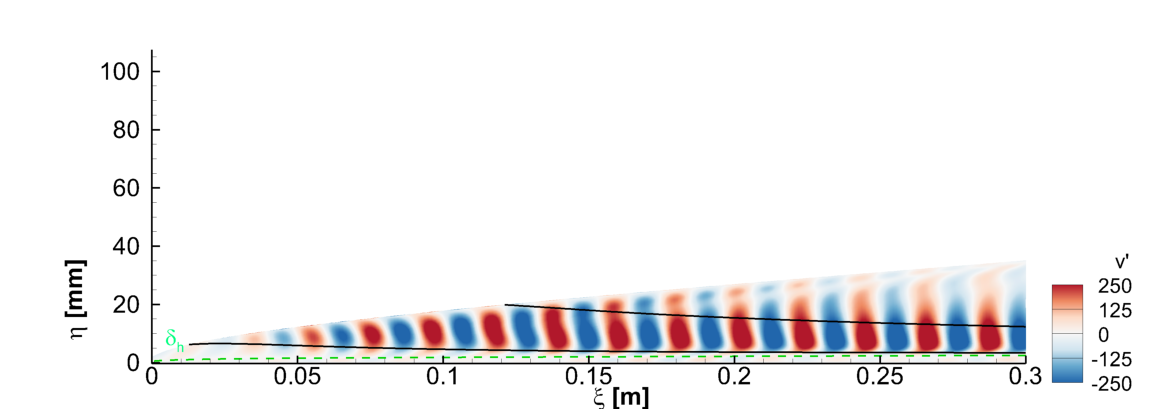} & \includegraphics[trim=4 4 4 220, clip,width = 0.45\textwidth]{ 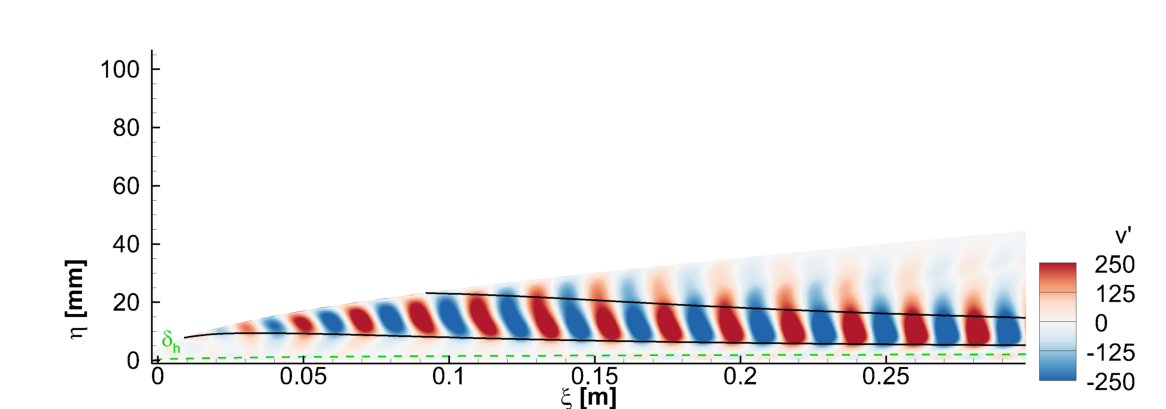} \\
 (c) & (d)\\
\includegraphics[trim=4 4 4 220, clip,width = 0.45\textwidth]{ 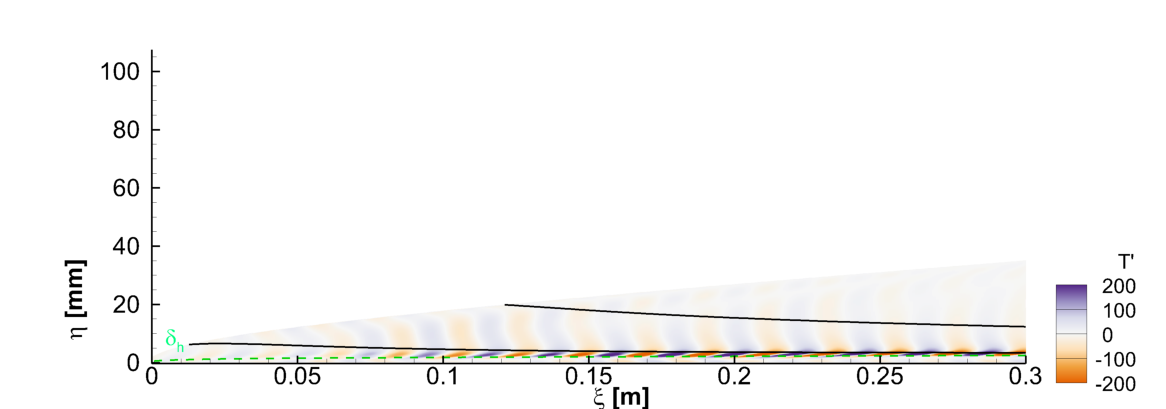} & \includegraphics[trim=4 4 4 220, clip,width = 0.45\textwidth]{ 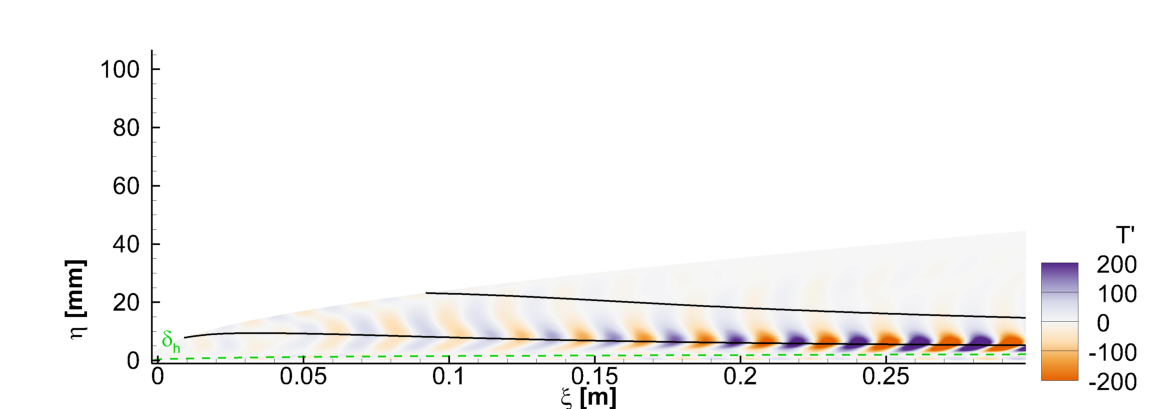}\\
(e) & (f) \\
\includegraphics[trim=4 4 4 220, clip,width = 0.45\textwidth]{ 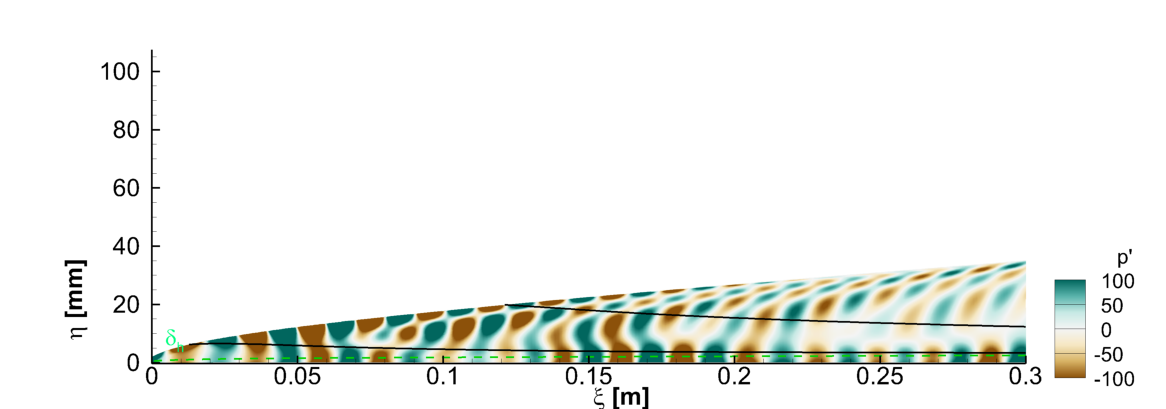} & \includegraphics[trim=4 4 4 220, clip,width = 0.45\textwidth]{ 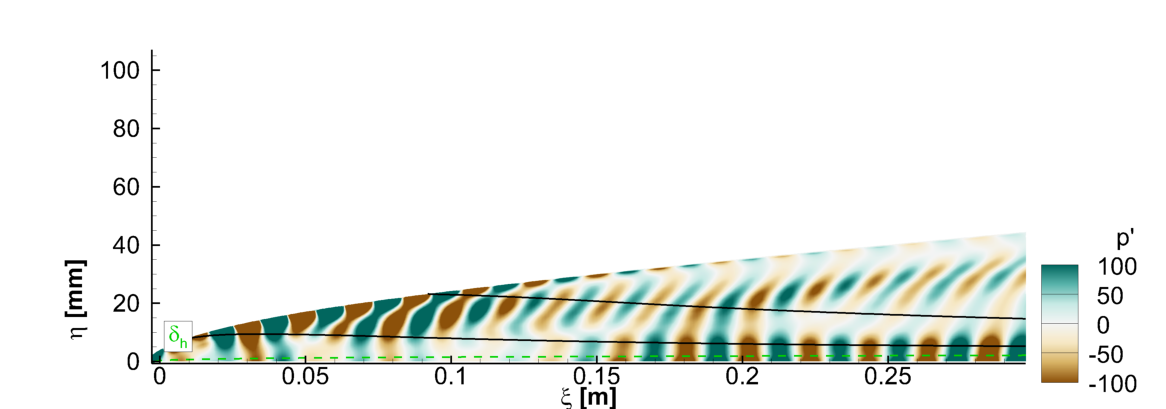} 
\end{tabular}
\caption{\label{fig:b3p6_f40_entropylayer} Contours of spatially amplifying (a)--(b) velocity, (c)--(d) temperature, and (e)--(f) pressure for the (a), (c), (e) $R_N = 3.6 \unit{\milli\metre}$ blunt cone and (b), (d), (f) $R_N = 7.2 \unit{\milli\metre}$ blunt cone at 40 kHz. The solid streamlines are extracted at the boundaries of the injected velocity packet, and the dashed lines show the boundary layer edge. }
\end{figure*}

The two-dimensional slices through the fluctuating velocity and temperature fields for the 7.2 mm cone in Figure \ref{fig:b3p6_f40_entropylayer}(b) and (d) tell a story similar to that of the 3.6 mm cone. The disturbances are injected above the entropy layer generation region and then amplify inside the entropy layer region, closely following the bounds of the streamlines. In this case, even though the shock to wall distance is greater, due to the increased shock standoff distance, the acoustic waves still reach the boundary layer, which leads to a similar beating effect. 

It appears that the nature of the entropy layer mechanism is twofold \cite{Cook2022}. First, the flow of the entropy layer is inviscid but rotational, which leads to a vorticity and entropy tilting effect, which in turn amplifies upstream tilted structures as they convect with the mean flow downstream. Secondly, the convergence of the streamlines in the supersonic entropy layer region leads to flow deceleration, further amplifying the disturbances through streamwise compression. The presence of the blunt tip is the dominant effect creating the shock curvature and the rotational effect, whereas the angle of the cone frustum is thought to contribute to the compression and deceleration part of the mechanism. 

The overall receptivity process for these two blunt cones begins with free-stream vorticity waves with a peak around $\psi = \ang{20}$, which is optimal for the maximum transmission of vorticity through the shock. This injected vorticity is amplified above the boundary layer by rotation and deceleration of the mean velocity. The entropy layer for the 7.2 mm cone leads to a roughly two-fold increase in maximum entropy layer instability amplitude. In contrast to the high-frequency cases, where the entropy layer instability interacts with a discrete boundary layer mode, this discrete mode is not destabilized at this low frequency. Instead, a weakly decaying acoustic beating pattern is observed in the boundary layer. 
}

\subsection{\label{subsec:summary} Effects of nose bluntness and frequency}
\dac{In order to summarize the overall receptivity of sharp and blunt cones to realizable free-stream disturbances, we show the leading $D_1$ input distributions for H-IO analyses performed across frequencies from 0 kHz to 100 kHz, in 10 kHz increments. Not every distribution from the leading two directions is shown. Instead, the most dominant coherent mechanisms are highlighted by showing their corresponding $D_1$ and $D_2$ inputs. The $D_1$ and $D_2$ forcing distributions for all frequencies considered are shown in Figure \ref{fig:inputs_vs_freq_b0p2_b3p6} for each flow. This affords a more comprehensive understanding of the receptivity of sharp and blunt cones and the instabilities present in the flow. 
\begin{figure}[t!]
\begin{tabular}{l}
 (a) \\
 \includegraphics[trim=4 4 4 4, clip,width = 0.45\textwidth]{ 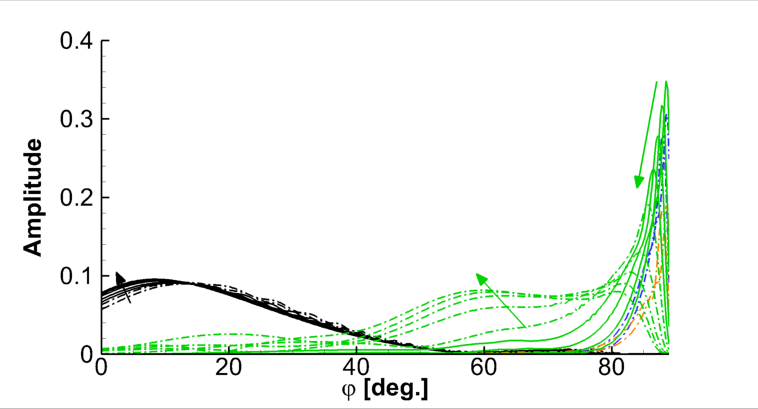} \\
 (b) \\
 \includegraphics[trim=4 4 4 4, clip,width = 0.45\textwidth]{ 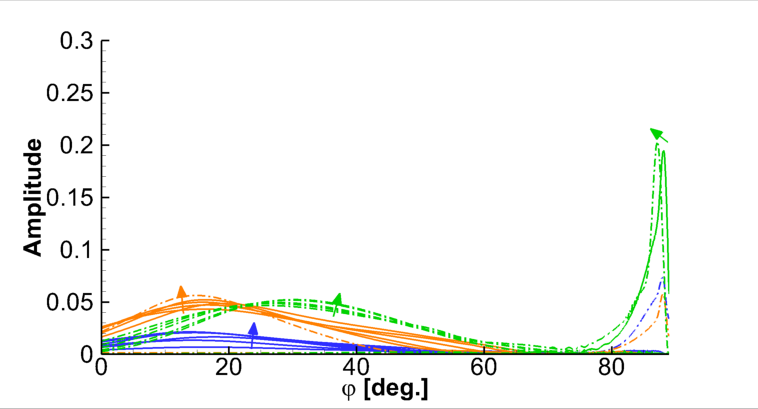} \\
 (c) \\
 \includegraphics[trim=4 4 4 4, clip,width = 0.45\textwidth]{ 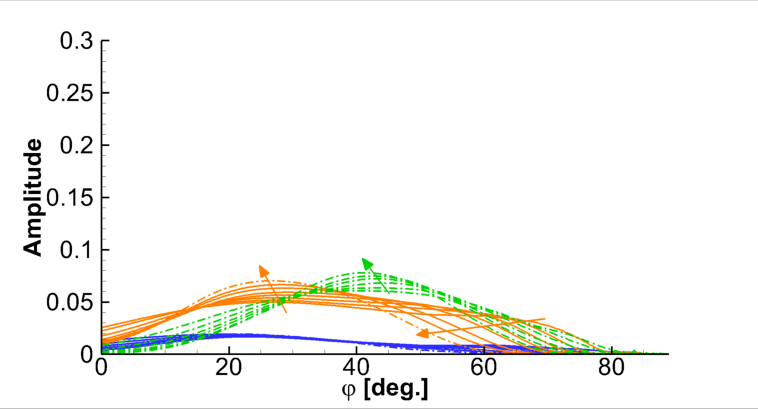}
 
\end{tabular}
\caption{\label{fig:inputs_vs_freq_b0p2_b3p6}  $D_1$ and $D_2$ forcing directions at frequencies from 10--90 kHz from H-IO analysis of the (a) sharp, (b) 3.6 mm blunt, and (c) 7.2 mm blunt cones. Dot-dashed lines are the $D_2$ forcing directions and solid lines are the $D_1$ forcing direction. Arrows denote increasing frequency.}
\end{figure}

For the sharp cone, the dominant receptivity mechanisms have two primary distributions corresponding to each of the dominant physical mechanisms. At frequencies below 40 \unit{\kilo\hertz}, the first mode is most receptive to $w$-vorticity waves with very high incidence angles. These waves create highly oblique acoustics and vorticity behind the shock, which enter the boundary layer and activate the Mack first mode instability. The optimal forcing distribution corresponding to the Mack first mode can be seen in Figure \ref{fig:inputs_vs_freq_b0p2_b3p6}(a). Figure \ref{fig:inputs_vs_freq_b0p2_b3p6} includes the $D_2$ forcing distributions, in order to demonstrate that the H-IO is capturing coherent mechanisms across frequencies. As the frequency changes, the $D_1$ distribution may jump between one or more physical mechanisms, so including higher directions shows what happens to those physical mechanisms. For frequencies above 50 \unit{\kilo\hertz}, the Mack second mode is the dominant instability mechanism and is most receptive to slow acoustic waves at shallow incidence angles with a peak around $\psi = \ang{10}$. The $u$-$v$ vorticity waves are present, though they are a less significant part of the receptivity process.

The addition of nose-tip bluntness fundamentally alters the dominant receptivity processes. The 10 kHz first mode in the 3.6 mm blunt cone boundary layer shares a receptivity mechanism with the first mode for sharp cones---highly oblique incident $w$-vorticity waves. This distribution is shown in Figure \ref{fig:inputs_vs_freq_b0p2_b3p6}(b). Above 20 kHz, the Mack second mode is absent from the leading H-IO directions. Instead, the entropy layer instability is the dominant mechanism. This mechanism is most receptive to vorticity and entropy waves at shallow incidence angles with a peak just below $\psi = \ang{20}$, which is the optimal angle for transmitting vorticity through the shock into the entropy layer. The entropy layer receptivity is more broadband in a range from $f = 30\text{--}80 \unit{\kilo\hertz}$, as shown by the distributions of entropy and vorticity waves present in Figure \ref{fig:inputs_vs_freq_b0p2_b3p6}(b). These entropy waves amplify in the entropy layer via a rotation and deceleration mechanism. For higher frequencies at which the boundary layer supports a damped F mode, the entropy layer can interact with the F mode and destabilize it upstream of its synchronization with the continuous spectra. For frequencies at which no F mode is supported in the boundary layer, the entropy layer simply grows and convects on top of the boundary layer, upstream of the entropy swallowing point. 

The increase in nose-tip bluntness from 3.6 mm to 7.2 mm at low frequency fully stabilizes both the Mack first and second modes. At 10 kHz, there is a combined acoustic wave shock interaction and entropy layer instability. At frequencies above 10 kHz, the entropy layer is the dominant mechanism and is most receptive to free-stream vorticity waves at incidence angles from \ang{0} to \ang{80}. For lower to mid-range frequencies, the absence of boundary layer modes that can be destabilized by the entropy layer means that the largest growth occurs outside the boundary layer and has very little signature at the wall. 

Receptivity N-factors, defined in \S \ref{sec:METHODOLOGY}, are shown in Figure \ref{fig:HIO_Receptivity_Nfactors} for each of the H-IO cases presented in this chapter. At 10 kHz, shown in \ref{fig:HIO_Receptivity_Nfactors}(a), the Mack first mode is the theoretically dominant mechanism, reaching N-factors as high as $N_r = 8$ for the sharp cone. The 3.6 mm cone is less receptive, but still shows a strong first mode response. The most blunt cone does not contain any first mode and is instead a very low amplitude multi-modal mechanism. The large N-factors associated with the first mode are a direct result of the high receptivity coefficients and efficient forcing discovered by this H-IO analysis. Whether or not this effect could be observed in an experiment would depend on the presence of highly oblique vorticity fluctuations in the free-stream environment. What the N-factor shows is that it is possible that the first mode instability could play a very dominant role, even when the free-stream disturbances are restricted to realizable planar waves.
\begin{figure}[t!]
\begin{tabular}{l}
 (a) \\
 \includegraphics[trim=4 4 4 4, clip,width = 0.345\textwidth]{ 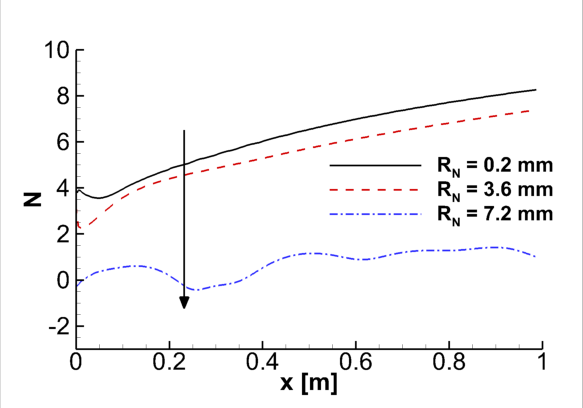} \\
 (b) \\
 \includegraphics[trim=4 4 4 4, clip,width = 0.345\textwidth]{ 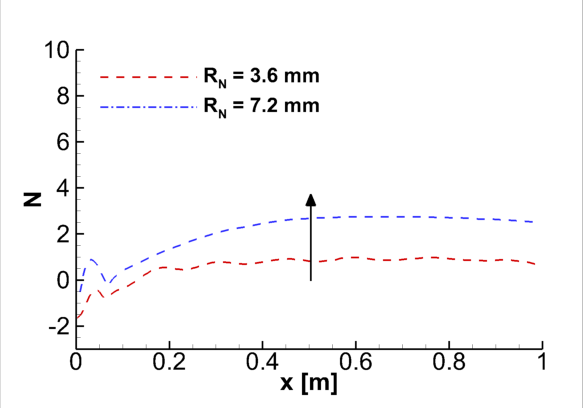} \\
 (c) \\
 \includegraphics[trim=4 4 4 4, clip,width = 0.345\textwidth]{ 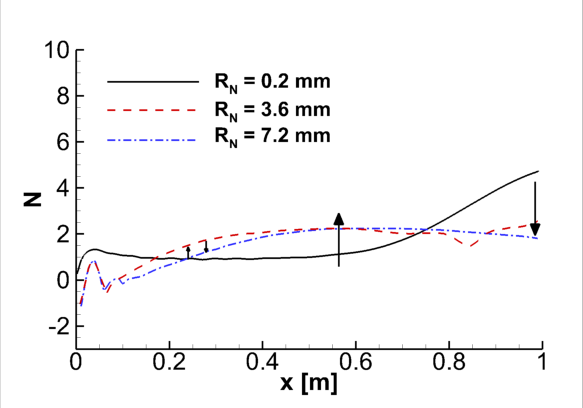}
 
\end{tabular}
\caption{\label{fig:HIO_Receptivity_Nfactors} Receptivity N-factors computed from the H-IO responses to the $D_1$ input directions for each cone at (a) 10 kHz, (b) 40 kHz, and (c) 70 kHz. The initial amplitude is determined by the peak of the free-stream forcing wave packet. Arrows indicate increasing nose-tip bluntness.}
\end{figure}

At 40 kHz, the N-factor associated with the entropy layer instability is greater than three for the 7.2 mm cone, a notable increase from that reached by the 3.6 mm cone. At 70 kHz, the sharp cone boundary layer is dominated by the second mode, reaching $N_r = 4.5$ at the end of the domain. The Mack second mode is quickly stabilized by the addition of nose-tip bluntness, while the upstream growth of the entropy layer instability is quickly destabilized. At higher frequencies, both cones reach the same slightly lower N-factors around $N_r = 2$. This suggests that there may be some frequency tuning of the entropy layer instability as well. Larger bluntness cones are more receptive to lower frequency entropy layer instabilities. Both the stabilization of the modal growth and the destabilization of the entropy layer are consistent with experiments and observations, although for the cases considered, the N-factors achieved by the entropy layer instability are not large enough that transition to turbulence would be expected. This may be due to the low Reynolds number, relative to experiments in which transition reversal was observed. The possibility of future research into these effects is addressed in the following chapter. }

\section{\label{sec:CONCLUSION}Conclusions and Future Work}
In this paper we developed and advanced several techniques for the global linear analysis \jwnresolved{of} flows over hypersonic sharp and blunt cones at $M = 5.8$. First, we constructed a framework with input-output analysis for studying receptivity to three-dimensional planar waves. This was accomplished by modifying the classical I/O framework to treat the free-stream as a forced boundary condition to the pre-shock state. This, in combination with an input matrix that maps amplitude distributions to the free-stream state allowed us to pose the receptivity question in terms of input forcing which satisfies the free-stream dispersion relation for acoustic, vortical, and entropic waves. Second, we developed hierarchical input-output (H-IO) analysis through the azimuthal Fourier decoupling of the global dynamics. Once the flow is parameterized with respect to the azimuthal wavenumber, H-IO analysis uses rank-compressed reduced-order models at each of the wavenumbers before re-coupling the terms and performing the final optimization problem in three dimensions. The combination of these two techniques provides a powerful tool for understand\jwnresolved{ing} the receptivity of hypersonic flows to realistic free-stream environments.

We verified our approach by applying it to $M = 5.8$ flow over a sharp cone and comparing our result to the global linear response of the flow to single free-stream waves at various incidence angles. H-IO not only predicted that the sharp cone boundary layer is most receptive to slow acoustic waves, but is also successfully predicted the optimal incidence angle to which the Mack 2\ts{nd} mode is most receptive: $\psi = \ang{10}$. 

\dac{H-IO analysis was then applied to the same flow over one meter long geometries, this time including a sharp cone and two blunt cones with 3.6 mm and 7.2 mm tip radii. At frequencies below 40 kHz, the sharp cone boundary layer was found to be most receptive to $w$-vorticity waves with very high incidence angles. These free-stream vorticity waves create highly oblique acoustics and vorticity behind the shock, which enter the boundary layer and activate the first mode instability. For frequencies above 50 \unit{\kilo\hertz}, the second mode is the dominant instability mechanism and is most receptive to slow acoustic waves at shallow incidence angles with a peak around $\psi = \ang{10}$. The $u\text{--}v$-vorticity waves are present in the receptivity distributions, though they are a less significant part of the receptivity process for the second mode. 

At low frequency, the 3.6 mm blunt cone boundary layer shares a receptivity mechanism with the first mode for sharp cones: highly oblique incident $w$-vorticity waves. While there is still some receptivity to slow acoustic waves present around 20 \unit{\kilo\hertz}, the entropy layer instability is most receptive to vorticity and entropy waves at shallow incidence angles with a peak just below $\psi = \ang{20}$, which is the optimal angle for transmitting vorticity through the shock into the entropy layer. The entropy layer receptivity is more broadband in a range from $f = 40\text{--}80 \unit{\kilo\hertz}$. These entropy waves amplify in the entropy layer via a rotation and deceleration mechanism. For higher frequencies, at which the boundary layer supports a damped F mode, the entropy layer can interact with and destabilize the F mode upstream of its synchronization with the continuous spectra. If no F mode is supported in the boundary layer, the entropy layer simply amplifies and convects on top of the boundary layer, upstream of the entropy swallowing point. 

The increase in nose-tip bluntness from 3.6 mm to 7.2 mm at low frequency leads to a nearly full stabilization of the first mode instability. At 10 kHz, there is a combined acoustic-shock interaction and entropy layer instability. At frequencies above 10 kHz, the entropy layer is the dominant mechanism and is most receptive to free-stream vorticity waves at incidence angles from \ang{0} to \ang{80}. For lower to mid-range frequencies, the absence of boundary layer modes that can be destabilized by the entropy layer means that the largest growth occurs outside the boundary layer and has very little signature at the wall. Increasing nose-tip bluntness also destabilizes the entropy layer instability, leading to an increase in N-factor with increasing bluntness. This destabilization with increasing bluntness is not predicted by modal stability analysis, but is captured by the receptivity based H-IO analysis.}

A \jwnresolved{natural and important extension of this work would be} to consider higher Reynolds numbers. The Reynolds numbers considered in this study \jwnresolved{were chosen so that both the sharp and blunt cone boundary layers supported first- and second- mode instabilities generating significant growth by the end of the domain, while simultaneously allowing all of the underlying waves to be well-resolved.  While this was sufficient to reveal new receptivity physics associated with different types of instability, and the effects of nose-tip bluntness on those physics, these Reynolds numbers} are not high enough such that we would expect a transition reversal to occur. \jwnresolved{Efforts are currently underway to apply H-IO analysis to higher-Reynolds-number flows, and in particular to} a subset of Stetson's cones for which transition reversal was observed. This would provide valuable insight into whether there are three-dimensional, globally linear mechanisms by which transition reversal could occur and the free-stream environmental factors to which those mechanisms may be receptive, \jwnresolved{but this remains beyond the scope of the current paper.}

Furthermore, recent experiments over ogive-cylinders utilizing high speed Schlieren show the presence of a low-frequency mechanism in addition to Mack 2\ts{nd} mode instability \cite{Hill2021} as well as a wisp structure outside of the boundary layer. Applying H-IO in this context would be an excellent case study for examining the observed mechanisms and how they are receptive to the wind tunnel environment in which the tests were done. 

Additionally, we foresee an extension of the methodology to more complex flows without axisymmetry, such as blunt cones at angle of attack, \jwnresolved{or blunt cones with swept fins~\cite{Araya2022}.} While this paper considered the decoupling of the global Jacobian via an azimuthal Fourier decomposition, this is not the only way to decompose a flow into sections such the an H-IO analysis could be used to rank-compress and reconstruct the global response. Geometric domain decomposition is one way in which we could obtain sub-sections for which H-IO might provide a way to \jwnresolved{overcome} the complexities and cost associated with the 3D problem.

\section*{Acknowledgments}

Support for this research from ONR grant number N00014-23-1-2460 is gratefully acknowledged.

\section*{Author declarations}

The authors have no conflicts of interest to disclose.

\appendix

\section{Derivation of the shock-kinematic boundary condition in three dimensions \label{AppendixA}}
To model shock/perturbation interaction in a Cartesian frame, consider a stationary shock aligned normal to the $x$-axis, subject to small perturbations, as depicted in Fig.~\ref{fig:skbc_schematic}. The baseflow passes through the shock from left to right, although there may be an oblique component with respect to the $y$ and $z$ directions. The jump conditions across the shock are governed by the Rankine--Hugoniot equations,
\begin{align}
    \label{eq:RH}
    \rho_1 u_1 &= \rho_2 u_2, \\
    p_1 + \rho_1 u^2_1 &= p_2 + \rho_2 u^2_2, \\
    v_1 &= v_2, \\
    w_1 &= w_2, \\
    h_1 + \frac{1}{2} u^2_1 &= h_2 + \frac{1}{2} u^2_2.
\end{align}
In response to small unsteady perturbations, the instantaneous position of the shock will shift a small distance upstream or downstream.  Let $X(y,z,t) << 1$ represent the instantaneous $x$-position of the shock relative to its mean position.  This function defines a local coordinate system along the shock,
\begin{equation}
    \hat{n} = (1, -X_y, X_z), \:\: \hat{t}_1 = (X_y, 1, 0), \:\: \hat{t}_2 = (-X_z, 0, 1),
\end{equation}
where $\hat{n}$, $\hat{t}_1$, and $\hat{t}_2$ define the normal and two tangential directions, respectively.  In this coordinate system,
\begin{align}
    u_n &= u - X_y v +  X_z w, \\
    u_t &= X_y u + v, \\
    u_p &= -X_z u + w, \\
    u_s &= X_t,
\end{align}
where $u_s$ is the instantaneous shock velocity. In this equation, subscripts denote partial differentiation with respect to the subscript variable. This allows the reformulation of the Rankine-Hugoniot equations in the moving frame of the shock, such that
\begin{align}
    \label{eq:RHframe}
    \rho_1 \left(u_{n_1} - u_s\right) &= \rho_2 \left(u_{n_2} - u_s\right), \\
    p_1 + \rho_1 \left(u_{n_1} - u_s\right)^2 &= p_2 + \rho_2 \left(u_{n_2} - u_s\right)^2 \\
    u_{t_1} &= u_{t_2}, \\
    u_{p_1} &= u_{p_2}, \\
    h_1 + \frac{1}{2} \left(u_{n_1} - u_s\right)^2 &= h_2 + \frac{1}{2} \left(u_{n_2} - u_s\right)^2.
\end{align}
For an ideal gas, $h_i = \frac{\gamma}{\gamma-1} \frac{p_i}{\rho_i}$, and we can replace the enthalpy equation by the shock adiabat,
\begin{equation}
    \frac{\rho_2}{\rho_1} = \frac{p_2(\gamma+1) + p_1(\gamma-1)}{p_1(\gamma+1) + p_2(\gamma-1)}.
\end{equation}

Linearization of the Rankine--Hugoniot equations in the reference frame of the moving shock yields
\begin{align}
    \left[\bar{\rho} u' + \bar{u} \rho'\right] - \left[\bar{\rho}\right] \left(X_t + \bar{v} X_y + \bar{w} X_z\right) &= 0, \label{eq:linear_RH0}
    \\
    \left[p' + 2 \bar{\rho} \bar{u} u' + \bar{u}^2 \rho' \right] &= 0, \\
    \left[v'\right] + \left[\bar{u}\right] X_y &= 0, \\
    \left[w'\right] - \left[\bar{u}\right] X_z &= 0,
    \label{eq:linear_RH}
\end{align}
and
\begin{align}
      \big(-\bar{\rho}_2(\gamma+1)&+\bar{\rho}_1(\gamma-1)\big) p'_1 \\
    + \big(\bar{p}_2(\gamma+1)&+\bar{p}_1(\gamma-1)\big) \rho'_1 \\
    = &\big(-\bar{\rho}_1(\gamma+1)+\bar{\rho}_2(\gamma-1)\big) p'_2 \\
    + &\big(\bar{p}_1(\gamma+1)+\bar{p}_2(\gamma-1)\big) \rho'_2.
\end{align}
The square brackets in Eq. \ref{eq:linear_RH0}--\ref{eq:linear_RH} denote a jump condition $[q] = q_1 - q_2$ across the shock. In deriving the momentum equation, we have used $[\bar{\rho} \bar{u}]=0$ to eliminate the terms dependent on shock motion.

The linearized Rankine--Hugoniot equations for a perturbed shock can be written compactly in the following form~\cite{Robinet2001},
\begin{equation}
    \label{eq: linRH}
    \mathbf{A_2 Z_2} = \mathbf{A_1 Z_1} + \mathbf{\xi} X_t + \mathbf{\zeta} X_y + \mathbf{\beta} X_z,
\end{equation}
where 
\begin{equation*}
    \mathbf{A_1} = 
    \begin{pmatrix}
    0 & \bar{\rho}_1 & 0 & 0 & \bar{u}_1 \\
    1 & 2 \bar{\rho}_1 \bar{u}_1 & 0 & 0 & \bar{u}^2_1 \\
    0 & 0 & 1 & 0 & 0 \\
    0 & 0 & 0 & 1 & 0 \\
    \scriptstyle \left(-\bar{\rho}_2(\gamma+1)+\bar{\rho}_1(\gamma-1)\right) & 0 & 0 & 0 & 
    \scriptstyle \left(\bar{p}_2(\gamma+1)+\bar{p}_1(\gamma-1)\right),
    \end{pmatrix},   
\end{equation*}
\begin{equation*}
    \mathbf{A_2} = 
    \begin{pmatrix}
    0 & \bar{\rho}_2 & 0 & 0 & \bar{u}_2 \\
    1 & 2 \bar{\rho}_2 \bar{u}_2 & 0 & 0 & \bar{u}^2_2 \\
    0 & 0 & 1 & 0 & 0 \\
    0 & 0 & 0 & 1 & 0 \\
    \scriptstyle \left(-\bar{\rho}_1(\gamma+1)+\bar{\rho}_2(\gamma-1)\right) & 0 & 0 & 0 & 
    \scriptstyle \left(\bar{p}_1(\gamma+1)+\bar{p}_2(\gamma-1)\right),
    \end{pmatrix},   
\end{equation*}
\begin{equation*}
    \mathbf{\xi} = 
    \begin{pmatrix}
    \bar{\rho}_2 - \bar{\rho}_1 \\
    0 \\
    0 \\
    0 \\
    0
    \end{pmatrix},\
    \mathbf{\zeta} = 
    \begin{pmatrix}
    (\bar{\rho}_2 - \bar{\rho}_1)\bar{v} \\
    0 \\
    \bar{u_1} - \bar{u_2} \\
    0 \\
    0
    \end{pmatrix},\
    \mathbf{\beta} = 
    \begin{pmatrix}
    (\bar{\rho}_1 - \bar{\rho}_2)\bar{w} \\
    0 \\
    0 \\
    \bar{u_2} - \bar{u_1} \\
    0
    \end{pmatrix},
\end{equation*}
and $\mathbf{Z_i} = \left[p'_i, u'_i, v'_i, w'_i, \rho'_i\right]^T$ is the perturbation state vector on either side of the shock.  In the frequency domain, Eq.  \ref{eq: linRH} constrains the perturbations upstream and downstream of the shock together with the unsteady displacement of the shock.  While this is sufficient to perform frequency response analysis (see e.g., ~\cite{McKenzie1968,Robinet2001}), Eq.  \ref{eq: linRH} also may be applied in the time-domain as a boundary condition for simulations which require accurate transmission of linear perturbations through shocks. Because our overall approach relies upon extracting Jacobians numerically from time-dependent simulations, we choose to develop the time-domain version of this shock-kinematic boundary condition (SKBC). To meet this goal, it is helpful to think in terms of characteristics in the direction normal to the shock rather the primitive perturbation variables.  The perturbation state vector is related to the characteristics normal to the shock by
\begin{equation}
    \mathbf{Z_i} = \mathbf{Q_i} \Phi_i,
\end{equation}
where
\begin{equation}
    \mathbf{Q_i} = 
    \begin{pmatrix}
    1 & 0 & 0 & 0 & 1 \\
    -\frac{1}{\bar{\rho}_i \bar{c}_i} & 0 & 0 & 0 & \frac{1}{\bar{\rho}_i \bar{c}_i} \\
    0 & 0 & 1 & 0 & 0 \\
    0 & 0 & 0 & 1 & 0 \\
    \frac{1}{\bar{c}^2_i} & 1 & 0 & 0 & \frac{1}{\bar{c}^2_i}
    \end{pmatrix},\
    \Phi_i = 
    \begin{pmatrix}
    \phi^-_i \\
    \phi^s_i \\
    \phi^v_i \\
    \phi^w_i \\
    \phi^+_i
    \end{pmatrix}.    
\end{equation}
Here, $\phi^-_i$ and $\phi^+_i$ are the amplitudes of the slow and fast acoustic waves, respectively, and $\phi^s_i$, $\phi^v_i$, and $\phi^w_i$ are the amplitudes of the entropy and vorticity waves.  Upstream of the shock, all five waves are traveling downstream, and so are impinging on the shock.  Downstream of the shock, the flow normal to the shock is subsonic, and so the slow acoustic wave travels toward the shock, while the other four waves travel downstream, away from the shock.  In total, there are four outgoing characteristics and six incoming characteristics with respect to the shock. In the time-domain, we solve for the four post-shock outgoing characteristics $\phi^s_2, \phi^v_2, \phi^w_2, \phi^+_2$ and the time rate-of-change $X_t$ of the shock position in terms of the six incoming characteristics $\phi^-_1, \phi^s_1, \phi^v_1, \phi^w_1, \phi^+_1, \phi^-_2$ and the local shock inclinations $X_y$ and $X_z$.  Explicitly splitting $\Phi_2 = \Phi^+_2 + \Phi^-_2$, we rewrite Eq.  \ref{eq: linRH} in terms of characteristics as
\begin{equation}
    \mathbf{A_2 Q_2} \Phi^+_2 - \xi X_t = \mathbf{A_1 Q_1} \Phi^+_1 - \mathbf{A_2 Q_2} \Phi^-_2  + \zeta X_y + \beta X_z.
\end{equation}
Because
\begin{equation}
    \Phi^+_2 = 
    \begin{pmatrix}
    0 \\
    \phi^s_2 \\
    \phi^v_2 \\
    \phi^w_2 \\
    \phi^+_2
    \end{pmatrix},\
    \xi X_t = 
    \begin{pmatrix}
    (\bar{\rho}_2 - \bar{\rho}_1) X_t \\
    0 \\
    0 \\
    0 \\
    0
    \end{pmatrix},
\end{equation}
we define a new matrix $\mathbf{A_3}$ where the first column of $\mathbf{A_2 Q_2}$ is replaced by $\xi$ and then solve the linear problem 
\begin{equation*}
    \mathbf{A_3} \Psi = \mathbf{A_1 Q_1} \Phi^+_1 - \mathbf{A_2 Q_2} \Phi^-_2 + \zeta X_y + \beta X_z,
\end{equation*}
for $\Psi$, where 
\begin{equation}
    \Psi = 
    \begin{pmatrix}
    X_t \\
    \phi^s_2 \\
    \phi^v_2 \\
    \phi^w_2 \\
    \phi^+_2
    \end{pmatrix}.
\end{equation}

%

\end{document}